\documentclass[letterpaper,11pt]{article}

\usepackage{jheppub}
\usepackage{subfig}
\usepackage{diagbox}

\def\Fig#1{fig.~{\ref{#1}}}

\DeclareRobustCommand{\Sec}[1]{sec.~\ref{#1}}
\DeclareRobustCommand{\Eq}[1]{eq.~(\ref{#1})}

\newcommand{\ga}{\gamma}

\newcommand{\de}{\delta}

\newcommand{\si}{\sigma}

\def\supnu{{[n]}}
\def \as {\relax\ifmmode\alpha_s\else{$\alpha_s${ }}\fi}

\newcommand{\df}{\mathrm{d}}
\newcommand{\nn}{\nonumber}

\allowdisplaybreaks[2]

\definecolor{darkgreen}{rgb}{0.13,0.55,0.13}
\definecolor{MJpurple}{rgb}{0.7,0.4,1}

\begin{document}

\title{Energy Correlators on Tracks:\\ Resummation and Non-Perturbative Effects}

\author[1,2]{Max Jaarsma,}
\author[3]{Yibei Li,}
\author[4]{Ian Moult,}
\author[1,2]{Wouter Waalewijn,}
\author[5]{Hua Xing Zhu}
\affiliation[1]{Nikhef, Theory Group,
	Science Park 105, 1098 XG, Amsterdam, The Netherlands}
\affiliation[2]{Institute for Theoretical Physics Amsterdam and Delta Institute for 
 Theoretical Physics, University of Amsterdam, Science Park 904, 1098 XH Amsterdam, The Netherlands}
\affiliation[3]{Zhejiang Institute of Modern Physics, Department of Physics, Zhejiang University, Hangzhou, Zhejiang 310027, China}
\affiliation[4]{Department of Physics, Yale University, New Haven, CT 06511, USA\vspace{0.5ex}}
\affiliation[5]{School of Physics, Peking University, Beijing 100871, China}

\emailAdd{m.jaarsma@uva.nl}
\emailAdd{yblee777@zju.edu.cn}
\emailAdd{ian.moult@yale.edu}
\emailAdd{w.j.waalewijn@uva.nl}
\emailAdd{zhuhx@pku.edu.cn}

\abstract{Energy correlators measured inside high-energy jets at hadron colliders have recently been demonstrated to provide a new window into both perturbative and non-perturbative Quantum Chromodynamics. 
A number of the most interesting features of these correlators, namely their universal scaling behavior and the ability to image the confinement transition, require precise angular resolution, necessitating the use of tracking information in experimental measurements.
Theoretically, tracking information can be incorporated into the energy correlators using track functions, which are non-perturbative functions describing the fragmentation of quarks and gluons into charged hadrons.  
In this paper, we apply our recently developed track function formalism to energy correlators, and study in detail the interplay of track functions with perturbative resummation and non-perturbative power corrections. 
We provide resummed results for the energy correlators at collinear next-to-leading-logarithmic accuracy and compare with parton shower Monte Carlo simulations.  
For the two-point correlator the use of tracking has a minimal effect throughout the entire distribution, but it has a significant effect for higher point correlators.
Our results are crucial for the theoretical interpretation of recent experimental measurements of the energy-energy correlators.
}

\maketitle

%%%%%%%%%%%%%%%%%%%%%%
\section{Introduction}\label{sec:intro}
%%%%%%%%%%%%%%%%%%%%%%

High-energy collider experiments provide unique insights into the microscopic dynamics of the Standard Model, particularly Quantum Chromodynamics (QCD). The solutions to many long-standing problems, ranging from the dynamics of confinement, the precise mass of the top quark, to the phase structure of the quark gluon plasma (QGP), are encoded in the asymptotic energy flux at colliders. Decoding this energy flux, and relating it to the microscopic properties of the underlying quantum field theory, allows one to maximally utilize the experimental data. This is similar to the situation in cosmology, where we learn about early-time microscopic physics from the cosmic microwave background, or large-scale structure.

One of the main developments at the LHC has been the ability to study the detailed structure of energy flow within high-energy jets, a field referred to as jet substructure  \cite{Larkoski:2017jix,Kogler:2018hem,Marzani:2019hun}. A key recent advance in this area has been the introduction of the energy correlator observables \cite{Basham:1979gh,Basham:1978zq,Basham:1978bw,Basham:1977iq} to jet substructure \cite{Dixon:2019uzg,Chen:2020vvp,Lee:2022ige}. This has opened the door to the use of sophisticated theoretical techniques developed in the study of conformal field theories (CFTs) \cite{Hofman:2008ar,Kologlu:2019mfz}, and provided new ways to extract physics from jets. For recent phenomenological applications, see refs.~\cite{Komiske:2022enw,Holguin:2022epo,Craft:2022kdo,Andres:2022ovj,Andres:2023xwr,Devereaux:2023vjz,Liu:2022wop,Liu:2023aqb,Cao:2023rga,Chen:2022swd,Chen:2019bpb}. The energy correlators were first studied inside jets at hadron colliders using Open Data \cite{Komiske:2022enw}, and were recently measured at both ALICE and STAR \cite{talk1,talk2,talk3}. 

A key aspect in the ability to precisely measure jet substructure observables in the hadron collider environment has been the use of tracking information. For applications of tracking to jet substructure see e.g.~\cite{STDM-2010-14,STDM-2014-17,STDM-2017-16,STDM-2017-33,STDM-2018-57} from ATLAS, \cite{CMS-SMP-20-010,CMS:2023ovl} from CMS, \cite{ALargeIonColliderExperiment:2021mqf,ALICE:2019ykw,ALICE:2021njq,ALICE:2021aqk} from ALICE and \cite{LHCb:2017llq,LHCb:2019qoc} from LHCb.  Tracks provide much better angular resolution, and will be key to maximizing the physics potential of energy correlators in jet substructure. It is therefore crucial to understand how the use of tracking information modifies the structure of the energy correlator observables. This will ensure that they are correctly interpreted, both for qualitative studies, and ultimately for precision measurements.

Tracks can be incorporated into theoretical calculations using the track function formalism \cite{Chang:2013rca,Chang:2013iba}. While track functions were introduced nearly a decade ago, it is only within the last year that they have become practical for jet substructure calculations. This is due both to an improved understanding of the non-linear structure of the renormalization group (RG) evolution equations \cite{Li:2021zcf,Jaarsma:2022kdd,Chen:2022pdu,Chen:2022muj}, and to the development of energy correlator observables which are only sensitive to (a few) moments of the track functions, enabling them track functions to be interfaced with higher order perturbative calculations \cite{Chen:2020vvp,Li:2021zcf}. Despite this recent progress, the improved understanding of the track function formalism has only been applied to event-based observables for $e^+e^-$, see in particular early work on thrust \cite{Chang:2013iba}, and not yet to jet substructure observables. This is particularly timely due to the recent measurements of the energy correlators \cite{talk1,talk2,talk3} using tracks.

In this paper we apply these recent advances in the track function formalism to study the collinear limit of the energy correlator observables measured on tracks. We focus in particular on the multi-point projected energy correlators \cite{Chen:2020vvp}, including both their perturbative, and non-perturbative aspects. We find that the two-point energy correlator is almost identical (up to an overall normalization) whether measured on tracks or all particles, throughout the entire distribution. This enables a clean interpretation of recent measurements of the confinement transition on tracks. By contrast, the (projected) higher-point correlators differ by a factor that has a non-trivial scale dependence. In the perturbative region, the scaling behavior of energy correlators is modified by the logarithmic running of the track functions, which can be computed systematically in perturbation theory.  We present resummed results for the scaling behavior of the projected energy correlators, and also extract the leading non-perturbative corrections to these observables. We will see that  the factorization theorem for the energy correlator on tracks takes a simple form, depending only on moments of the track functions, illustrating that the track function formalism can be combined with factorization formulae in a systematically improvable manner.

The outline of this paper is as follows: In \Sec{sec:fact_theorem}, we review the energy correlator observables, the track function formalism, and present a factorization theorem for the track-based energy correlators in the collinear limit. We provide a general overview of differences between track-based energy correlators and all-hadron energy correlators in \Sec{sec:fact_pheno}. We believe that the conclusions of this section are particularly important for the interpretation of recent and forthcoming experimental measurements of the energy correlators on tracks. In \Sec{sec:resum}, we study perturbative resummation for the energy correlators in the collinear limit with tracks, showing how the anomalous dimensions of the track functions slightly modify their scaling behavior. In \Sec{sec:NP}, we investigate the structure of non-perturbative power corrections for track-based energy correlators, and show an interesting relation between the leading non-perturbative power correction for the energy correlators with and without tracks. We conclude in \Sec{sec:conc}.

%%%%%%%%%%%%%%%%%%%%%%
\section{Energy Correlators, Tracks, and Factorization}\label{sec:fact_theorem}
%%%%%%%%%%%%%%%%%%%%%%  

In this section we review both the energy correlator observables and the track function formalism. We then combine them to present a factorization theorem describing the small-angle limit of the energy correlator measured on tracks. Apart from the factorization theorem, this section is primarily a review, and can be skipped by those familiar with these prerequisites.

%%%%%%%%%%%%%%%%%%%%%%
\subsection{Review of Energy Correlators}\label{sec:review_correl}
%%%%%%%%%%%%%%%%%%%%%%  

Energy correlators are a class of observables that probe correlations in the asymptotic energy flux in collider experiments. The two-point energy correlator was first defined in $e^+e^-$ collisions as~\cite{Basham:1978bw,Basham:1978zq}
%%%
\begin{align}
  \label{eq:EECdef}
  \frac{\df \sigma}{\df z}= \sum_{i,j}\int \df \sigma\ \frac{E_i E_j}{Q^2}\, \delta\Bigl(z - \frac{1 - \cos z_{ij}}{2}\Bigr) \,,
\end{align}
%%%
where $z_{ij}$ is the angle between particles $i$ and $j$.
This definition can be straightforwardly generalized to multi-point correlation functions \cite{Chen:2019bpb,Yan:2022cye,Yang:2022tgm}.  From a field-theoretic point of view, the energy correlators are particularly convenient since they can be formulated in terms of matrix elements of energy flow operators \cite{Sveshnikov:1995vi,Tkachov:1995kk,Korchemsky:1999kt,Bauer:2008dt,Hofman:2008ar,Belitsky:2013xxa,Belitsky:2013bja,Kravchuk:2018htv} 
%%%
\begin{align}\label{eq:ANEC_op}
\mathcal{E}(\hat n) = \lim_{r\to \infty}  \int_0^\infty \df t\, r^2\, n^i T_{0i}(t,r \hat n)\,,
\end{align}
%%%
where $n^\mu = (1,\hat n)$ is a light-like vector.
These operators are also referred to as ANEC (Average Null Energy Condition), or lightray operators \cite{Kravchuk:2018htv} in the literature. Using this formulation of the energy correlators, it was first conjectured in ref.~\cite{Hofman:2008ar} that energy correlators should exhibit universal scaling behavior in the small-angle limit. This has since been rigorously proven in CFTs~\cite{Kologlu:2019mfz,Chang:2020qpj}, and within perturbation theory in QCD~\cite{Chen:2020adz,Chen:2021gdk}. Furthermore, it has been observed inside high-energy jets at the LHC \cite{Komiske:2022enw,talk1,talk2,talk3}.

In ref.~\cite{Chen:2020vvp}, it was shown how the scaling behavior of higher-point correlators can be extracted in an experimentally convenient way using the ``projected $N$-point energy correlators". The projected $N$-point energy correlators are defined by integrating out the shape information of higher-point correlators, remaining differential only in the largest angular size, $x_L$. In terms of a measurement on particles, the projected energy correlators are defined as
%%%
\begin{align}
  \label{eq:projected_mom}
  \frac{\df \sigma^{[N]}}{\df x_L} & = \sum_m\sum_{1\leq j_1,\ldots, j_N \leq m} 
\int\! \df \sigma_{e^+e^- \to X_m}
\frac{\prod_{k=1}^N E_{j_k}}{Q^N}
\, \delta\bigl(x_L - \max \{z_{j_1 j_2}, z_{j_1 j_3}, \ldots , z_{j_{N-1} j_N } \}\bigr) \,.
\end{align}
%%%
Here $X_m$ denotes a $m$-particle final state, $j_k$ numbers the particles, $E_{j_k}$ is the particle energy and 
$z_{jk} = (1 - \vec{n}_j \cdot \vec{n}_k)/2 = (1 -  \cos\theta_{jk})/2$ is the two-particle angular distance. The understanding of these observables in the small-angle limit, when measured on tracks, will be the focus of this paper.

%%%%%%%%%%%%%%%%%%%%%%
\subsection{Review of Track Functions}\label{sec:review_track}
%%%%%%%%%%%%%%%%%%%%%%  

Measurements at hadron colliders often use tracking information to improve their resolution. Since track-based measurements are sensitive to quantum numbers (other than energy) of the final-state hadrons, they are not infrared and collinear safe \cite{Kinoshita:1962ur,Lee:1964is} and cannot be calculated purely in perturbation theory. In refs.~\cite{Chang:2013rca,Chang:2013iba}, a factorization theorem for track-based observables was developed in terms of a universal non-perturbative function called a ``track function". Track functions allow for a systematic separation of perturbative and non-perturbative physics for jet observables measured on tracks.

Loosely speaking, the track functions, $T_q$ ($T_g$) describe the energy fraction of all hadrons with some property, $R$, arising from the fragmentation of a quark (gluon). The most used experimental case is when $R$ is the set of electrically-charged hadrons. The track functions are defined as
%%%
\begin{align} \label{T_def}
T_q(x)&=\!\int\! \df y^+ \df ^{d-2} y_\perp e^{ik^- y^+/2} \sum_X \delta \biggl( x\!-\!\frac{P_R^-}{k^-}\biggr)  \frac{1}{2N_c}
\text{tr} \biggl[  \frac{\gamma^-}{2} \langle 0| \psi(y^+,0, y_\perp)|X \rangle \langle X|\bar \psi(0) | 0 \rangle \biggr]\,,
 \\
T_g(x)&=\!\int\! \df y^+ \df^{d-2} y_\perp e^{ik^- y^+/2} \sum_X \delta \biggl( x\!-\!\frac{P_R^-}{k^-}\biggr) \frac{-1}{(d\!-\!2)(N_c^2\!-\!1)k^-}
 \langle 0|G^a_{- \lambda}(y^+,0,y_\perp)|X\rangle \langle X|G^{\lambda,a}_- (0)|0\rangle. 
\nn \end{align} 
%%%
The matrix elements describe the production of an unpolarized quark or gluon, whose large light-cone momentum component $k^-$ is fixed by the Fourier transform of $y^+$. The delta function encodes the measurement of the momentum fraction $x$, with $P_R$ the momentum of the states with quantum number $R$. The above definition is valid in light-cone gauge and in general Wilson lines need to be included.
Note that, compared with standard fragmentation functions, track functions describe the total energy fraction carried by \emph{all} hadrons with property $R$, instead of a single hadron. This enables them to be used to compute correlations on the energy flux of the hadrons with property $R$. This also means that they contain in their RG evolution equations all the $N$-hadron fragmentation function evolution equations, as shown in refs.~\cite{Chen:2022pdu,Chen:2022muj}.

Much like standard fragmentation functions, track functions are non-perturbative, but exhibit a perturbatively calculable RG. Due to the fact that they are sensitive to the energy fraction on all hadrons of type $R$, this RG is non-linear. This has required developing techniques to go beyond the standard DGLAP~\cite{Gribov:1972ri,Dokshitzer:1977sg,Altarelli:1977zs} paradigm, both for computing, and solving the evolution equations. The evolution equation for the track functions is expressed in terms of perturbatively calculable kernels, $K_{i\to i_1 \cdots i_k}$, which describe the mixing between a track function $T_i$ and a product of track functions $T_{i_1} \cdots T_{i_k}$. These kernels were computed at next-to-leading order (NLO) in refs.~\cite{Li:2021zcf,Jaarsma:2022kdd,Chen:2022pdu,Chen:2022muj}, and the RG evolution to this order reads
%%%
\begin{align}
 \!\!\!\frac{\df }{\df \ln\mu^2}T_i(x) &= a_s \Bigl(
 K^{(0)}_{i\to i} T_i(x)  +
[K^{(0)}_{i\to i_1i_2}\otimes T_{i_1}  T_{i_2}](x) \Bigr)
  \\ & \quad
 +
a_s^2\Bigl(
K^{(1)}_{i\to i} T_i(x) 
+
[K^{(1)}_{i\to i_1i_2}\otimes T_{i_1}  T_{i_2}](x)
    + [K^{(1)}_{i\to i_1i_2i_3}\otimes T_{i_1} T_{i_2} T_{i_3}](x)
  \Bigr) + \mathcal{O}(a_s^3)\,,
\nn\end{align}
%%%
where $a_s = \alpha_s(\mu)/(4\pi)$ is the coupling and $i,i_k$ denote quarks and gluon. Here one clearly sees the non-linear evolution governed by the kernels $K$. The explicit form of these equations can be found in refs.~\cite{Li:2021zcf,Jaarsma:2022kdd,Chen:2022pdu,Chen:2022muj}, and a numerical code solving these integro-differential RG equations was provided in refs.~\cite{Chen:2022pdu,Chen:2022muj}, enabling their use in phenomenology. 

A particularly convenient aspect of the energy correlator observables that we will study is that they are only sensitive to the low moments of the track functions. Therefore, instead of requiring a non-perturbative \emph{function}, they only require a handful of non-perturbative \emph{numbers}. Furthermore, the moments of the track function satisfy simpler RG equations than the full function. We note that for track-based observables there is no further complication when modifying the energy weighting in \eqref{eq:projected_mom} from e.g.~$E_{j_k} \to E_{j_k}^2$. This change was exploited in ref.~\cite{Holguin:2022epo} to suppress the contribution from soft radiation. However, when all particles are summed over, this change causes the measurement to no longer be infrared- and collinear safe. 

We define the $n$-th moment of the track function $T_i$ as
%%%
\begin{align} \label{eq:T_mom}
T_i(n,\mu)=\int_0^1 \df x~ x^n~ T_i (x,\mu)\,
\end{align}
%%%
where $a$ denotes parton flavors. 
The zeroth moment is fixed by the normalization condition
%%%
\begin{align}\label{T0}
T_i(0,\mu)=1\,,
\end{align}
%%%
and as such is scale independent. Higher moments mix with products of lower moments, exhibiting a non-linear evolution. Explicitly, we have
%%%
\begin{align} \label{T_evo_n}
  \frac{\df}{\df \ln \mu^2} T_i(n) &= \sum_N \sum_{\{i_k\}} \sum_{\{m_k\}} \ga_{i \to \{i_k\}}(\{m_k\})\,  \prod_{j=1}^N T_{i_j}(m_j,\mu) \,,
  \nn \\
  \ga_{i \to \{i_k\}}(\{m_k\}) &= \begin{pmatrix} & n & \\ m_1 & m_2 & \cdots \end{pmatrix}
  \biggl[\prod_{j=1}^N \int_0^1\! \df z_j\, z_j^{m_j}  \biggr]\de\Bigl(1 - \sum_{j=1}^N z_j\Bigr) P_{i \to \{i_k\}}(\{z_k\})
\,.\end{align}
%%%
Here, the sum of the moments of the track functions on the right-hand side must equal $n$, i.e.~$\sum_k m_k = n$. Explicit anomalous dimensions for the first six moments, which are needed to obtain our numerical results, can be found in refs.~\cite{Li:2021zcf,Jaarsma:2022kdd}.

%%%%%%%%%%%%%%%%%%%%%%
\subsection{Energy Correlators on Tracks in the Collinear Limit}\label{sec:fact_collinear}
%%%%%%%%%%%%%%%%%%%%%%  

The track function formalism provides the necessary non-perturbative matrix elements to achieve factorization for energy correlators measured on a subset $R$ of hadrons. In ref.~\cite{Li:2021zcf} this was used to calculate the two-point energy correlator in fixed-order perturbation theory on tracks, where it was shown that the track function formalism interfaces nicely with perturbative calculations. In the small-angle limit of the energy correlator perturbative resummation of logarithms of $z$ is required, which in turn requires a factorization formula. A factorization formula describing this limit was derived in conformal field theories~\cite{Kologlu:2019mfz,Korchemsky:2019nzm}, and for general field theories~\cite{Dixon:2019uzg}. Moreover, the factorization formula was extended to general $N$-point projected correlators in ref.~\cite{Chen:2020vvp}. Here we  incorporate the effects of performing these measurements on tracks, showing that the track function formalism also interfaces nicely with factorization theorems.

There are two approaches to deriving a factorization theorem for the projected correlators on tracks in the small angle limit. One way is to start with a factorization theorem for the general angle energy correlator on tracks. This factorization into perturbatively calculable matrix elements and track functions was used to perform the fixed-order calculation of the energy correlator on tracks in ref.~\cite{Li:2021zcf}. The perturbative matrix elements appearing in this factorization theorem can then be refactorized in the small angle limit. Alternatively, since the track function formalism applies to any IRC safe observable, in particular jet functions, one can first apply collinear factorization to derive a factorization into a jet and hard function. The jet function can then be factorized into a perturbative component and a non-perturbative component described by  moments of the track functions. The two approaches result in the same factorization theorem. However, the second approach allows for a more uniform notation for all $n$-point projected correlators, so we will follow this approach here.

We follow the factorization for the small angle limit of the energy correlator presented in ref.~\cite{Dixon:2019uzg}, and generalized to $n$-point correlators in ref.~\cite{Chen:2020vvp}. Since the use of tracks only modifies the infrared behavior of the measurement, it does not modify the hard-collinear factorization. We can therefore factorize the cross section into an inclusive hard matching coefficient that depends on the source, and a jet function that describes the measurement of the energy correlators on a highly boosted quark or gluon state.

To avoid the use of distributions, it is convenient to work in terms of the cumulant of the projected $n$-point correlator
%%%
\begin{align}
  \label{eq:cumulant}
\Sigma^{[n]}\Bigl(x_L, \ln \frac{Q^2}{\mu^2} , \mu\Bigr)\
\equiv\ \frac{1}{\sigma_0}
\int^{x_L}_0 \df z' \, 
\frac{d\sigma^{[n]}}{dx_L} \Bigl(z', \ln\frac{Q^2}{\mu^2}, \mu\Bigr) \,.
\end{align}
%%%
Here the superscript $n$ denotes that this is an $n$-point projected correlator, and $\si_0$ is the Born cross section.
The hard-collinear factorized expression for the cumulant is 
%%%
\begin{align}
  \label{eq:fac_nu}
  \Sigma^{[n]} \Bigl( x_L, \ln\frac{Q^2}{\mu^2},\mu \Bigr) = \int_0^1 \! \df x\, 
x^n \vec{J}\,^{[n]} \Bigl(\ln \frac{x_L x^2 Q^2}{\mu^2},\mu \Bigr) \cdot \vec{H}\Bigl(x, \ln\frac{Q^2}{\mu^2},\mu\Bigr) \,,
\end{align}
%%%
where $\vec H$ is the standard hard function for inclusive fragmentation, and $\vec J^{[n]}$ is the jet function, which contains the dependence on the energy correlator measurement. Both $\vec H$ and $\vec J^{[n]}$ are vectors in flavor space, i.e. $\vec H=\{ H_q, H_g\}^\text{t}$ and $\vec{J}^{[n]}=\{J^{[n]}_q,J^{[n]}_g\}$ with $\text{t}$ giving the usual transpose of a matrix. The hard function obeys the RG equation
%%%
\begin{align}
  \label{eq:hard_evo}
  \frac{\df \vec{H}\bigl(x, \ln\frac{Q^2}{\mu^2},\mu\bigr)}{\df \ln \mu^2}
= - \int_x^1\! \frac{\df y}{y}\, \widehat{P}(y,\alpha_s) \cdot \vec{H} \Bigl( \frac{x}{y},
\ln\frac{Q^2}{\mu^2},\mu\Bigr) \,,
\end{align}
%%%
where $\widehat{P}(y,\alpha_s)$ is the singlet timelike splitting matrix $\{\{P_{qq},2 n_f P_{qg}\},\{P_{gq},P_{gg}\}\}$. 
The hard-collinear factorization in  \eqref{eq:fac_nu} implies that overall RG evolution of the jet function is DGLAP
%%%
\begin{align}
  \label{eq:jet_evo}
  \frac{\df \vec{J}\,^{\supnu} \bigl( \ln \frac{x_L Q^2}{\mu^2},\mu \bigr)}{\df \ln\mu^2}
=
\int_0^1 \! \df y\, y^n \vec{J}\,^{\supnu} \Bigl( \ln \frac{x_L y^2 Q^2}{\mu^2},\mu\Bigr) 
\cdot 
\widehat{P}(y,\alpha_s) \,,
\end{align}
%%%
due to renormalization group consistency, regardless of whether the measurement is performed on tracks or not.

Since the jet functions are IRC safe, we can use the track function formalism to factorize them into perturbatively calculable coefficients multiplying moments of the track function.  To write the jet functions, we follow the notation of ref.~\cite{Jaarsma:2022kdd}, where we define a vector ${\bf T}_n$ of all the products of track function moments of a fixed total weight $n$. For example, for $n=2$, $ {\bf T}_2=\{T_g(2),T_q(2), T_q(1)T_q(1), T_g(1)T_q(1),$ $T_g(1) T_g(1) \}$, where for simplicity of notation, we consider the case in which quarks with different flavors have the same track function $T_q$. We also make the assumption $T_q=T_{\bar q}$, which is true for track functions of electric charge, but not for other quantum numbers. 

We can now write the jet function of the projected $n$-point energy correlator on tracks as
%%%
\begin{align}
J_i^{[n]}=  {\bf{j}}_i^{[n]} \cdot  {\bf T}_n\,,
\end{align}
%%%
where ${\bf{j}}_i^{[n]}$ are perturbatively calculable coefficients. Note that for each parton index $i$, these coefficients are vectors in the track function space, which we will indicate by writing them in bold. 
This provides a rigorous factorization theorem for the small angle limit of the energy correlators, separating perturbative from non-perturbative physics. As compared to previous factorization theorems for track-based observables \cite{Chang:2013iba}, it involves only the moments of the track function, which are numbers, as opposed to functions. 

To illustrate the perturbative calculation of the matching coefficients in the energy correlator jet functions, we present the results for the $n$-point correlators ($n=2,3,4,5,6$) at one-loop order. These coefficients will be used in our next-to-leading logarithmic (NLL) resummation for the energy correlators. We write the jet function constants (i.e.~the non-logarithmic terms, as indicated by the subscript 0) as 
%%%
\begin{align}
J_{i,0}^{\supnu}\bigl(\alpha_s(\mu), {\bf T}_n(\mu)\bigr)=\sum_k \Bigl(\frac{\alpha_s(\mu)}{4\pi}\Bigr)^k J_{i,0}^{\supnu,(k)}\bigl( {\bf T}_n(\mu)\bigr)\,,
\end{align}
%%%
where, as before, $i=q,g$ is a flavor index, the superscript $n$ denotes that this is the jet function for an $n$-point projected correlator, and $k$ denotes the order in the $\alpha_s$ expansion. The arguments show that the $\mu$-dependence arises from the running coupling as well as the track functions. (More complete expressions are given in \Sec{sec:solve}.). The logarithmic terms can be derived from the renormalization group evolution of the $ {\bf{j}}_i^{[n]}$, which are fixed by the known RGs of $ \vec{J}^{[n]}$ and $ {\bf T}_n$; then, with the up to order-$\alpha_s^k$ jet function constant, beta function and DGLAP splitting kernels, as well as the track function evolution kernels (in Mellin space) up to $(k+1)$-loop order, one can achieve the (next-to-)$^k$leading logarithmic (N$^k$LL for short) accuracy for the track jet function, which is the essential and sufficient ingredient for the N$^k$LL energy correlators on tracks with the corresponding hard function. We will use this fact in \Sec{sec:solve} to perform the perturbative resummation in the small-angle limit. 

Without accounting for tracks, the one-loop jet function constants were presented in ref.~\cite{Chen:2020vvp} for any $\nu$-point energy correlator, while the two-loop constants for the 2-point correlator were given in ref.~\cite{Dixon:2019uzg}, and for the 3-point correlator in ref.~\cite{Chen:2023zlx}. With tracks, following the normalization convention in Ref.~\cite{Chen:2020vvp}, 
the first order constants read
%%%
\begin{align}\label{eq:initial_J00}
    J_{i,0}^{[n],(0)}=T_i(n)\frac{1}{2^n}\,,
\end{align}
%%%
At the one-loop order, we obtain for quarks
\begin{align}\label{jq2vsqbar}
J_{q,0}^{[2],(1)}=&\,-C_F\frac{37}{12} T_g(1)T_q(1)\,,\nn\\
J_{q,0}^{[3],(1)}=&\,-C_F\biggl(\frac{611}{1200}T_g(2) T_q(1)
+\frac{541}{300} T_g(1) T_q(2)\biggr)\,,\nn\\
J_{q,0}^{[4],(1)}=&\,-C_F\biggl(
\frac{89}{600} T_g(3)T_q(1) + \frac{43}{150}T_g(2)T_q(2) + \frac{91}{90}T_g(1)T_q(3)
\biggr)
\,,\nn\\
J_{q,0}^{[5],(1)}=&\,-C_F\biggl(
\frac{839}{15680}T_g(4) T_q(1)
+\frac{461}{5880}T_g(3) T_q(2)
+\frac{2831}{17640}T_g(2) T_q(3)
%\nn\\
%&
+\frac{811}{1470}T_g(1) T_q(4)
\biggr)
\,,\nn\\
J_{q,0}^{[6],(1)}=&\,-C_F\biggl(
T_g(5) T_q(1)\frac{629}{29400}
+T_g(4) T_q(2)\frac{2519}{94080}
+T_g(3) T_q(3)\frac{3013 }{70560}
+T_g(2) T_q(4)\frac{8311}{94080}
\nn\\
&
+T_g(1)T_q(5)\frac{1987}{6720}
\biggr)
\,,
\end{align}
and for the case of gluons, 
\begin{align}
J_{g,0}^{[2],(1)}=&\,
-C_A \frac{449}{150}T_g(1) T_g(1)
-n_fT_F \frac{7}{25}T_q(1)T_{q}(1)\,, \nn\\
J_{g,0}^{[3],(1)}=&\,
-C_A
\frac{449}{200} T_g(2)T_g(1)
-n_f T_F 
\frac{21}{100} T_q(2)T_q(1)\,,
\nn\\
J_{g,0}^{[4],(1)}=&\,
-C_A\biggl(
\frac{8293}{7350} T_g(3)T_g(1)
+\frac{1083}{3920}T_g(2)T_g(2)
\biggr)
-n_f T_F\biggl(
\frac{1028}{11025} T_q(3)T_q(1)\nn\\
&
+\frac{1031}{29400}T_q(2)T_q(2)
\biggr)
\,,
\nn\\
J_{g,0}^{[5],(1)}=&\,
-C_A\biggl(
\frac{27757}{47040} T_g(4)T_g(1)
+\frac{361}{1568} T_g(3)T_g(2)
\biggr)
-n_fT_F\biggl(
\frac{1027}{23520} T_q(4)T_q(1)\nn\\
&
+\frac{1031}{35280} T_q(3)T_q(2)
\biggr)
\,,
\nn\\
J_{g,0}^{[6],(1)}=&\,
-C_A\biggl(
\frac{81931}{264600} T_g(5)T_g(1)
+\frac{93991}{846720} T_g(4)T_g(2)
+\frac{26107}{635040}T_g(3)T_g(3)
\biggr)
\nn\\
&
-n_f T_F\biggl(
\frac{4409}{211680} T_q(5)T_q(1)
+\frac{29}{2160} T_q(4)T_q(2)
+\frac{719}{127008}T_q(3)T_q(3)
\biggr)
\,.
\end{align}
Here we have suppressed the renormalization scale dependence of the track function moments, and assumed $T_q=T_{\bar{q}}$ for simplicity. The full $T_q$ vs.~$T_{\bar{q}}$ dependence can easily be restored using the charge conjugation invariance of the gluon state, and the quark flavor information can be recovered through changing $n_f$ to $\sum_q$. We see that the only difference  compared to the all-hadron jet function is the appearance of moments of the track functions. However, the matching of the jet functions can still be performed systematically in perturbation theory.

%%%%%%%%%%%%%%%%%%%%%%
\section{Phenomenological Aspects of Track-Based Energy Correlators}\label{sec:fact_pheno}
%%%%%%%%%%%%%%%%%%%%%%  

Our main motivation for studying the structure of the energy correlator observable on tracks is recent (and hopefully future!) experimental measurements of the energy correlators on tracks  \cite{talk1,talk2,talk3}. These measurements focus both on the universal scaling behavior of the energy correlators as a test of precision \emph{perturbative} QCD, as well as on imaging the \emph{non-perturbative} confinement transition. The interpretation of these measurements relies crucially on our understanding of how tracks modify the behavior of the energy correlators. 

We therefore begin by discussing \emph{qualitative} features of the track-based energy correlators in both the perturbative and non-perturbative regimes. We will show that many aspects of the energy correlators on tracks can be understood in terms of basic properties of the moments of the track functions, providing a general view of how the use of tracks modifies the observables. Quantitative studies of the modifications in the perturbative and non-perturbative regions will then be given in the forthcoming sections.

\renewcommand{\arraystretch}{1.5}
\begin{table}
\centering
\begin{tabular}{|c|cccccc|}
\hline
%\diagbox{}{}{}
& 
$T_i(1)$ & 
$T_i(2)$ & 
$T_i(3)$ &
$T_i(4)$ &
$T_i(5)$ &
$T_i(6)$ 
\\ \hline
$T_q(N)$ &
$0.62$ & 
$0.42$ &
$0.29$ &
$0.22$ & 
$0.16$ &
$0.13$ 
\\
$T_g(N)$ & 
$0.62$ & 
$0.40$ &
$0.26$ &
$0.18$ & 
$0.13$ &
$0.090$ 
\\ \hline 
\end{tabular}
\caption{A summary of the numerical values of low moments of the track functions in QCD at $\mu=250$ GeV, derived from the initial condition at $\mu_0=100$~GeV extracted from \textsc{Pythia}~\cite{Chang:2013rca}.  Differences between quarks and gluons appear primarily at higher moments.} \label{tab:mom_summary}
\end{table}
\renewcommand{\arraystretch}{1.0}

%%%%%%%%%%%%%%%%%%%%%%
\subsection{Non-Perturbative Transition}\label{sec:transition}
%%%%%%%%%%%%%%%%%%%%%%  

One of the most interesting observed aspects of the energy correlators is their ability to image the transition to confinement as an abrupt change in their scaling behavior \cite{Komiske:2022enw}. We therefore want to understand if and how this is modified through the use of tracks. This is non-trivial due to the fact that tracking information is itself non-perturbative.

The energy correlators are characterized by a perturbative region, a transition region, and a deeply non-perturbative region. In the perturbative region the track function formalism applies, but will introduce new non-perturbative power corrections that will in general also modify the transition region. In the deeply non-perturbative region, the track function formalism completely breaks down, as it assumes perturbative energy scales. This makes it particularly crucial to understand in detail how the perturbative, deep non-perturbative and transition region are modified by the use of tracks, to ensure that the experimental measurements of the energy correlators on tracks are indeed measuring aspects of the confinement transition in the energy correlators, and not artifacts from the use of tracks.

\begin{figure}
  \begin{center}
    \subfloat[]{
    \includegraphics[scale=0.36]{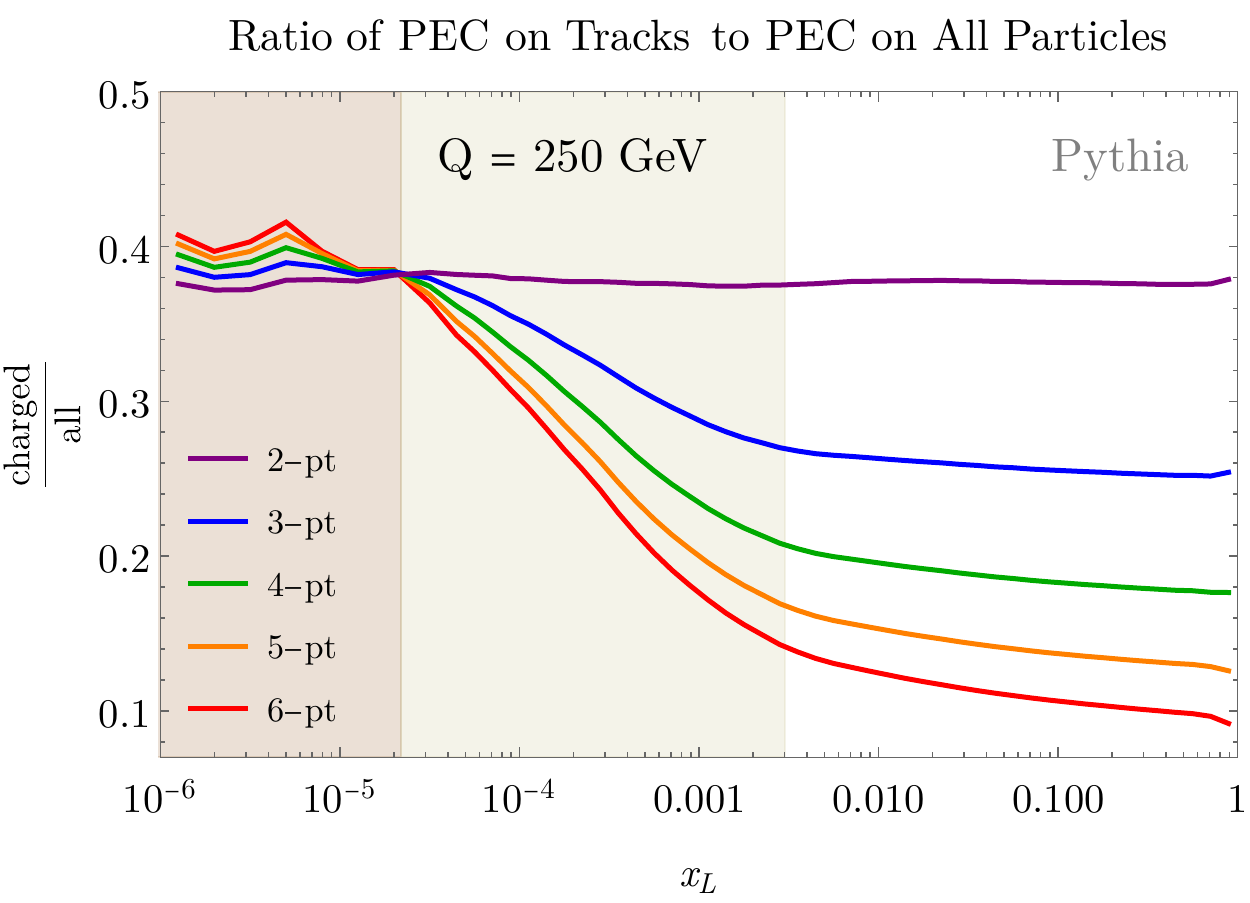}
    \label{fig:ratioNto2_pythia_250_xL_reg_a}
    }
    \subfloat[]{ \includegraphics[scale=0.36]{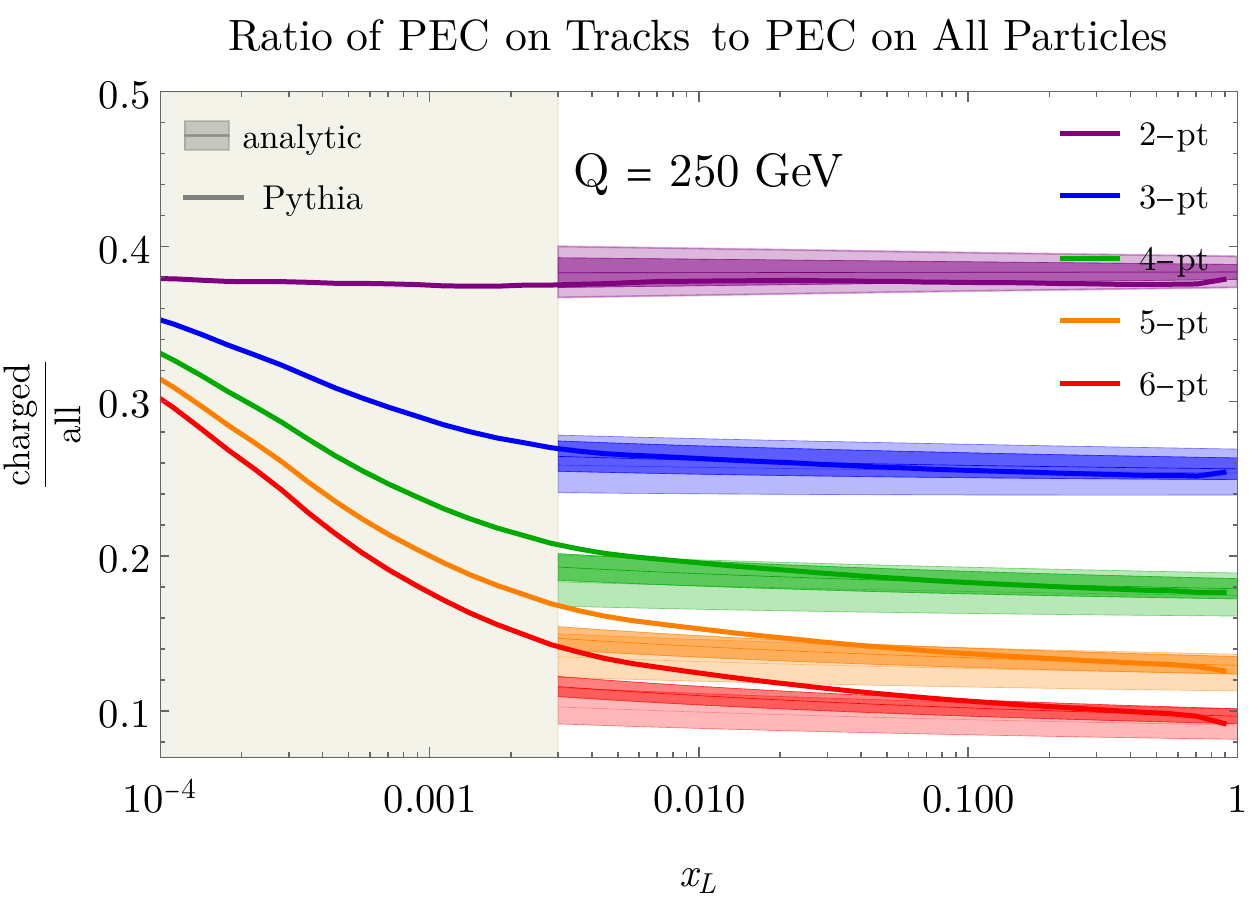}
    \label{fig:ratioNto2_pythia_250_xL_reg_b}
    }
  \end{center}
  \caption{The ratio of the $N$-point projected energy correlators as computed on charged hadrons vs.~all hadrons. (a) Results from \textsc{Pythia} using a c.o.m energy of $Q=250$~GeV and $N=2,3,4,5,6$. The ratio shows a non-trivial transition between the perturbative region (far right), the transition region (middle, shaded), and the deep non-perturbative region (left, dark shaded). (b) Comparison with our analytic results at LL (light shaded) and NLL (dark shaded). The deviation between our analytic results and \textsc{Pythia} as the curves approach the transition region is due to non-perturbative power corrections discussed in \Sec{sec:NP}. }
  \label{fig:ratioNto2_pythia_250_xL_reg}
\end{figure}

To gain a preliminary understanding of the projected energy correlators in the transition from the perturbative to the non-perturbative regime, we use the \textsc{Pythia} Monte Carlo parton shower~\cite{Sjostrand:2014zea,Sjostrand:2007gs} to simulate $e^+e^-$ collisions. The simulation includes hadron decays and we consider all charged particles for our track-based predictions. We will interchangeably use the words hadrons and particles when describing our results, even though we always include non-hadronic particles.
In \Fig{fig:ratioNto2_pythia_250_xL_reg} we plot the ratio of the track-based energy correlators to the all-hadron energy correlators, for two-point up to six-point correlators. In these plots, we see three regions as we go from right to left: A perturbative region, a transition region, and a free hadron region. With the exception of the two-point correlator, this ratio is not flat in any of the three regions (although it is quite flat in the perturbative region due to the slow perturbative running of the track functions, as will be discussed), illustrating that the use of tracking modifies the behavior of the observable. Remarkably, we will be able to understand all the qualitative features of this plot using simple arguments. A quantitative description will then be provided in the next sections.

We first begin with the perturbative region. In this region, the use of tracks only modifies the scaling behavior of the energy correlators through the perturbatively calculable logarithmic scaling of the track functions. This scaling (which we will describe in more detail in \Sec{sec:resum}) is shown in  \Fig{fig:ratioNto2_pythia_250_xL_reg_b}. The low moments of the track function have slow RG evolution \cite{Li:2021zcf,Jaarsma:2022kdd}, so that they only provide a weak modification of the scaling, leading to a very flat ratio in the perturbative regime. The logarithmic scaling of the track functions increases for higher moments, and therefore becomes sizable for higher-point projected correlators. However, it can be reliably computed in perturbation theory. Therefore, apart from the overall normalization, the use of tracks in the perturbative regime provides only a minor, and calculable modification of the scaling behavior. The overall normalization of the ratio in the perturbative regime is governed by the non-perturbative parameter $\sim T(N-1) T(1) $ , for an $N$-point correlator. The values of $T(N)$ are summarized in table~\ref{tab:mom_summary}, and in \Fig{fig:track_evo} and \Fig{fig:track_evo_compare} we show plots of the products of the track functions explicitly\footnote{In these plots we focus on the representative example of $T(N-1)T(1)$. The full perturbative result is a sum of $T(N-k)T(k)$'s ($1\leq k\leq N-1$) and contributions involving more than two track functions. The latter give a smaller contribution, and the numerical sizes of $T(N-k)T(k)$'s for $1< k < N-1$ and $T(N-1)T(1)$ are approximately equal in QCD for the higher point cases, and so these plots should be meant as representative that the size of $T(N-1)T(1)$ sets the behavior of the ratio. In our full perturbative results, all combinations are included with appropriate perturbative coefficients.}. 

We next consider the behavior of the deeply non-perturbative (free-hadron region), and will argue that the ratio of the charged/all hadron correlators in this region should be independent of $N$. This forces there to be a transition region for the ratio between the perturbative and non-perturbative regime. In the free hadron region we can no longer use the track function formalism. Instead, we can imagine a gas of free hadrons, each of which has some probability of being charged. In the real world, this probability is $\approx 2/3$. This is also numerically approximately equal to the first moment of the track function, $T(1)$ (Note that the track functions have scale dependence, so that this relation is simply meant as an approximate numerical relation when the track function is evaluated at a reasonable scale.). In the extreme small angle limit, we should have a small probability of $N>2$ hadrons being together at small angles, and so we expect that the projected energy correlators are dominated by the correlation of two-particles. The value of the ratio in this region is therefore the probability that both of these particles are charged\footnote{We emphasize that in the deep non-perturbative region the track function formalism does not apply. Rather we have a sum over all correlations, which is dominated by pairs of hadrons. The only question is how many of these pairs are charged, which is $\approx (2/3)^2$. It is then convenient for comparing to the perturbative region to express this as $(2/3)^2 \approx T(1)^2$.}, namely $\approx (2/3)^2\approx T(1)^2$. Using the values in Table~\ref{tab:mom_summary}, we see that this agrees well with what is observed in \textsc{Pythia}.

This simple behavior in the deep non-perturbative regime has an interesting consequence for understanding the transition region for the track-based energy correlators: In the perturbative regime the energy correlators are determined by $T(N-k)T(k)$ for $k\geq 1$ and in the non-perturbative region by $T(1)^2$. However, with the exception of $N=2,k=1$, $T(N-k)T(k) \neq T(1)^2$, which means that generically the use of tracks modifies the transition region.  We can now see that to have an approximately flat ratio between track and all hadron based measurements of the $N$-point projected correlators requires the condition 
%%%
\begin{align}
(T(1) )^2\approx T(N-k)T(k),\qquad N\geq 2\,,
\end{align}
%%%
which is equivalent to the condition
%%%
\begin{align}
 \int_0^1 \df x\, \df y\, x y\, T(x)T(y)\approx \int_0^1\! \df x\, \df y\, x^{N-k} y^k\, T(x) T(y)\,, \qquad N\geq 2\,, \ \ 1\leq k \leq N-1.
\end{align}
%%%
This is only satisfied if 
%%%
\begin{align}
T(x)\approx\delta(x) \qquad \text{or}\qquad  T(x)\approx\delta(1-x)\,,
\end{align}
%%%
which is true as an equality if the subset of hadrons on which the energy correlators are measured is all hadrons or no hadrons. In both of these cases the measurement reduces to an IRC safe measurement. Higher-point projected correlators therefore ``see" the non-perturbative aspects associated with the track measurements, which modifies the transition region. Using the numerical values for the moments of the track functions in table~\ref{tab:mom_summary}, this also shows that the non-perturbative corrections using the track function formalism increase as a function of $N$, as expected. 

 In the case of $N=2$ there is an additional coincidental numerical relation between $T(2)$ (which appears in contact terms) and $T(1)^2$, which further suppresses modifications due to the use of tracks. If the track function is a delta function
$T(x) = \delta(x-a)\,,$ then it is infrared and collinear safe, as it just corresponds to a rescaling of the observable. However, in this case $T(2)=T(1)^2$, which are the moments probed by the two-point energy correlator. This is very nearly true in QCD, as shown in table~\ref{tab:mom_summary}. Therefore, we find that to a remarkable accuracy, the two-point correlator on tracks is almost identical to the two-point correlator on all particles, including deep in the perturbative region, and through the transition. This is also in agreement with the observation that for the use of tracks for the thrust observable is nearly an overall rescaling~\cite{Chang:2013iba}, since thrust is nearly (although not precisely) a two-point correlator.

%%%%%%%%%%%%%%%%%%%%%%
\subsection{Decays}\label{sec:decays}
%%%%%%%%%%%%%%%%%%%%%%  

\begin{figure}
  \begin{center}
    \subfloat[]{\label{fig:pi0_decay_a}
    \includegraphics[scale=0.34]{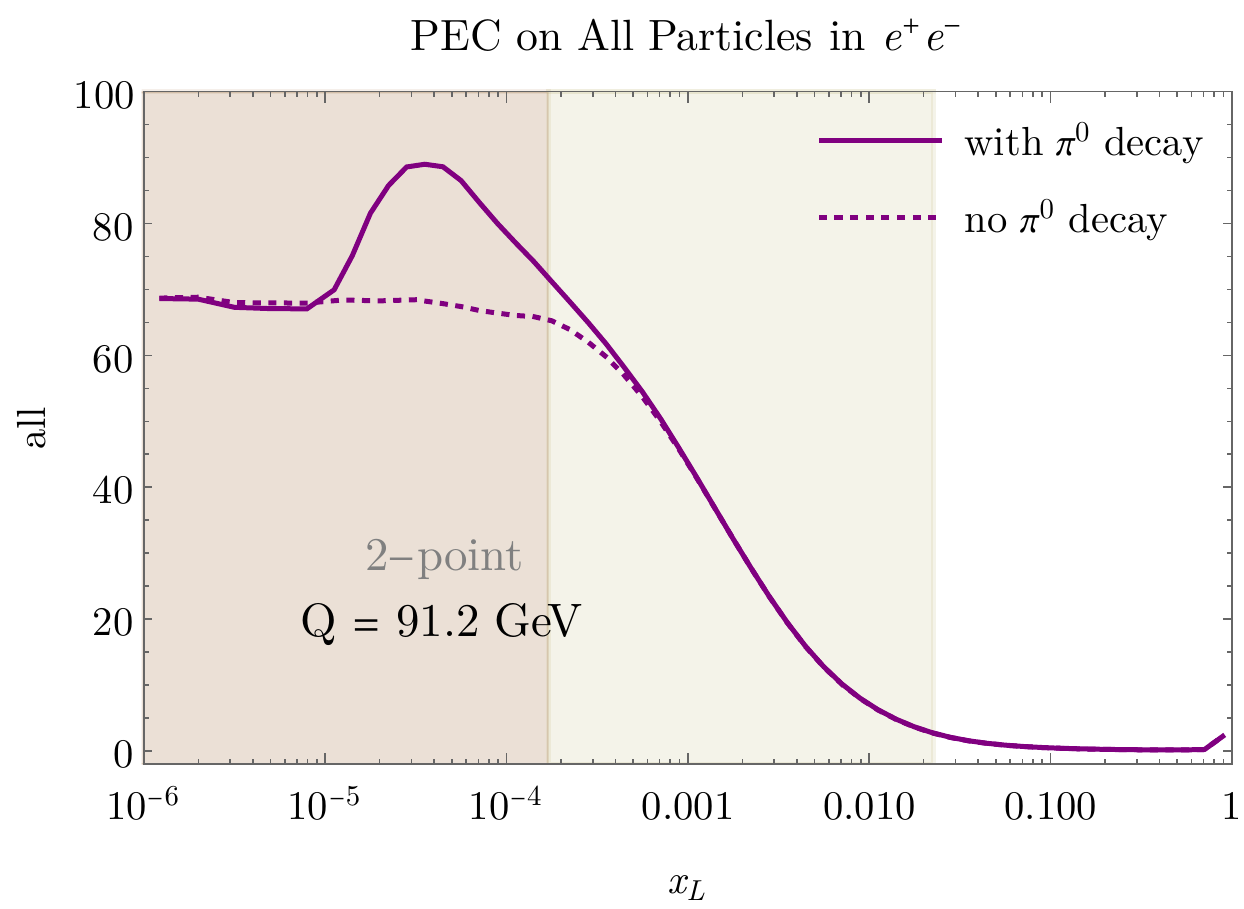}
    }
    %\quad 
    \subfloat[]{\label{fig:pi0_decay_b}
    \includegraphics[scale=0.33]{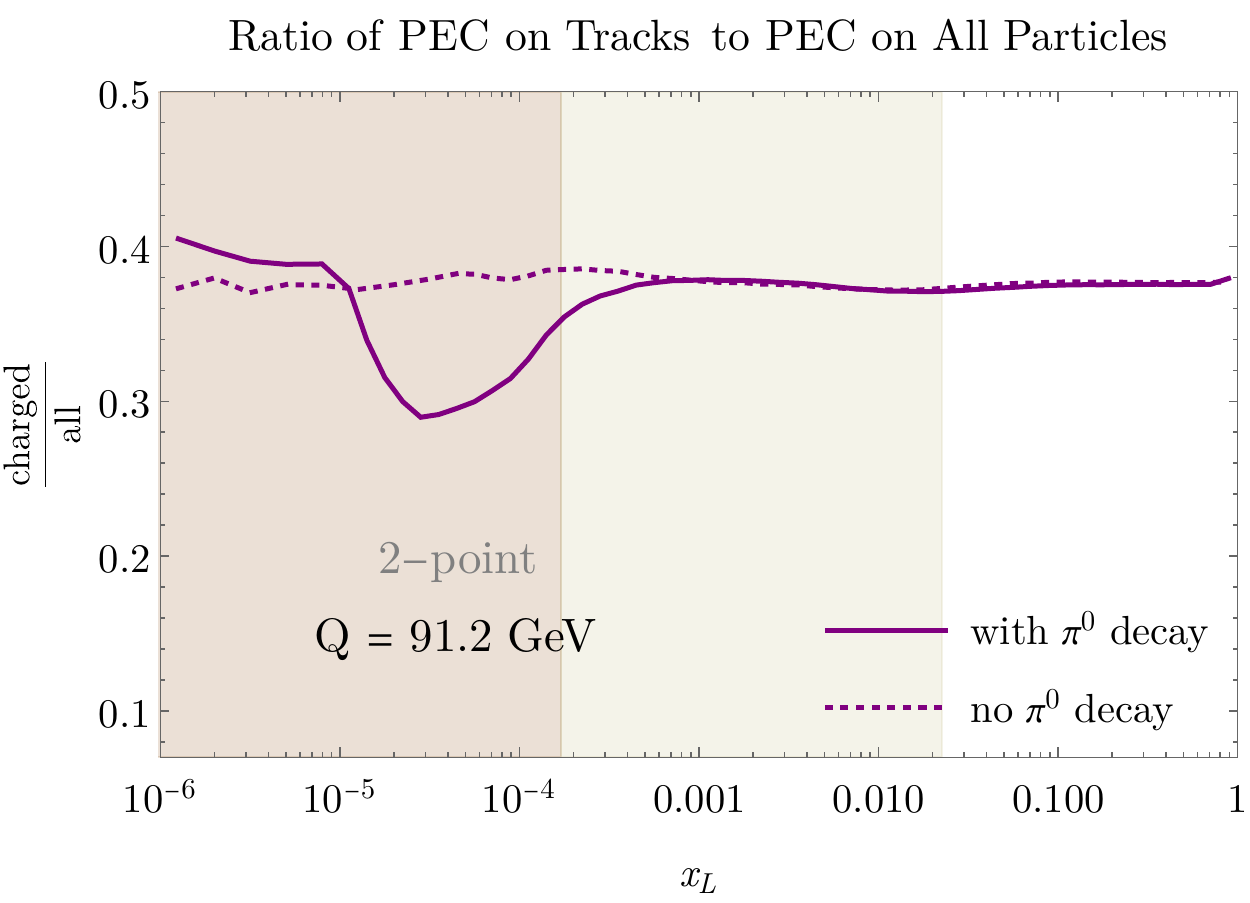}
    }
    \\
        \subfloat[]{\label{fig:pi0_decay_c}
    \includegraphics[scale=0.34]{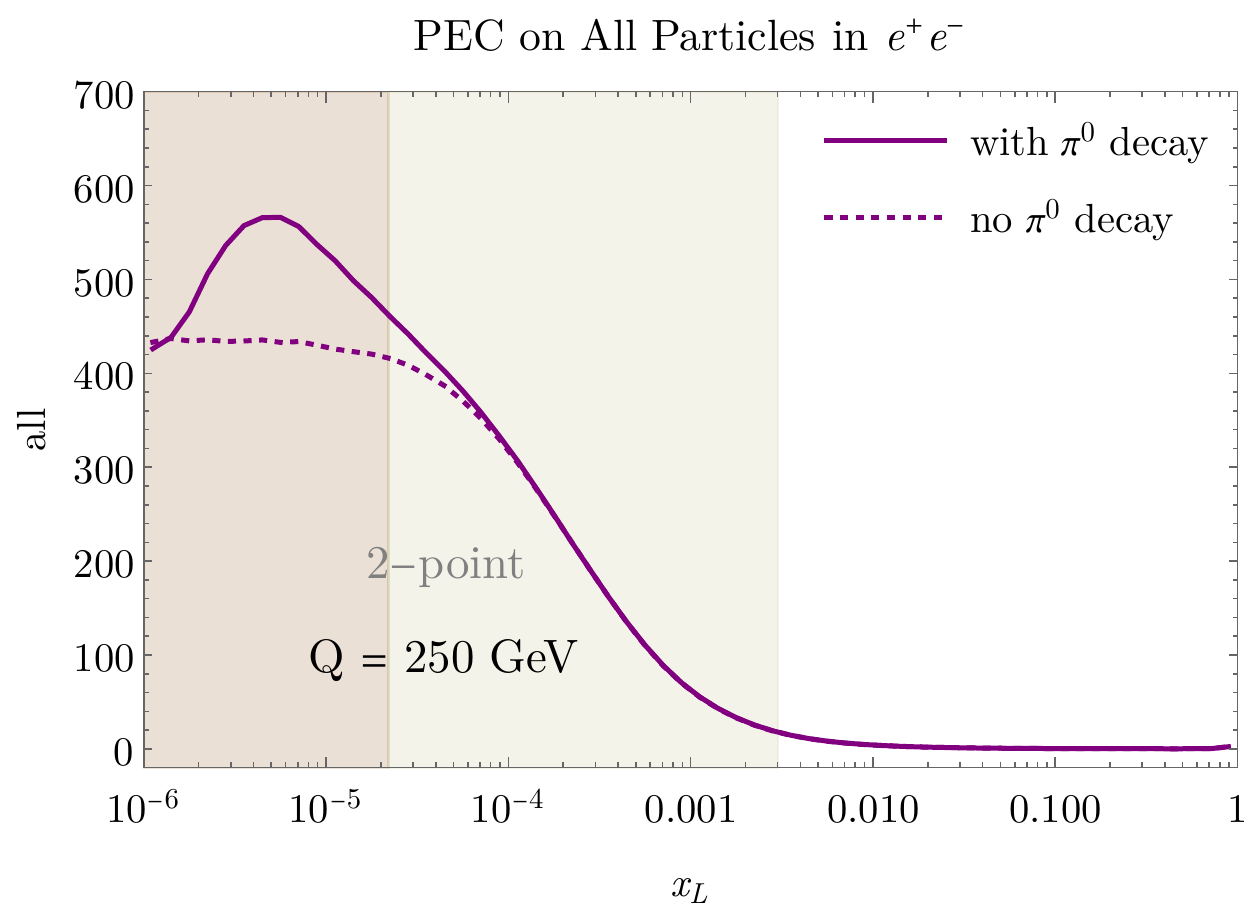}
    }
    %\quad 
    \subfloat[]{\label{fig:pi0_decay_d}
    \includegraphics[scale=0.33]{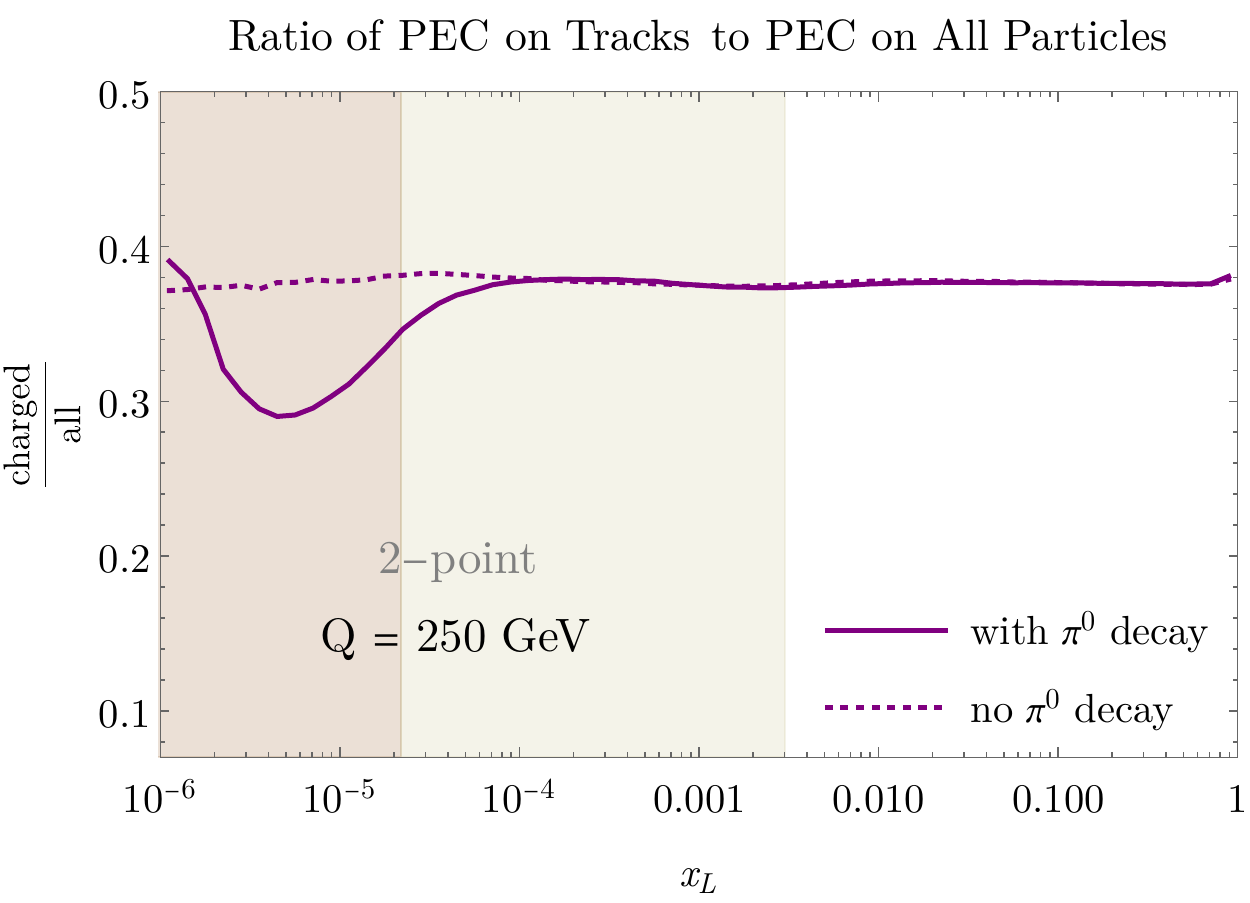}
    }
  \end{center}
  \caption{The projected two-point energy correlator on all hadrons in (a) and (c), and the corresponding ratio of charged to all hadrons in (b) and (d). Results are obtained using \textsc{Pythia}, with the c.o.m energy of $Q=91.2$~GeV (top) and 250 GeV (bottom).}
  \label{fig:pi0_decay}
\end{figure}

Another interesting difference between the track-based and all-hadron energy correlators is the treatment of decays of unstable hadronic states into neutral particles. These are not included in the track-based observable. A relevant example is $\pi^0 \to \gamma \gamma$. While decays in general are not included in our analytic calculations, they could be relevant experimentally, and so we comment on them briefly.

In \Fig{fig:pi0_decay_a} and \Fig{fig:pi0_decay_c} we show the hadronic-level two-point energy correlator measured in \textsc{Pythia}, with and without the decay of the $\pi^0$. In \Fig{fig:pi0_decay_b} and \Fig{fig:pi0_decay_d}, we show the corresponding ratio between the two-point energy correlator on tracks and on all hadrons. We clearly see a sharp peak at an angle corresponding to the mass of the pion\footnote{Regarding other figures in this paper, we turn off the $\pi^0$ decay channel for the Pythia hadronic-level data in order to focus on the flat behavior in the deeply non-perturbative region, as shown in fig.~\ref{fig:ratioNto2_pythia_250_xL_reg_a}. }.

Measurements of the energy correlators at small angles have so far been track based, and have therefore not seen this feature. It would be extremely interesting to somehow measure it, since it corresponds to a standard candle deep in the non-perturbative regime, allowing for a conversion between the angles of the energy correlator, and mass scales. 

%%%%%%%%%%%%%%%%%%%%%%
\section{Analytic Predictions for Track-Based Projected Energy Correlators}\label{sec:resum}
%%%%%%%%%%%%%%%%%%%%%%  

In this section we study the track-based energy correlators in the perturbative regime, showing that the corrections to the scaling behavior of the energy correlators are logarithmic, and can be systematically computed from the renormalization group evolution of the track functions.

%%%%%%%%%%%%%%%%%%%%%%
\subsection{Renormalization Group Analysis}\label{sec:solve}
%%%%%%%%%%%%%%%%%%%%%%  

\begin{figure}
  \begin{center}
    \subfloat[]{
    \includegraphics[scale=0.36]{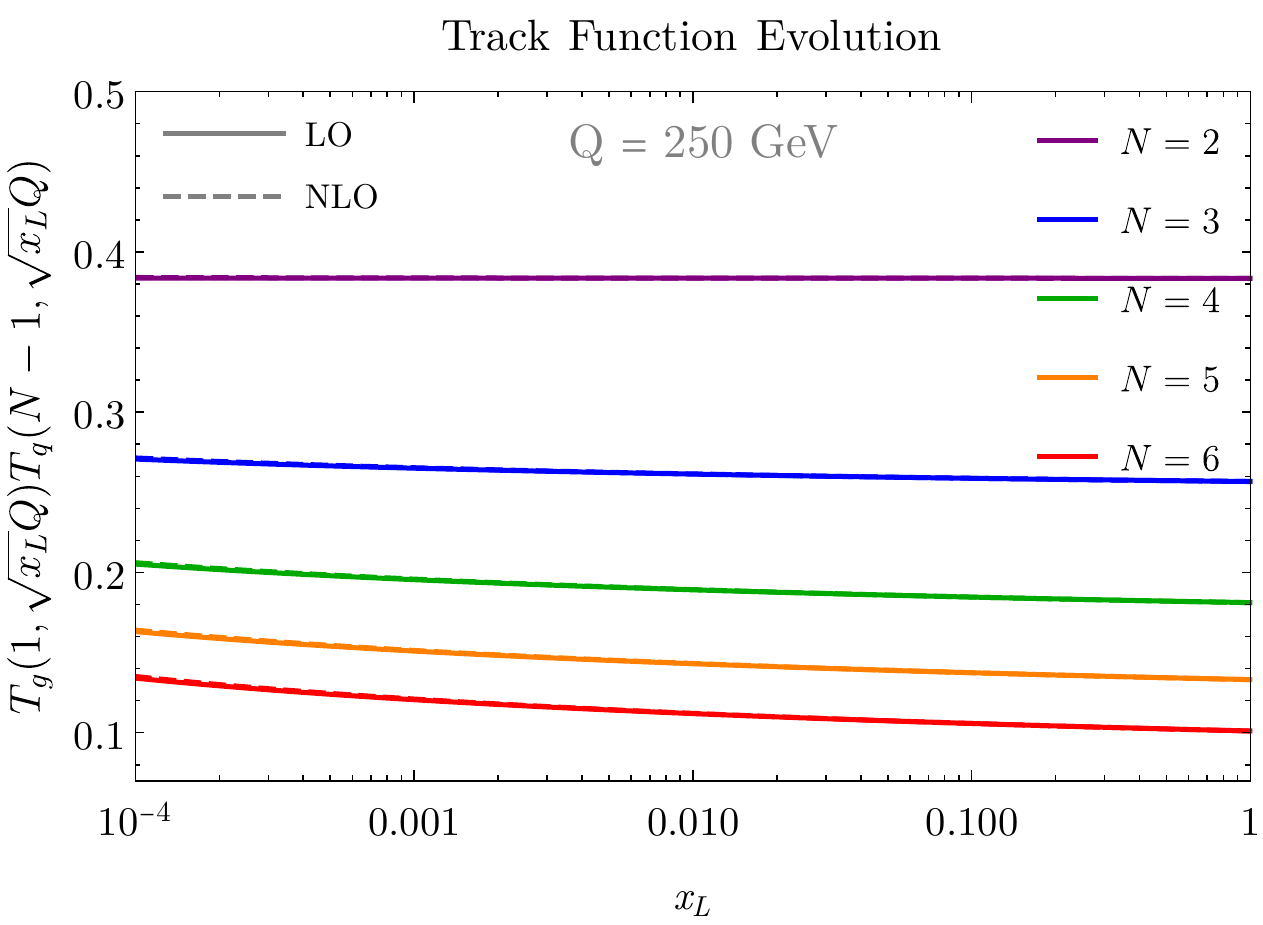}
    \label{fig:track_evo_1}
    }
    %\quad 
    \subfloat[]{
    \includegraphics[scale=0.36]{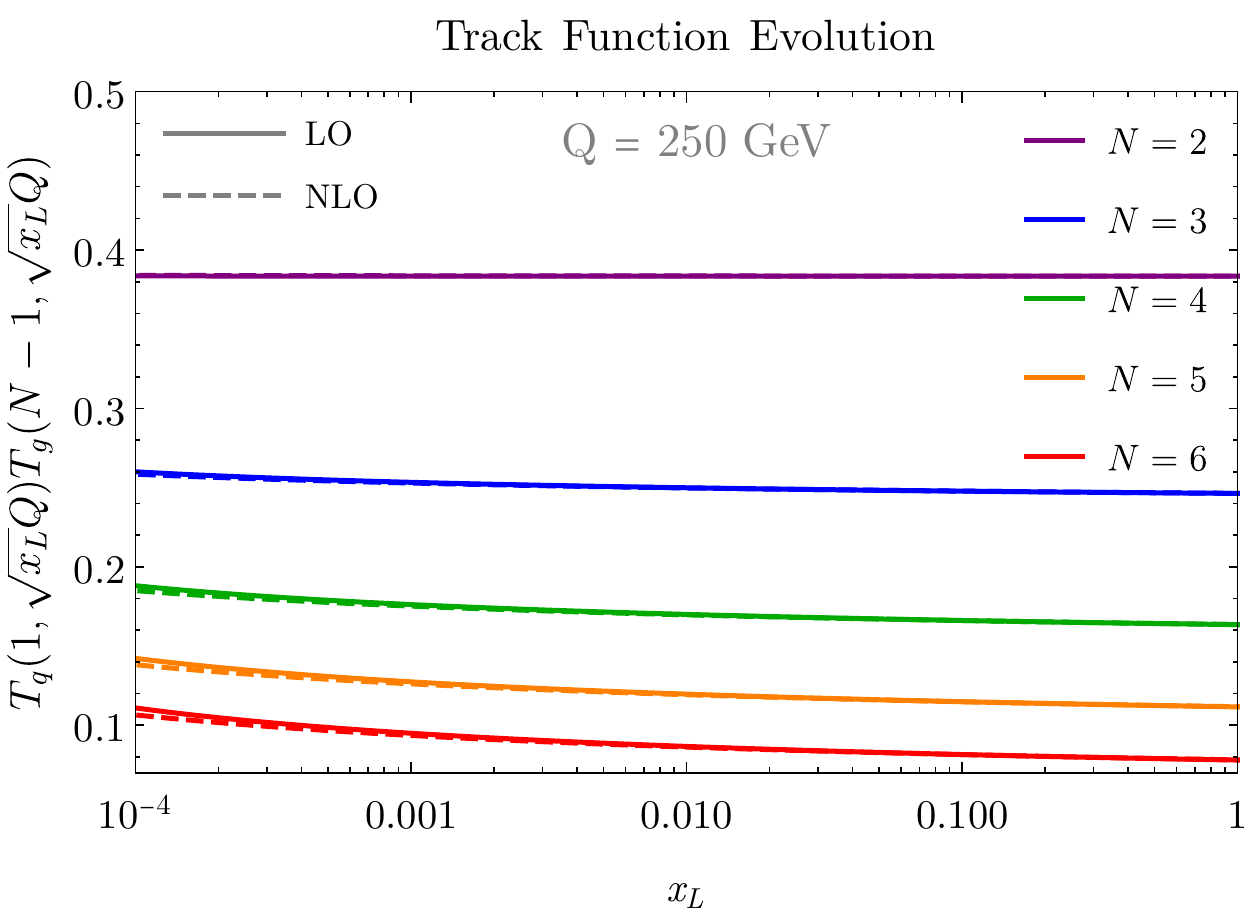}
    \label{fig:track_evo_2}
    }
  \end{center}
  \caption{The perturbative evolution at LO (solid) and NLO (dashed) of the track function combinations $T_g(1) T_q(N-1)$ in (a) and $T_q(1) T_g(N-1)$ in (b), that appear in the study of the energy correlators. Extremely slow RG evolution is observed, leading to minor modifications of the scaling behavior of the energy correlators in the collinear limit. }
  \label{fig:track_evo}
\end{figure}

In this section we discuss the perturbative resummation of the jet function in the small angle limit. The renormalization group evolution of the jet function, with or without tracks, is fixed to be DGLAP
%%%
\begin{align}
  \label{eq:jet_evo_2}
  \frac{\df \vec{J}\,^{\supnu} \bigl( \ln \frac{x_L Q^2}{\mu^2},\mu \bigr)}{\df \ln\mu^2}
=
\int_0^1 \! \df y\, y^n \vec{J}\,^{\supnu} \Bigl( \ln \frac{x_L y^2 Q^2}{\mu^2},\mu\Bigr) 
\cdot 
\widehat{P}(y,\alpha_s) \,,
\end{align}
%%%
After writing the jet function in terms of matching coefficients and track function moments
%%%
\begin{align}
J_i^{[n]}= {\bf{j}}_i^{[n]} \cdot  {\bf T}_n\,,
\end{align}
%%%
the known renormalization group evolution of the track functions, which we write schematically as
%%%
\begin{align}
\frac{\df}{\df\ln \mu^2}{ \bf T}_n=\widehat R_n~ {\bf T}_n\,,
\qquad
\widehat R_n\equiv \sum_{L=1}^\infty \Bigl(\frac{\alpha_s}{4\pi}\Bigr)^L \widehat R_n^{(L)}
\,,\end{align}
%%%
then fixes the renormalization group evolution of the matching coefficients ${\bf{j}}_i^{[n]}$.

To perform the RG evolution of these matching coefficients, we write an ansatz for the jet function as
%%%
\begin{align}\label{eq:ansatz_jet_track}
J_i^{\supnu} \biggl(\alpha_s(\mu), {\bf T}_n(\mu), \ln \frac{x_L Q^2}{\mu^2} \biggr)
=
\sum\limits_{L=0}^\infty \Bigl(\frac{\alpha_s}{4\pi}\Bigr)^L \biggl[\, \sum\limits_{m=0}^L  {\bf j}_{i,m}^{\,[n],(L)} \cdot {\bf T}_n ~ \ln^m \Bigl( \frac{x_LQ^2}{\mu^2}  \Bigr)   \biggl]\,.
\end{align}
%%%
Here $i=q,g$ is the flavor index, and both $ {\bf j}_{i,m}^{\,[n],(L)}$  and ${\bf T}_n$ are vectors in the track function space. This ansatz can then be substituted into \Eq{eq:jet_evo_2} to derive a recursive equation for ${\bf{j}}_{i,m}^{[n],(L)}$, given the boundary values presented in \Sec{sec:fact_collinear}.

\begin{figure}
  \begin{center}
    \includegraphics[scale=0.40]{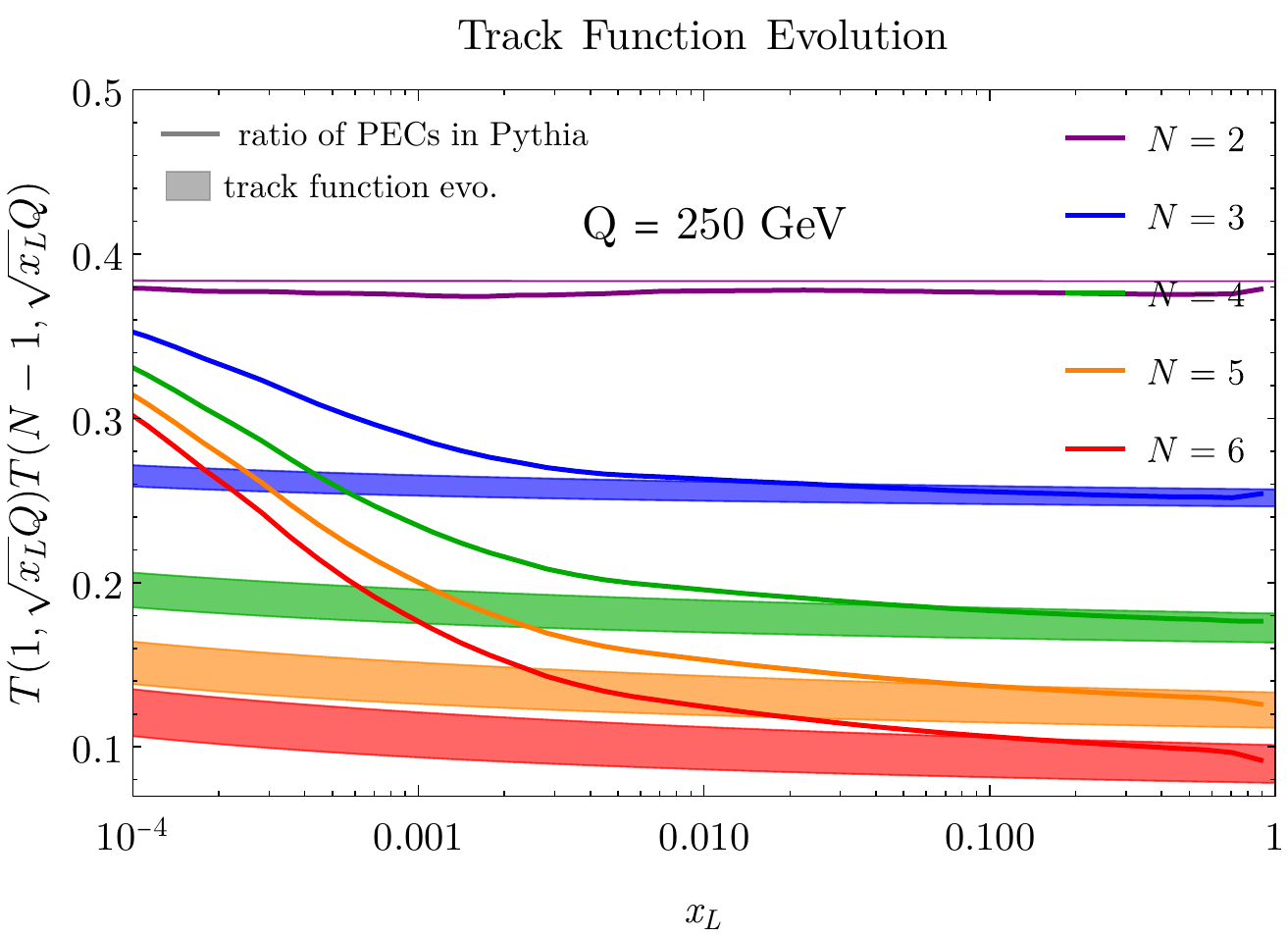}
    \label{fig:track_evo_2}
  \end{center}
  \caption{A comparison of the ratio between the projected energy correlators on tracks and on all hadrons obtained using \textsc{Pythia}, with (the evolution of) the product of track functions. The band for the track functions is computed by taking the envelope for $T_g(1,\sqrt{x_L}Q)T_q(N-1,\sqrt{x_L}Q)$ and $T_q(1,\sqrt{x_L}Q)T_g(N-1,\sqrt{x_L}Q)$ which are shown in fig.~\ref{fig:track_evo}.}
  \label{fig:track_evo_compare}
\end{figure}

We express our results in terms of the timelike splitting function, which we expand as
%%%
\begin{align}
P_{ij}(z)=\sum_{L=0}^\infty \Bigl(\frac{\alpha_s}{4\pi}\Bigr)^{L+1}P_{ij}^{(L)}(z)\,.
\end{align}
%%%
Expressions for $P_{ij}(z)$ to order-$\alpha_s^3$ can be found in ref.~\cite{Chen:2020uvt} (see also refs.~\cite{Mitov:2006wy,Mitov:2006ic,Moch:2007tx,Almasy:2011eq}). We similarly expand the $\beta$-function as
%%%
\begin{align}
\beta(\alpha_s)= -2 \alpha_s \sum\limits_{n=0}^\infty \beta_n
\Bigl(\frac{\alpha_s}{4\pi}\Bigr)^{n+1}
\,.
\end{align}
%%%

Inserting the above expansions into eq.~\eqref{eq:ansatz_jet_track} and matching the leading logarithmic terms, we find that for $L\geq 1$ 
%%%
\begin{align}
L\, {\bf{j}}_{i, L}^{[n],(L)}\cdot {\bf T}_n
&= 
{\color{red}{\bf{j}}_{i,L-1}^{[n],(L-1)} \cdot \widehat{R}_{n}^{(1)}\cdot {\bf T}_n
}\nn \\
& \quad
-(L-1) \beta_0 \, {\bf{j}}_{i, L-1}^{[n],(L-1)}\cdot {\bf T}_n
-\sum_r~ {\bf{j}}_{r, L-1}^{[n],(L-1)}\cdot {\bf T}_n \int_0^1 \df y\, y^n P^{(0)}_{ri}(y)
\,,
\end{align}
%%%
%\begin{align}
%L\, \vec{\bf{j}}_{i, L}^{[n](L)}
%&= 
%{\color{red}\vec{\bf{j}}_{i,L-1}^{[n](L-1)} \cdot \widehat{R}_{n}^{(1)}
%}
%-(L-1) \beta_0 \, \vec{\bf{j}}_{i, L-1}^{[n](L-1)}
%-\sum_r~\vec{\bf{j}}_{r, L-1}^{[n](L-1)} \int_0^1 \df y\, y^n P^{(0)}_{ri}(y)
%\,,
%\end{align}
%%%
with $i,r$ denoting the flavors. 
%Recall that this is a vector, which is then dotted with ${\bf T}_n$ to form the full jet function.
On the left hand side, we have ${\bf{j}}_{L}^{(L)}$ and on the right, we have ${\bf{j}}_{L-1}^{(L-1)}$, allowing us to iteratively solve for the perturbative coefficients of the jet function with the initial condition $J_{i,0}^{[n],(0)}$ (see eq.~\eqref{eq:initial_J00}). We have highlighted the term arising from the evolution of the track functions in red. Setting this term to zero and the track function moment vector ${\bf T}_n$ to $\{1,1,...,1\}^\text{t}$, the result reduces to the all-hadron projected energy correlators, where the evolution comes from the $\beta$-function, and DGLAP evolution. Here we see explicitly how the evolution of the track function modifies the single logarithmic coefficients in the series for the projected energy correlator jet functions.

We can derive a similar recursive equation at NLL. We find that for $L\geq 2$ 
%%%
\begin{align}
(L-1)\,  {\bf{j}}_{i,L-1}^{[n],(L)}\cdot {\bf T}_n
&={\color{red}
~{\bf j}_{i, L-2}^{[n],(L-2)}\cdot \widehat{R}_n^{(2)}\cdot {\bf T}_n
}
{\color{red} ~+~ {\bf{j}}_{i, L-2}^{[n],(L-1)} \cdot \widehat{R}_n^{(1)}\cdot {\bf T}_n} \nn \\
&\quad -\sum_r  {\bf{j}}_{r,L-2}^{[n],(L-2)}\cdot {\bf T}_n \int_0^1 \df y\, y^n P^{(1)}_{ri}(y)
-\sum_r {\bf{j}}_{r,L-2}^{[n],(L-1)}\cdot {\bf T}_n \int_0^1 \df y\, y^n P^{(0)}_{ri}(y)\nn \\
&\quad
-(L-1)\,\sum_r  {\bf{j}}_{r, L-1}^{[n],(L-1)}\cdot {\bf T}_n {\color{blue} \int_0^1 \df y\, y^n \ln(y^2) \, P^{(0)}_{ri}(y)} \nn \\
&\quad-(L-2)\beta_1 \,  {\bf{j}}_{i, L-2}^{[n],(L-2)}\cdot {\bf T}_n 
~-~(L-1)\beta_0 \,  {\bf{j}}_{i,L-2}^{[n],(L-1)}\cdot {\bf T}_n
\,.
\end{align}
%%%
%%%
%\begin{align}
%(L-1)\, \vec {\bf{j}}_{i,L-1}^{[n],(L)}
%&={\color{red}
%~\vec{\bf{j}}_{i, L-2}^{[n](L-2)}\cdot \widehat{R}_n^{(2)}
%}
%{\color{red} ~+~\vec {\bf{j}}_{i, L-2}^{[n](L-1)} \cdot \widehat{R}_n^{(1)}} 
%-(L-2)\beta_1 \, \vec {\bf{j}}_{i, L-2}^{[n](L-2)}
%~-~(L-1)\beta_0 \, \vec {\bf{j}}_{i,L-2}^{[n],(L-1)}
%\nn \\
%
%&\qquad -\sum_r \biggl[\vec {\bf{j}}_{r,L-2}^{[n](L-2)} \int_0^1 \df y\, y^n P^{(1)}_{ri}(y)
%+\vec {\bf{j}}_{r,L-2}^{[n](L-1)} \int_0^1 \df y\, y^n P^{(0)}_{ri}(y)\nn \\
%&\qquad
%+(L-1)\, \vec {\bf{j}}_{r, L-1}^{[n](L-1)} {\color{blue} \int_0^1 \df y\, y^n \ln(y^2) \, P^{(0)}_{ri}(y)} \biggr]
%\,.\end{align}
%%%
Here, we have separated terms coming from the DGLAP evolution of the jet function, the running coupling, and the contributions from the track function evolution (highlighted in red). In addition to the higher order $\beta$ function and splitting functions, we see the appearance of logarithmic moments of the splitting functions (highlighted in blue), as was first discussed in ref.~\cite{Dixon:2019uzg}.  With the evolution of the track function (in red) dropped and the track function moment vector ${\bf T}_n$ taken equal to $\{1,1,...,1\}^\text{t}$, the equation becomes the recurrence relation for the coefficients of the all-hadron jet function\footnote{Indeed, ${\bf j}_{i,m}^{[n],(L)}\cdot {\bf T}_n$ is the counterpart of the coefficient for the $a_s^L\ln^m(x_LQ^2/\mu^2)$ term in the all-particle jet function, as can be seen from eq.~\ref{eq:ansatz_jet_track}.}. This illustrates how the track function evolution slightly modifies the perturbative coefficients in the evolution of the jet function. One can derive similar recursive relations to any logarithmic order, which will be interesting for exploring higher order perturbative resummation of the track-based energy correlators. However, for our current purposes, NLL suffices.

In \Fig{fig:track_evo} we show the evolution of the combinations $T_g(1) T_q(N-1)$ and $T_q(1) T_g(N-1)$ that appear in the energy correlators. One can see that these exhibit a remarkably slow evolution (i.e. a weak scale dependence), due partly to cancellations in non-perturbative QCD parameters, as discussed in detail in ref.~\cite{Jaarsma:2022kdd}. This is particularly true for low moments. The higher moments exhibit a faster evolution (stronger scale dependence). 
Note that in the full perturbative result a sum of $T(N-k)T(k)$'s ($1\leq k\leq N-1$) and contributions involving more than two track functions appears. The latter give a smaller contribution, and the numerical sizes of $T(N-k)T(k)$'s for $1< k < N-1$ and $T(N-1)T(1)$ are approximately equal in QCD for the higher point cases, so these plots should be meant as representative that the size of $T(N-1)T(1)$ sets the behavior of the ratio, and that the running is slow. Therefore, while the evolution of the track functions is single logarithmic and thus must be included in precision calculations of the energy correlators, it has a small overall effect in modifying the scaling behavior in the collinear limit.

In \Fig{fig:track_evo_compare} we compare the ratio between the projected energy correlators on tracks and on all hadrons as computed in \textsc{Pythia}, with the product of track functions. The band for the track functions is computed by taking the envelope of values for quarks and gluons (i.e. the curves in the left and right plots of \Fig{fig:track_evo}). We see that this ratio provides a good description of the ratio in the perturbative region.  

%%%%%%%%%%%%%%%%%%%%%%
\subsection{Numerical Results}\label{sec:num}
%%%%%%%%%%%%%%%%%%%%%%  

\begin{figure}
  \begin{center}
    \subfloat[]{
    \includegraphics[scale=0.36]{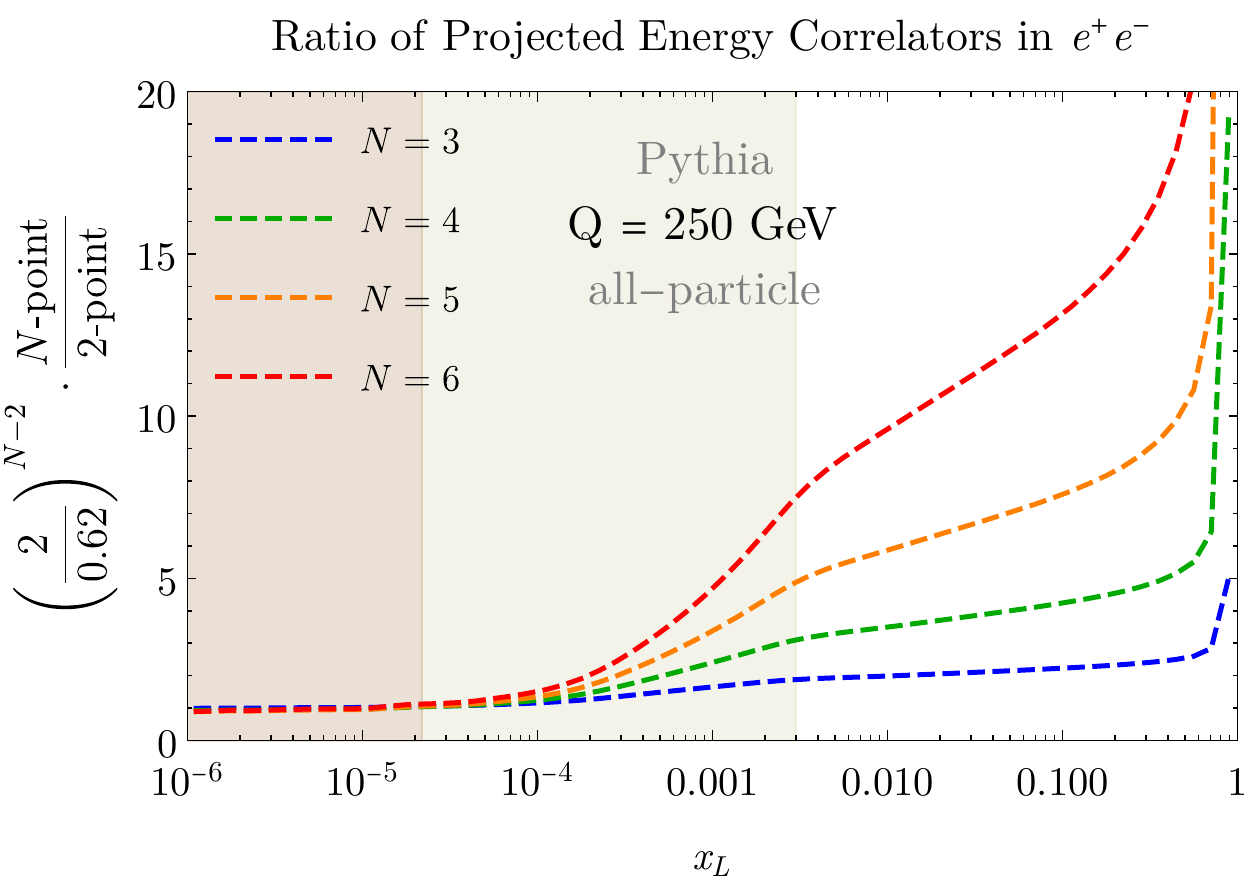}
    \label{fig:ratioNto2_pythia_250_xL_reg_normfreehad_a}
    }
    \quad 
    \subfloat[]{
    \includegraphics[scale=0.36]{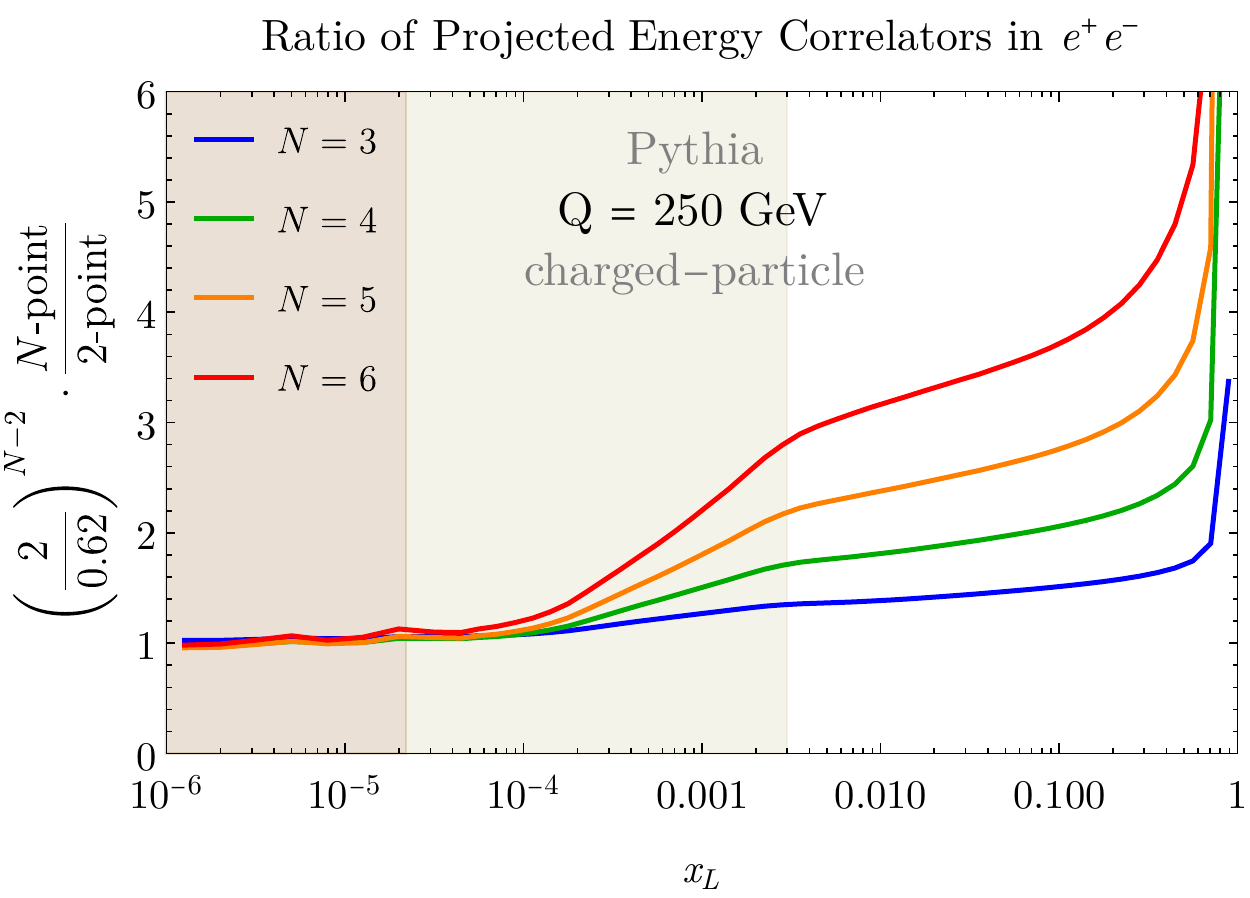}
    \label{fig:ratioNto2_pythia_250_xL_reg_normfreehad_b}
    }
    \\
    \subfloat[]{    
    \includegraphics[scale=0.36]{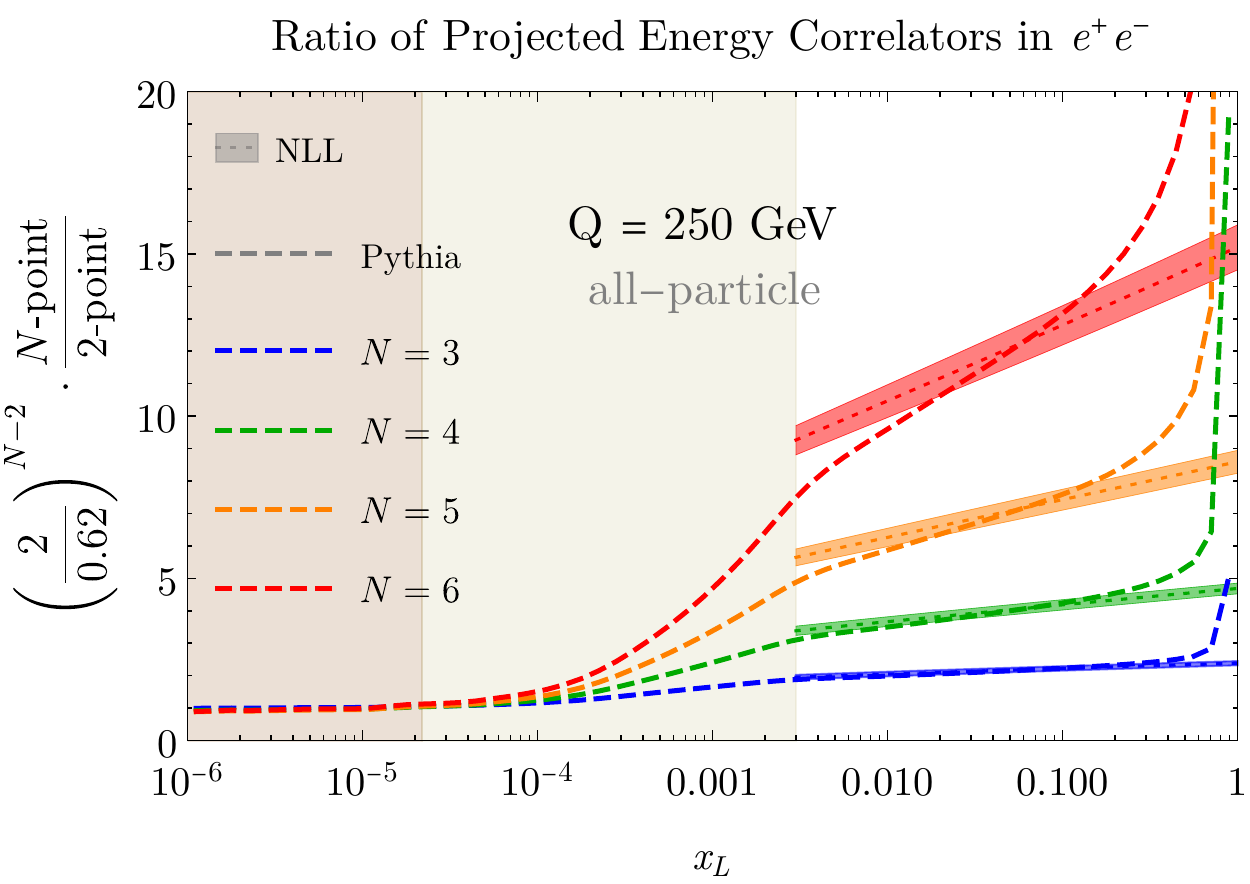}
    \label{fig:ratioNto2_NLLwithpythia_250_xL_reg_normfreehad_cut_a}
    }
    \quad 
    \subfloat[]{
    \includegraphics[scale=0.36]{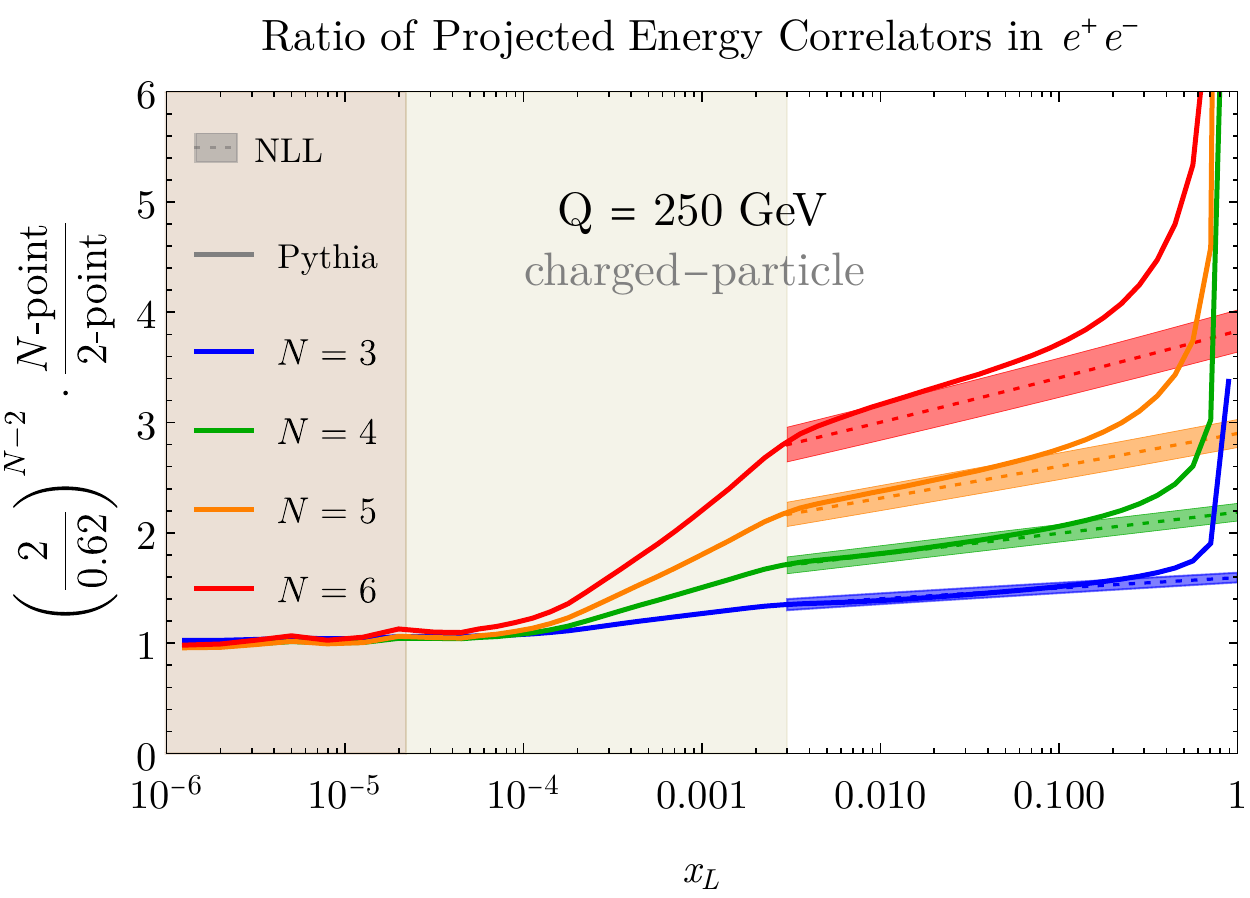}
    \label{fig:ratioNto2_NLLwithpythia_250_xL_reg_normfreehad_cut_b}
    }
  \end{center}
  \caption{The ratio of the $N$-point to two-point projected energy correlator on all particles in (a) and (c), and on charged particles in (b) and (d). In the (a) and (b) we show results from \textsc{Pythia}, 
  with a c.o.m energy $Q=250$~GeV and $N=3,4,5,6$. We have normalized the results in the free hadron region. In (c) and (d) we compare with our analytic results at NLL. Our analytic results are plotted only in the perturbative regime. }
  \label{fig:ratioNto2_pythia_250_xL_reg_normfreehad}
\end{figure}

\begin{figure}
\captionsetup[subfigure]{labelformat=empty}
  \begin{center}
  \subfloat[]{
  \includegraphics[scale=0.32]{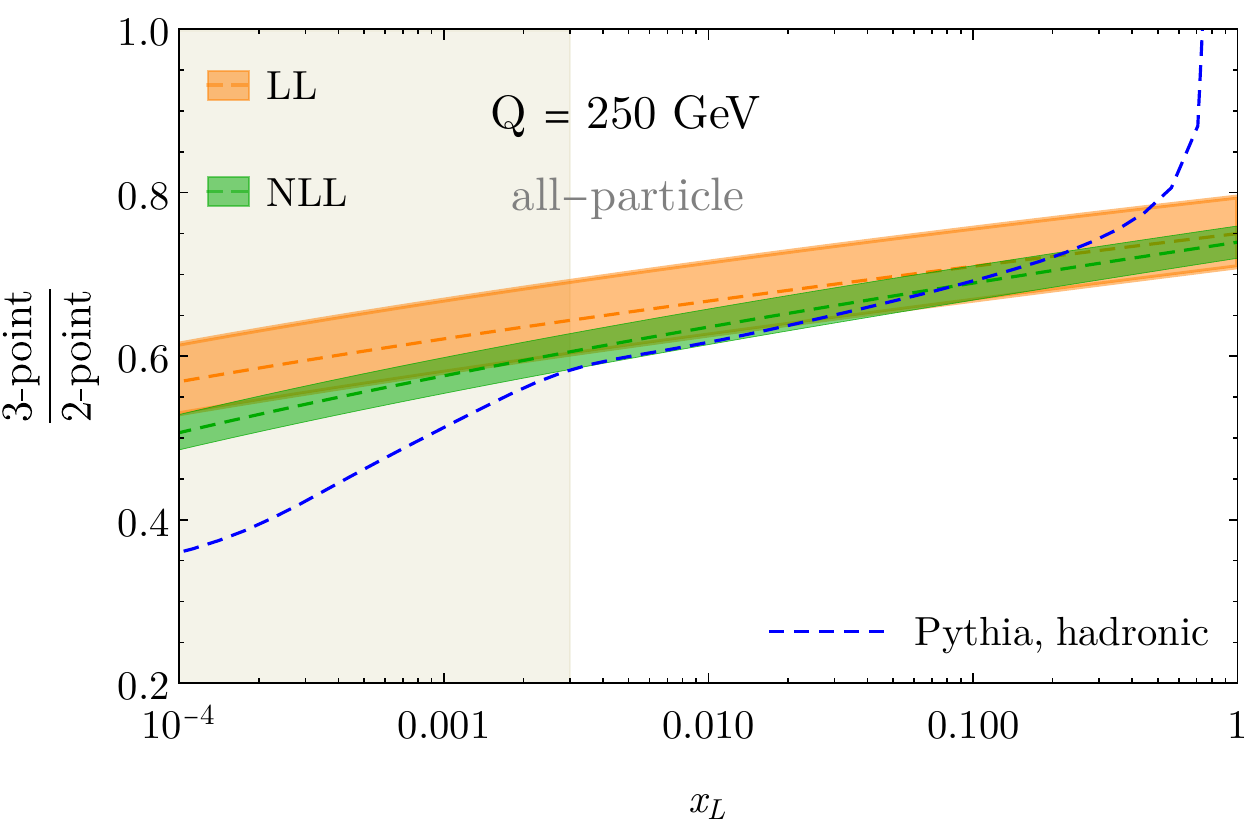}
  \label{fig:ratioNto2_3pt_a}
  }
  \quad 
  \subfloat[]{
  \includegraphics[scale=0.32]{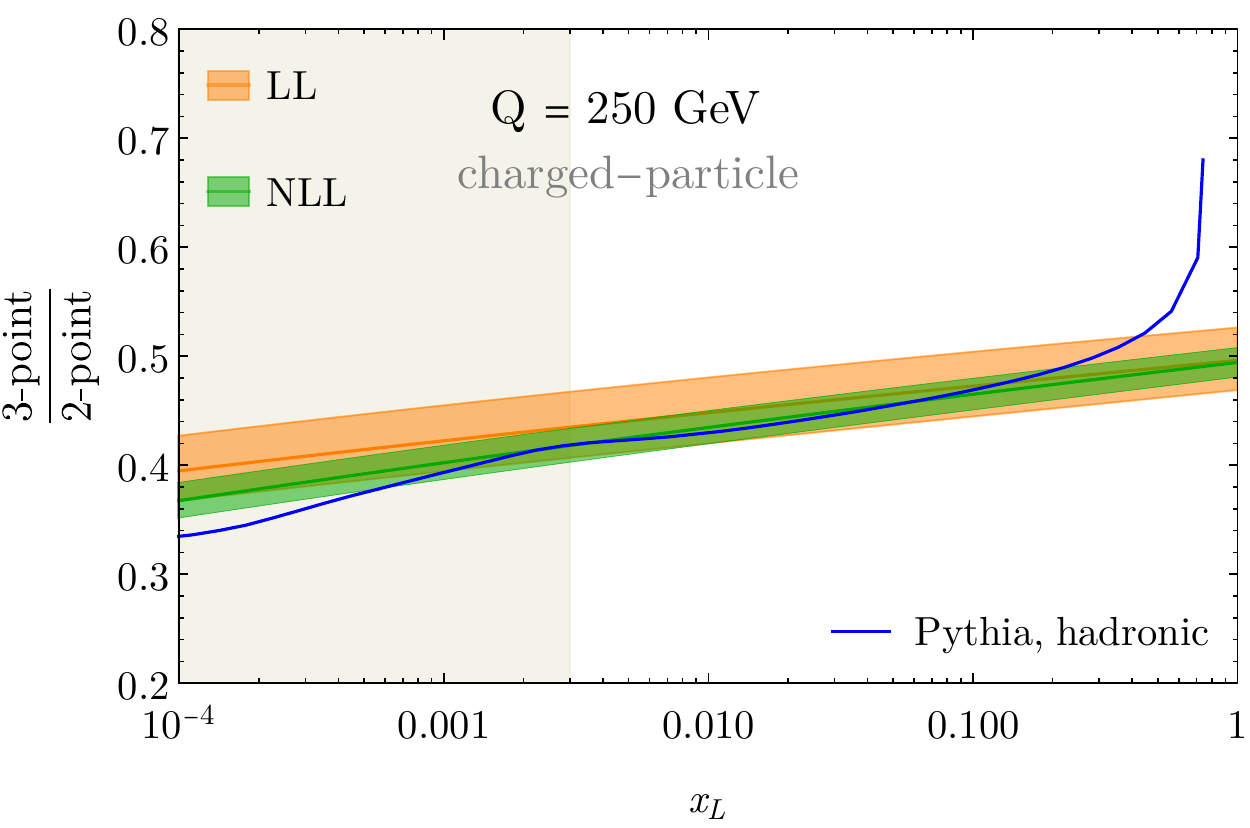}
  \label{fig:ratioNto2_3pt_b}
  }
  \\
 \vspace{-0.9cm} 
  \subfloat[]{
  \includegraphics[scale=0.32]{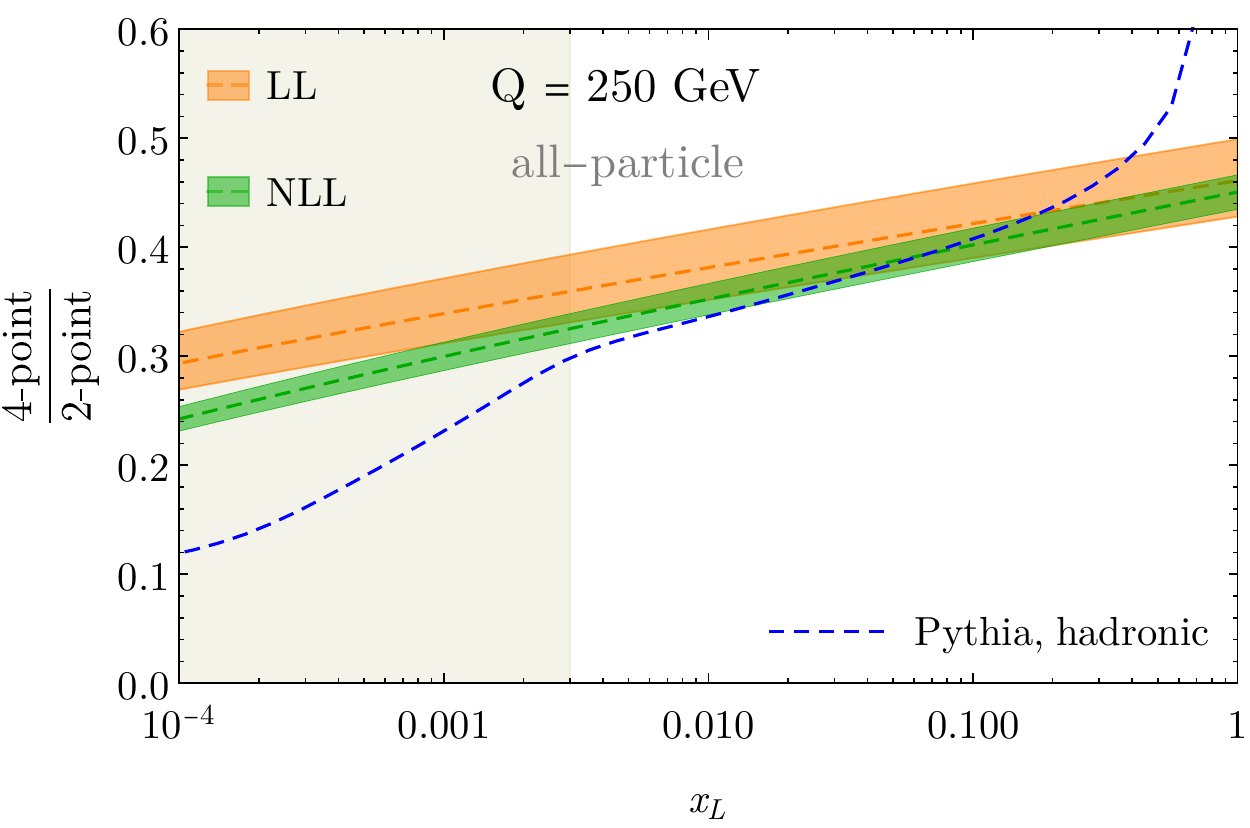}
  \label{fig:ratioNto2_4pt_a}
  }
  \quad 
  \subfloat[]{
  \includegraphics[scale=0.32]{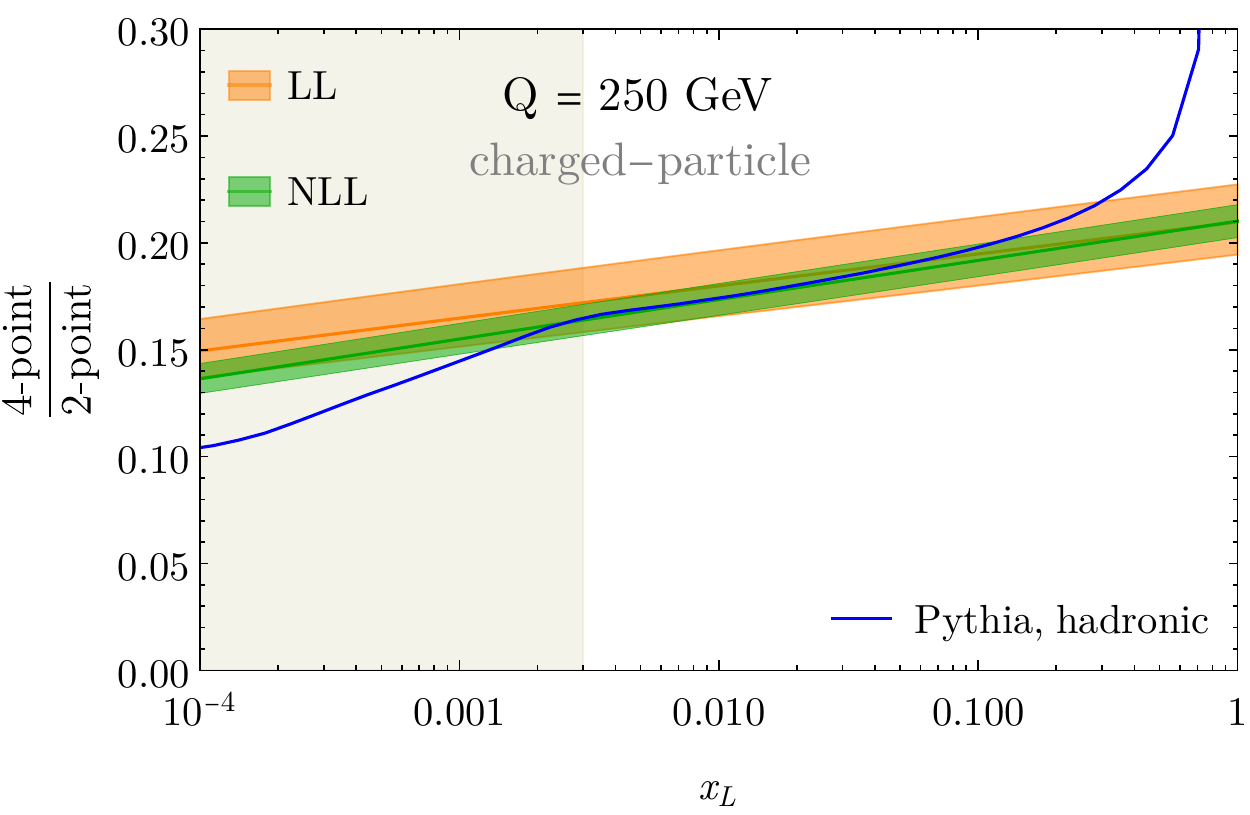}
  \label{fig:ratioNto2_4pt_b}
  }
  \\
 \vspace{-0.9cm}   
  \subfloat[]{
  \includegraphics[scale=0.32]{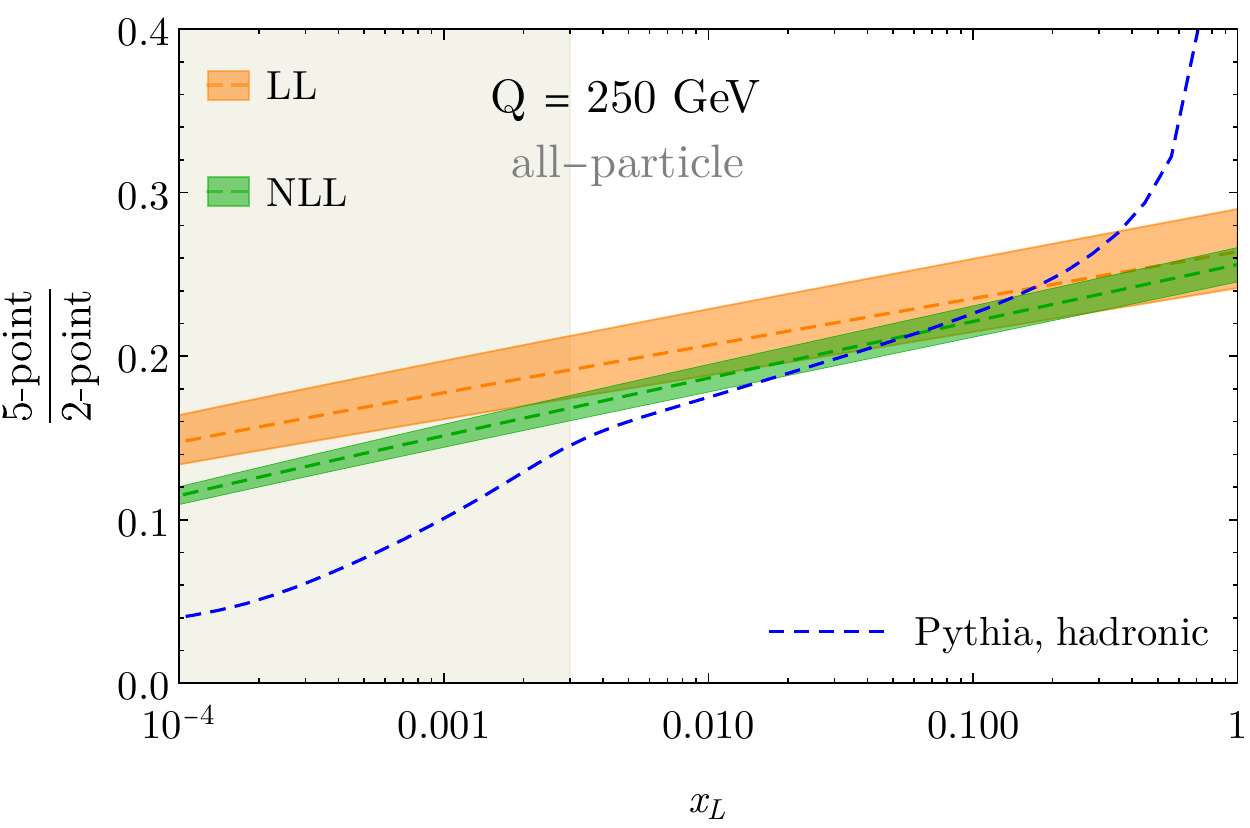}
  \label{fig:ratioNto2_5pt_a}
  }
  \quad 
  \subfloat[]{
  \includegraphics[scale=0.32]{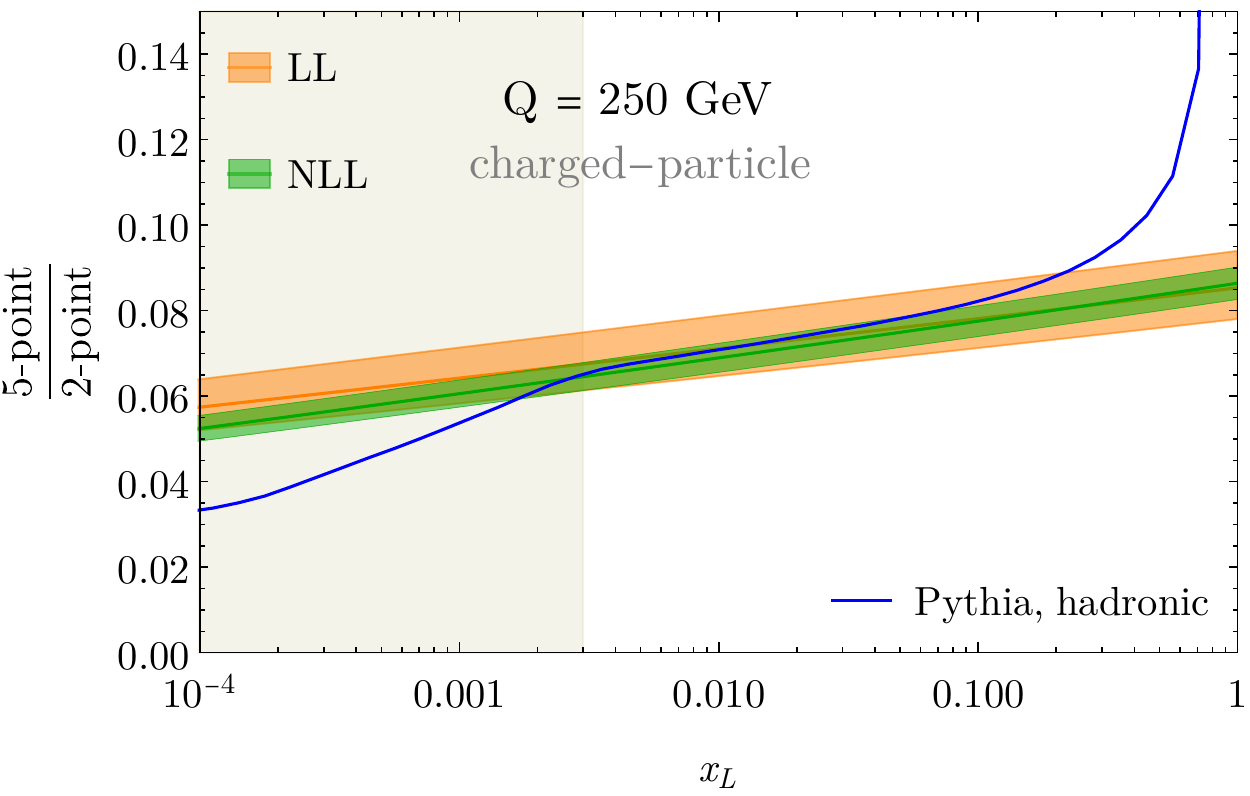}
  \label{fig:ratioNto2_5pt_b}
  }
  \\
   \vspace{-0.9cm} 
\subfloat[]{
    \includegraphics[scale=0.32]{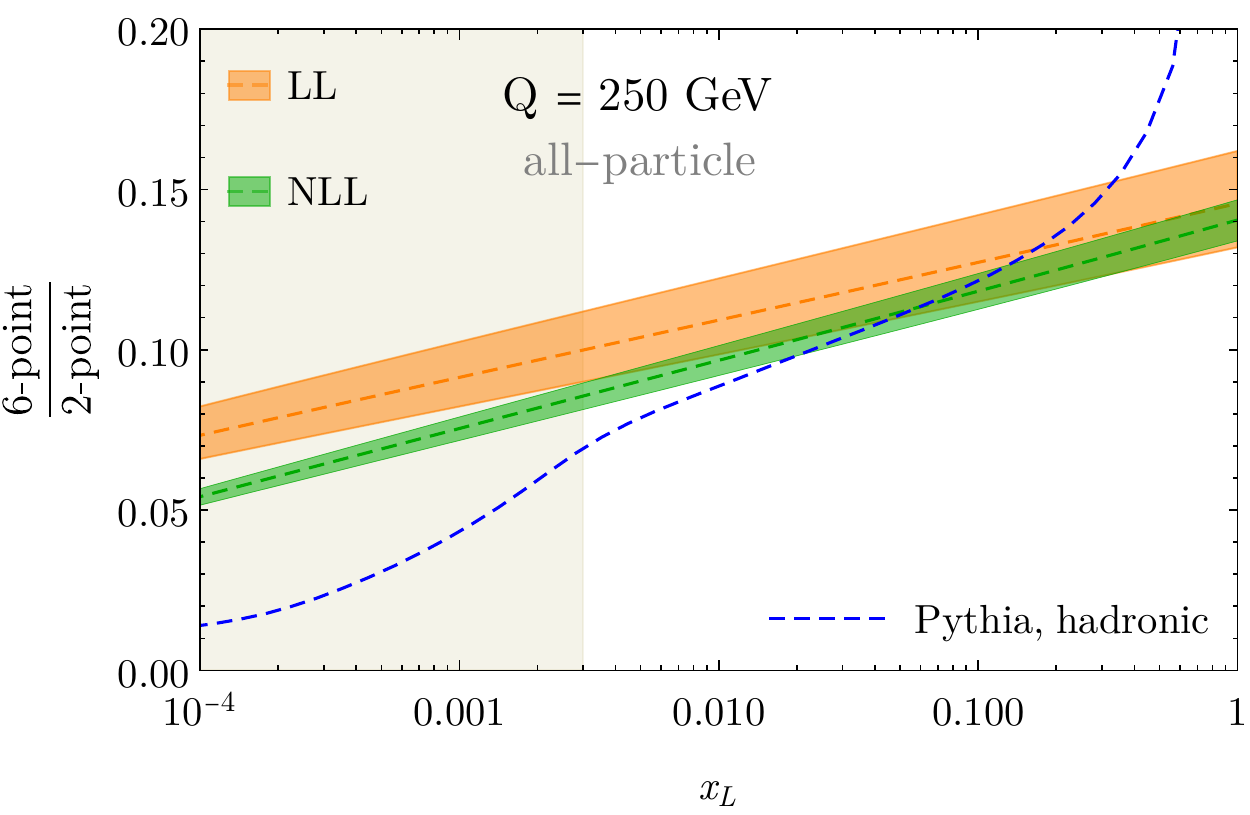}
    \label{fig:ratioNto2_6pt_a}
    }
    \quad 
    \subfloat[]{
    \includegraphics[scale=0.32]{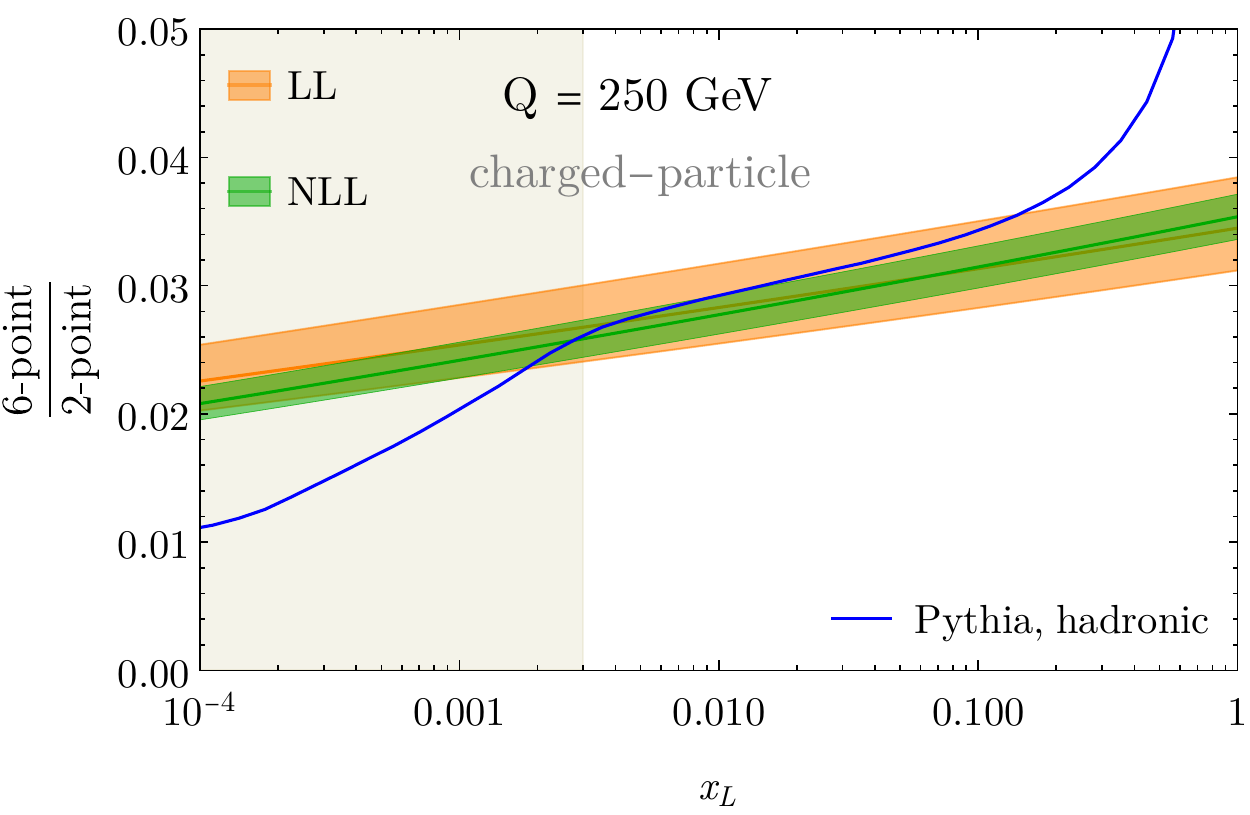}
    \label{fig:ratioNto2_6pt_b}
    }
  \end{center}
  \vspace{-1.0cm}
  \caption{The $N$-point/2-point ratio of energy correlators on all particles (left)  and charged particles (right) obtained using \textsc{Pythia} (blue dashed) and our analytic calculations at LL  (orange), NLL (green) order. In the shaded region the correlators are dominated by non-perturbative effects, and our results should not be trusted. }
  \label{fig:ratioNto2_pythia_ana}
  \end{figure}

 The ratios of the projected $N$-point correlators isolate the quantum scaling of the correlators in the small angle limit. In this section we use our perturbative results of \Sec{sec:solve} to study these  ratios on tracks.  In \Fig{fig:ratioNto2_pythia_250_xL_reg_normfreehad_a} and \Fig{fig:ratioNto2_pythia_250_xL_reg_normfreehad_b}  we show the results, as computed in \textsc{Pythia}, for the ratios of projected correlators on both all hadrons and on tracks. To compare the different $N$-point correlators, we have rescaled them so that they are equal to one in the deep non-perturbative region. In \Fig{fig:ratioNto2_NLLwithpythia_250_xL_reg_normfreehad_cut_a} and \Fig{fig:ratioNto2_NLLwithpythia_250_xL_reg_normfreehad_cut_b}, we compare with our analytic calculations at NLL, both with and without tracks. The shaded bands correspond to perturbative scale variations. In the panels of \Fig{fig:ratioNto2_pythia_ana} we show each of the two to six point correlators individually, along with the predictions at both LL and NLL, to demonstrate the convergence.

We see that due to the weak running of the track functions, the use of tracks does not significantly modify the elegant scaling behavior of the correlators. In particular, the increase in slope of the ratios as a function of $N$, arising from the monotonicity of the anomalous dimensions, occurs both with and without the use of tracks, as is clearly seen in \Fig{fig:ratioNto2_pythia_250_xL_reg_normfreehad}. However, at a quantitative level, the use of tracks modifies the logarithmic scaling in a computable fashion. Overall, we see good agreement between our analytic calculations and \textsc{Pythia}, particularly for smaller values of $N$. For larger $N$, due to the larger anomalous dimensions, we expect relatively larger higher-order corrections. Furthermore, in \Fig{fig:ratioNto2_pythia_ana} we see good convergence between the LL and NLL.  Overall, we find that perturbative region is well understood, and that the logarithmic modifications to the scaling behavior can be incorporated perturbatively using track functions. This will enable precision measurements of the scaling behavior to be used for extractions of the strong coupling constant. Calculations of the ratio of the three-point/two-point projected correlators on all hadrons were recently extended to NNLL, along with an investigation of prospects for extracting the strong coupling \cite{Chen:2023zlx}. It will be important to extend the track-based calculations presented here to NNLL as well.

  %

%%%%%%%%%%%%%%%%%%%%%%
\section{Non-Perturbative Power Corrections}\label{sec:NP}
%%%%%%%%%%%%%%%%%%%%%%  

In this section we perform a simple phenomenological study of non-perturbative power corrections to the track-based projected energy correlators. We begin by deriving the power-law scalings of the leading non-perturbative power corrections for track-based energy correlators, and fit for the values of these parameters using \textsc{Pythia}. We then apply these to present numerical results for the resummed track-based energy correlators including leading non-perturbative power corrections. This section is primarily to illustrate some simple observations we have made, and to draw attention to developing a better understanding of the non-perturbative corrections to energy correlators. We hope that more detailed investigations will be performed in the future, particularly once high-quality data is available.

%%%%%%%%%%%%%%%%%%%%%%
\subsection{Structure of Power Corrections and Fitting}\label{sec:NP_fit}
%%%%%%%%%%%%%%%%%%%%%%  

A standard (but overly simplistic) method of studying non-perturbative corrections to observables in the absence of data is to compare the observables at parton level and hadron level using a parton shower Monte Carlo program that implements a model of the hadronization process. Unfortunately, this cannot be done for track-based observables, since they are only defined at hadron level. This makes the comparison more subtle. Although one can compare perturbative calculations incorporating track functions to a hadron-level parton shower Monte Carlo, such comparisons can be misleading due to the fact that the perturbative ingredients used in analytic calculations and in the parton shower are not equivalent. Disagreements in these perturbative ingredients can then be misinterpreted as non-perturbative corrections. 

In the absence of data, we therefore take a pragmatic approach: we first adjust the factorization scale in our perturbative calculation of the energy correlators on all particles so as to achieve agreement with parton level \textsc{Pythia}. We then compare our track function-based calculation to hadron-level \textsc{Pythia}. While this is imperfect, we believe that it is the best we can do without data, and we will show that we are able to draw a consistent picture from this approach. We hope that in the future a more detailed analysis can be performed using higher-order perturbative calculations, and comparisons with real data.

For the all-hadron energy correlator it is known \cite{Belitsky:2001ij,Korchemsky:1999kt,Korchemsky:1997sy,Korchemsky:1995zm,Korchemsky:1994is} that the leading power correction is additive, and has a $1/x_L^{1.5}$ scaling law, namely
%%%
\begin{align}
\text{EEC}(x_L)=\text{EEC}_\text{pert}(x_L)+\frac{\Lambda_1}{x_L^{1.5}}\,.
\end{align}
%%%
This was recently calculated using renormalon techniques \cite{Schindler:2023cww}. Generalizing to the projected $N$-point correlator, which we denote with PNC,
%%%
\begin{align}
\text{PNC}(x_L)=\text{PNC}_\text{pert}(x_L)+\frac{\Lambda_1^{(n)}}{x_L^{1.5}}\,.
\end{align}
%%%
In addition to the relatively well-understood leading power correction, we expect there to be a subleading power correction proportional to the transverse momentum scale of the pair being correlated, corresponding to a scaling of $1/x_L$. This subleading power correction can potentially be  important when making comparisons to calculations with tracks, since we also expect $1/x_L$ power corrections to the track function formalism. When fitting for hadronization corrections for the projected energy correlators, we will therefore use the following functional form
\begin{align}
\text{PNC}(x_L)=\text{PNC}_\text{pert}(x_L)+\frac{\Lambda_1^{(n)}}{x_L^{1.5}}+\frac{\Lambda_2^{(n)}}{x_L}\,.
\label{eq:fit_form}
\end{align}

\begin{figure}
\captionsetup[subfigure]{labelformat=empty}
  \begin{center}
  \subfloat[]{
  \includegraphics[scale=0.25]{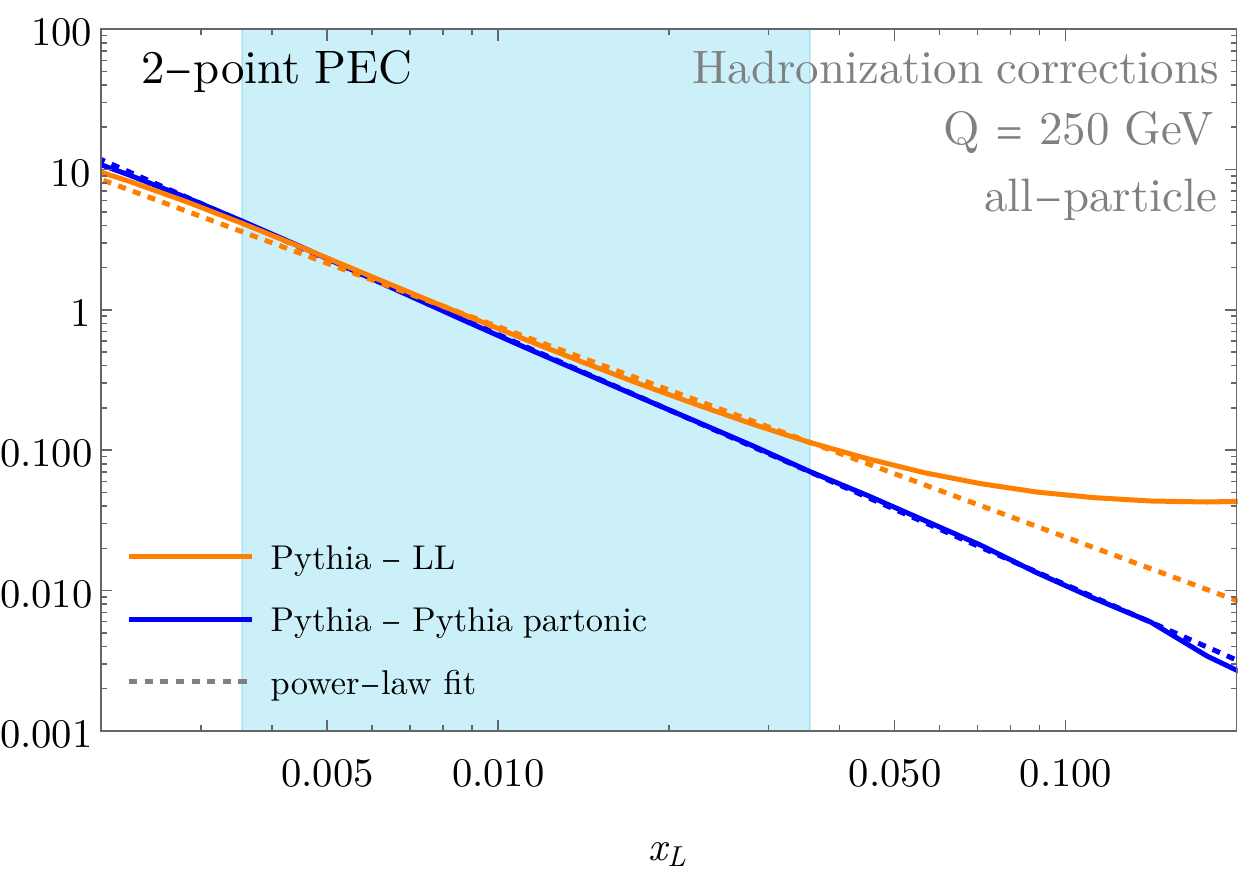}
  \label{fig:ratioNto2_2pt_a}
  }
  \qquad 
  \subfloat[]{
  \includegraphics[scale=0.25]{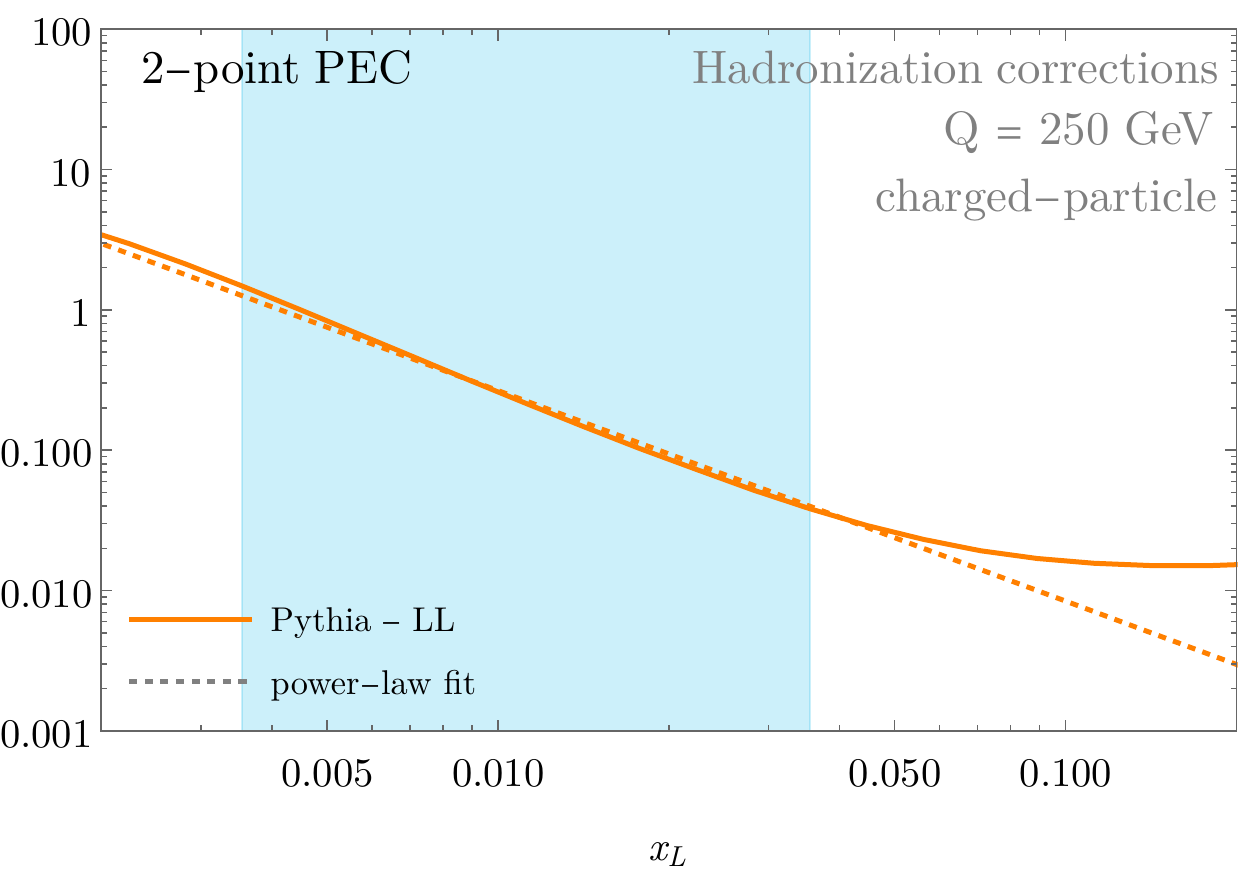}
  \label{fig:ratioNto2_2pt_b}
  }
  \\
 \vspace{-0.9cm} 
  \subfloat[]{
  \includegraphics[scale=0.25]{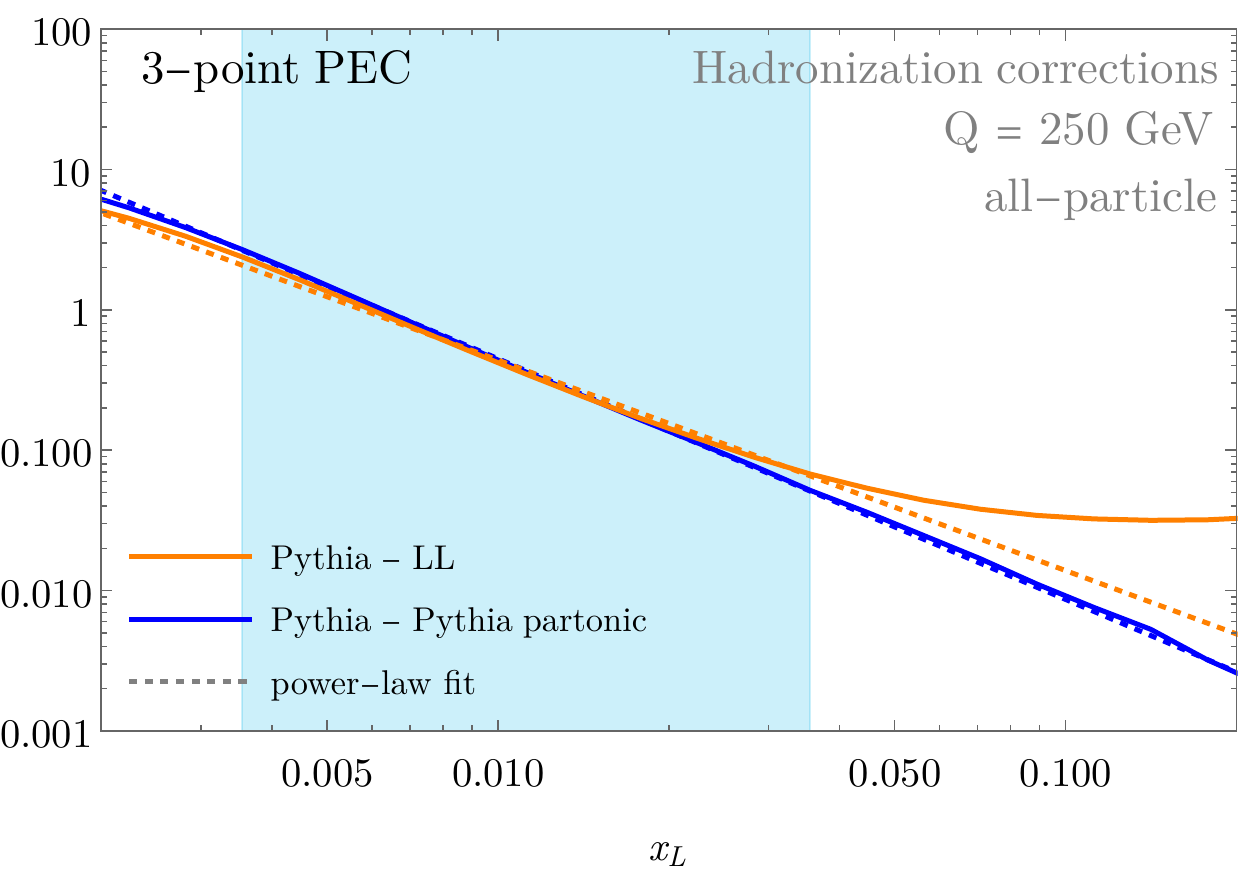}
  \label{fig:ratioNto2_3pt_a}
  }
  \qquad 
  \subfloat[]{
  \includegraphics[scale=0.25]{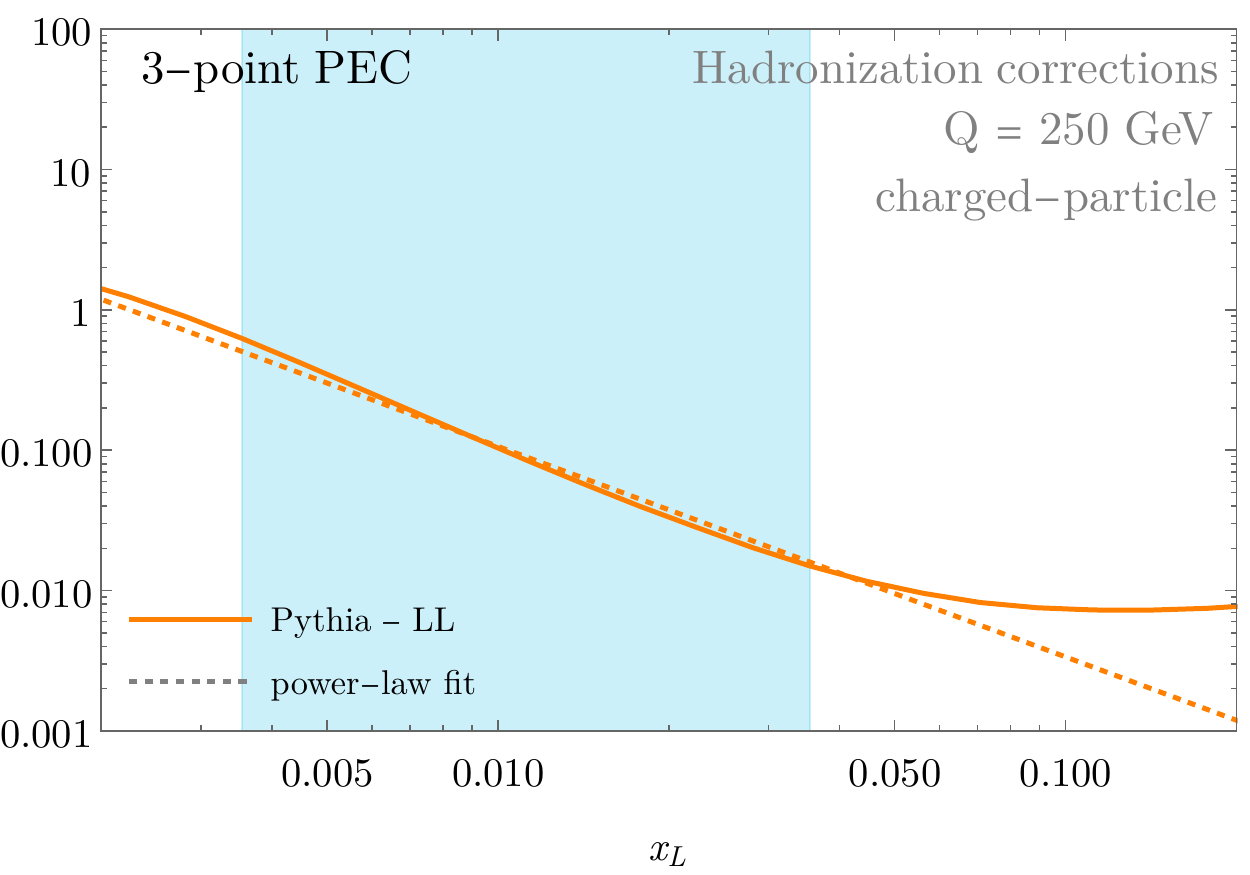}
  \label{fig:ratioNto2_3pt_b}
  }
  \\
 \vspace{-0.9cm}   
  \subfloat[]{
  \includegraphics[scale=0.25]{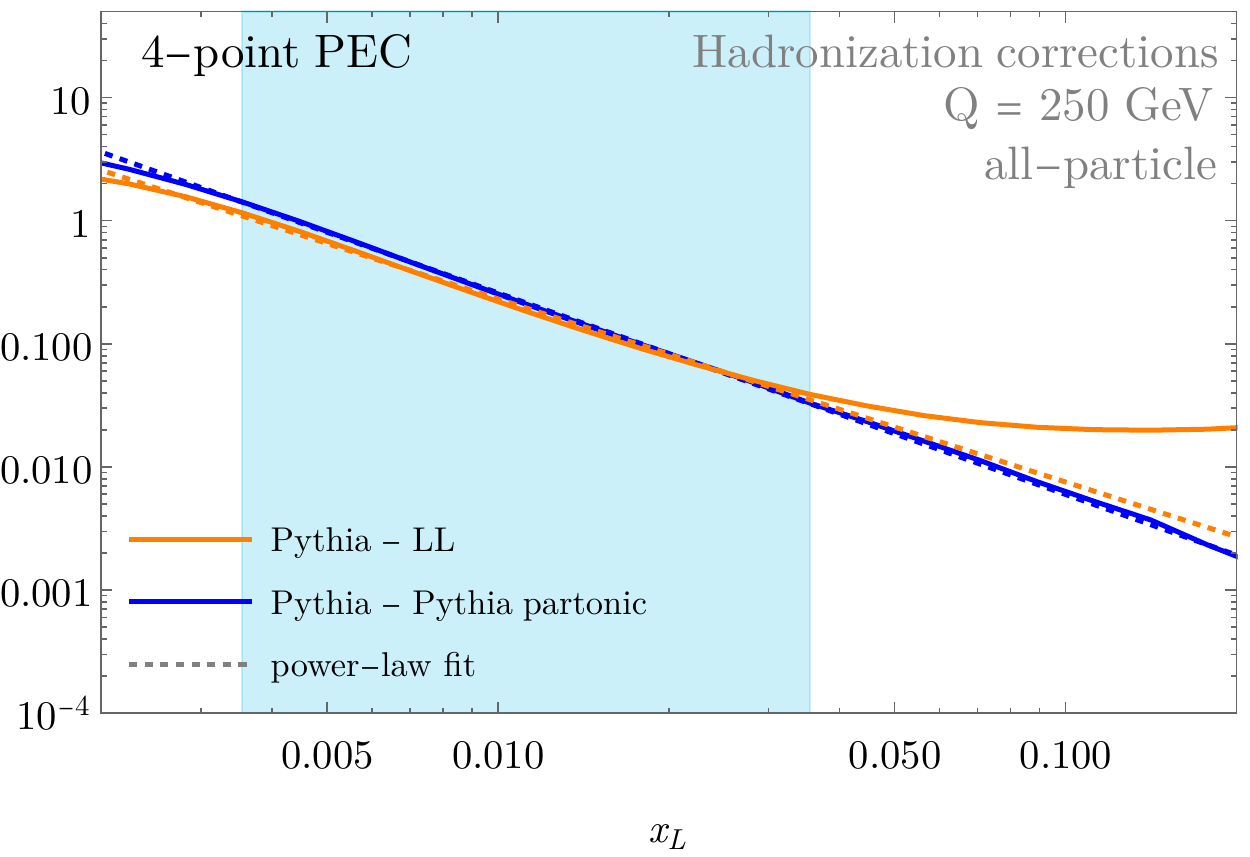}
  \label{fig:ratioNto2_4pt_a}
  }
  \qquad 
  \subfloat[]{
  \includegraphics[scale=0.25]{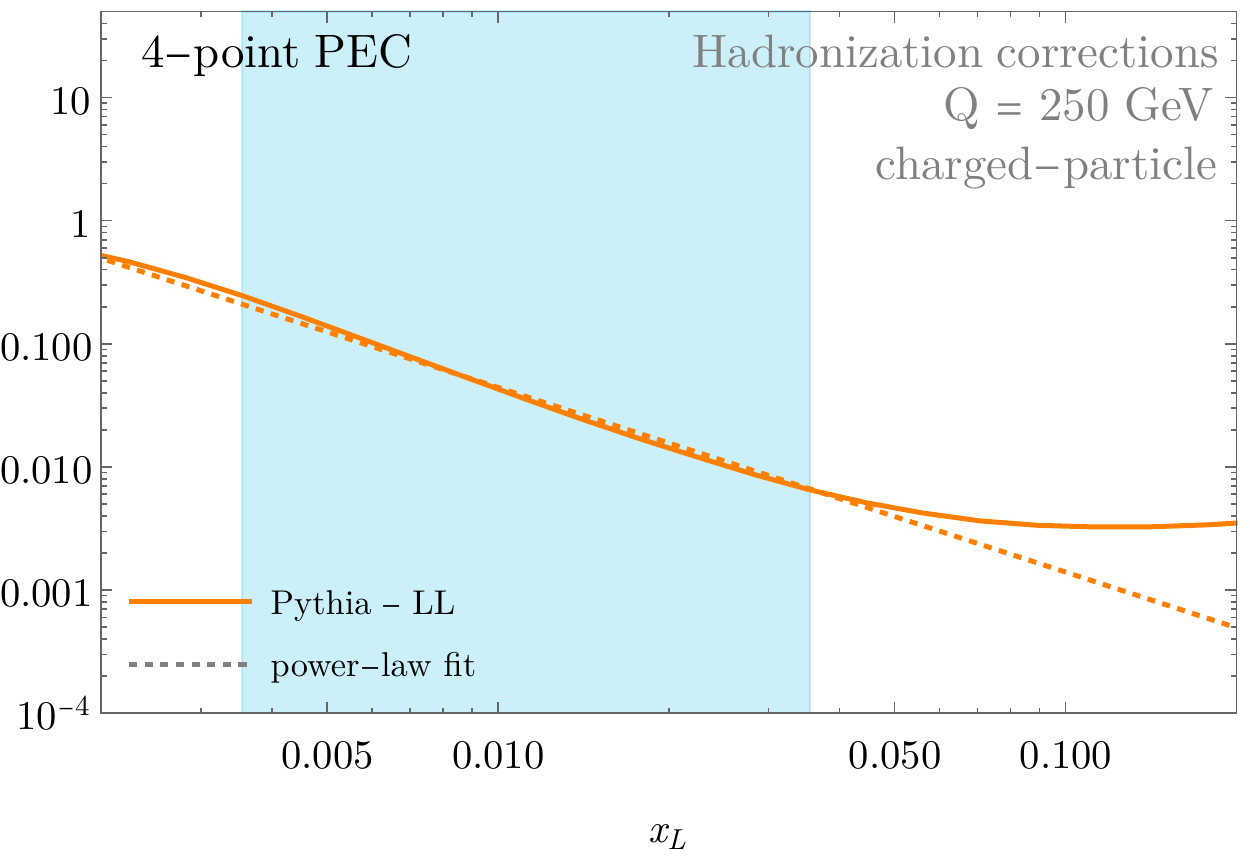}
  \label{fig:ratioNto2_4pt_b}
  }
  \\
   \vspace{-0.9cm} 
\subfloat[]{
    \includegraphics[scale=0.25]{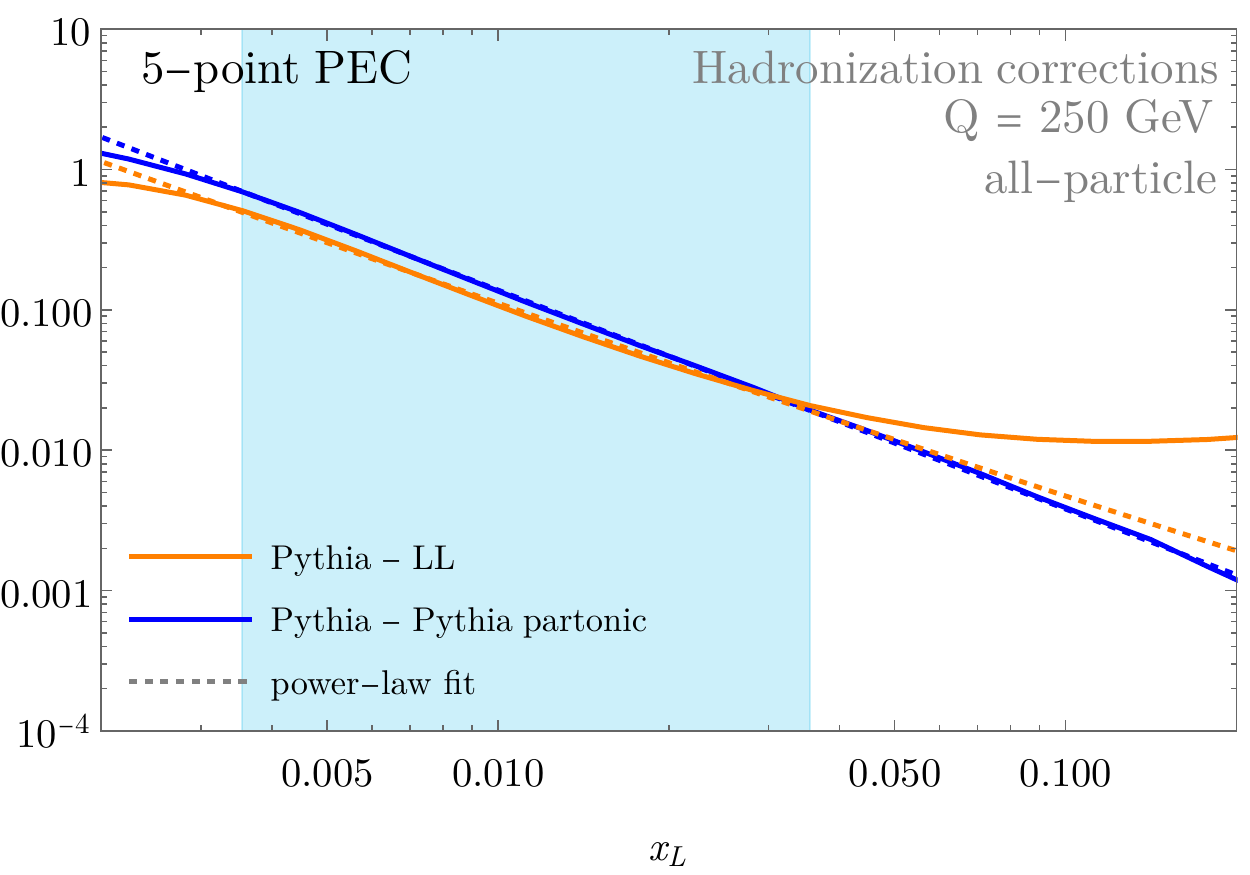}
    \label{fig:ratioNto2_5pt_a}
    }
    \qquad 
    \subfloat[]{
    \includegraphics[scale=0.25]{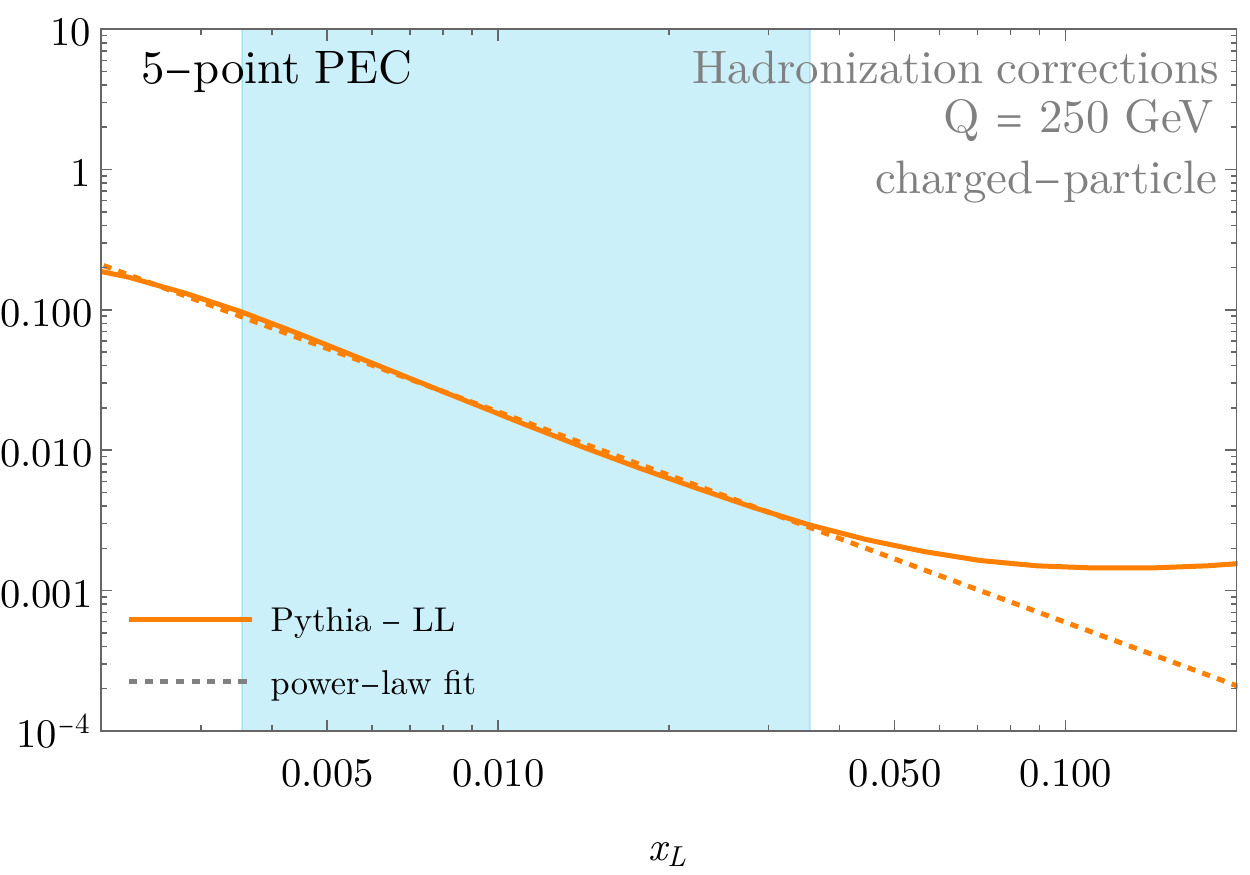}
    \label{fig:ratioNto2_5pt_b}
    }
  \\
   \vspace{-0.9cm} 
\subfloat[]{
    \includegraphics[scale=0.25]{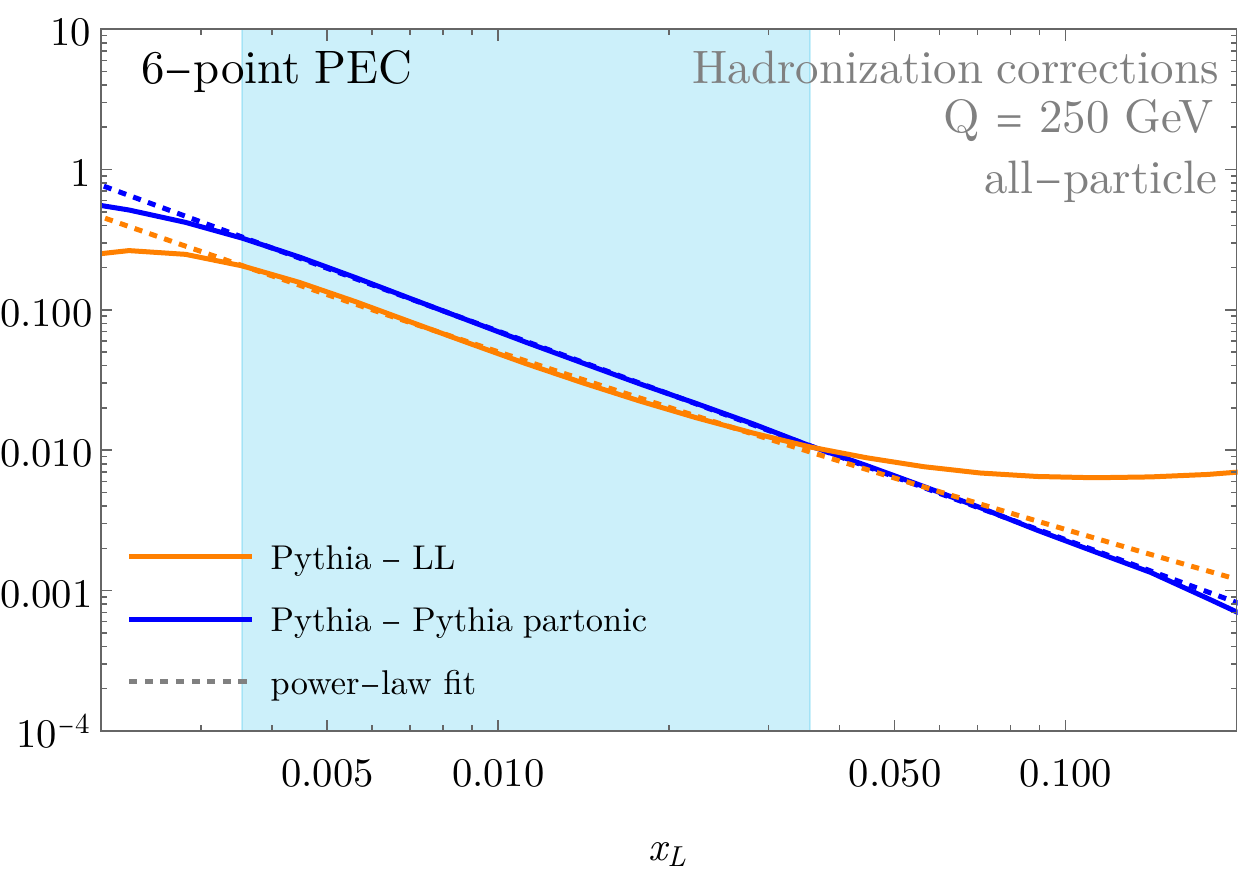}
    \label{fig:ratioNto2_6pt_a}
    }
    \qquad 
    \subfloat[]{
    \includegraphics[scale=0.25]{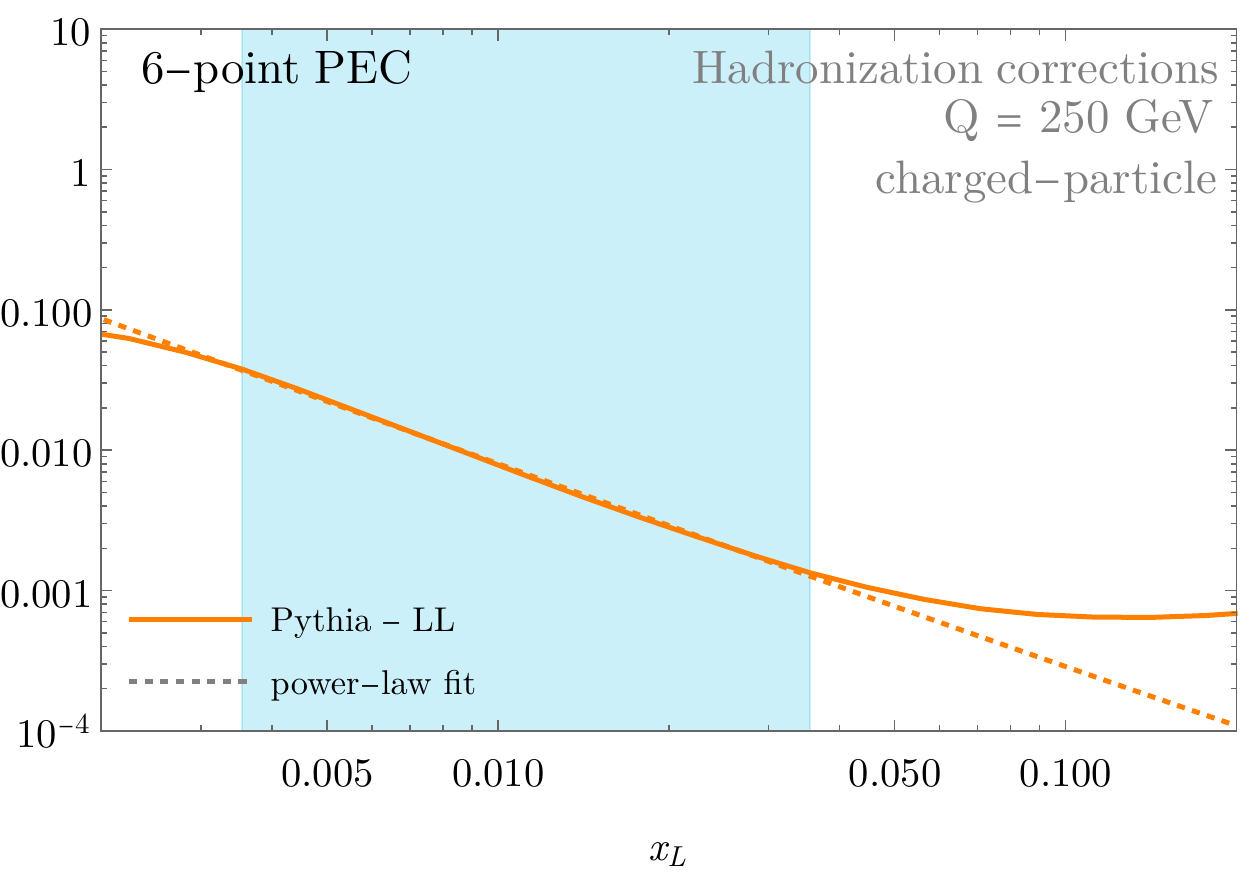}
    \label{fig:ratioNto2_6pt_b}
    }    
  \end{center}
     \vspace{-1.0cm} 
  \caption{Fits (dashed) to the leading non-perturbative corrections for the $N$-point projected energy correlators, obtained as the difference of \textsc{Pythia} and \textsc{Pythia} partonic (blue solid) or our leading logarithmic calculation (orange solid), for all particles (left) and on tracks (right). The fit is performed for the blue region, and the fitted parameters are given in table~\ref{tab:fitting_hadcor}.}
  \label{fig:NP_fits}
  \end{figure}

We now consider the case of the projected energy correlators on tracks. Since the scaling of the leading non-perturbative power correction is a result of boost invariance \cite{Belitsky:2001ij,Korchemsky:1999kt,Korchemsky:1997sy,Korchemsky:1995zm,Korchemsky:1994is}, we do not expect its power law scaling to be changed when measured on tracks, only the value of the non-perturbative parameter. Similarly, we expect there to be power corrections to the track function formalism that scale like the virtuality between the detected particles and modify the $1/x_L$ power correction. We therefore parameterize the leading non-perturbative power corrections to the projected energy correlators on tracks as
\begin{align}
\text{PNC}^\text{tr}(x_L)=\text{PNC}^\text{tr}_\text{pert}(x_L)+\frac{\Lambda_{\text{tr},1}^{(n)}}{x_L^{1.5}}+\frac{\Lambda^{(n)}_{\text{tr},2}}{x_L}\,.
\end{align}
In this case $\text{PNC}^\text{tr}_\text{pert}$ is not strictly perturbative, since it also contains the track functions.

\begin{figure}
  \begin{center}
    %\subfloat[]{
    \includegraphics[scale=0.50]{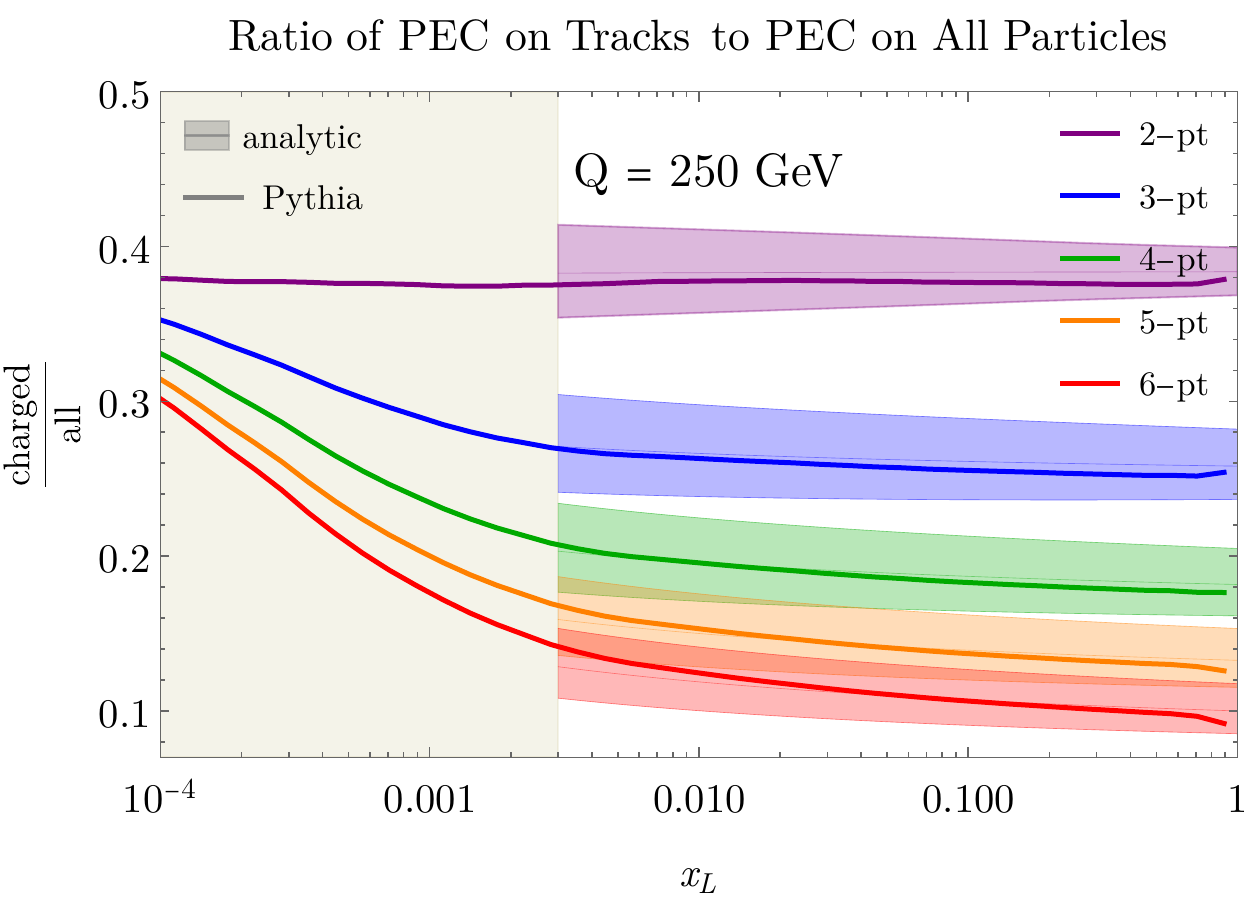}
    \label{fig:track_evo_1}
    %}
  \end{center}
  \caption{A comparison of the ratio of the energy correlators computed on tracks and all hadrons, obtained from \textsc{Pythia} and for our LL result with scale $\mu=Q/10$, as is used in our fits for non-perturbative corrections.}
  \label{fig:Q10_evo}
\end{figure}

The non-perturbative power correction scaling like $1/x_L$ has the same power law as the perturbative component of the energy correlators (up to logarithms). It is for this reason that it is essential that our perturbative results agree with parton level \textsc{Pythia}. Otherwise, these disagreements will be absorbed into this non-perturbative parameter.  We tune the hard scale in our analytic calculations to achieve agreement between parton-level \textsc{Pythia} and our analytic results. This is achieved by setting the hard scale to $\mu=Q/10$ for the all-particle case and, which is then used for the charged-particle case as well. One might worry that the use of such a low scale could complicate interpretations of our results in terms of moments of the track functions, which are also now evaluated at much lower scales. However, due to their extremely slow running, this is not a problem.  In \Fig{fig:Q10_evo}, we show a comparison of the ratio of the charged/all-hadron projected energy correlators using the scale $\mu=Q/10$, illustrating good agreement.

In \Fig{fig:NP_fits} we show results of our fits for the leading non-perturbative power corrections. For both all hadron and charged hadrons, we compute the difference between  \textsc{Pythia}  and our leading logarithmic calculation (which is denoted by $\text{Pythia}-\text{LL}$). We emphasize that we only use this scale in this section for the specific reason of achieving agreement with parton-level \textsc{Pythia} to extract non-perturbative parameters. Having achieved agreement between our analytic results and partonic level \textsc{Pythia}, the residuals in \Fig{fig:NP_fits} are then fit to the power law form in \Eq{eq:fit_form} to extract the values for the non-perturbative parameters. We performed this fit for the $N$-point projected correlator with $N=2$ to 6. The fitting range is $[0.00355,0.0355] = [10^{(-2.45)},10^{(-1.45)}]$ which is
within the perturbative region (the upper bound of the transition region is $x_L=0.003$). For the all-hadron case, we also show the difference \textsc{Pythia}$-$\textsc{Pythia} partonic.

As discussed previously, it is crucial for our fitting procedure that our perturbative result accurately reproduces the perturbative inputs of \textsc{Pythia}. Otherwise perturbative differences will appear as contributions to the non-perturbative parameters. We have found that the agreement  gradually gets worse as $N$ is increased, and we were only able to achieve sufficiently good agreement for $N=2,3,4$, see \Fig{fig:NP_fits}. For this reason, we restrict to these values of $N$ when studying the non-perturbative power corrections. It would be interesting to perform a more systematic study in the future.

\renewcommand{\arraystretch}{1.5}
\begin{table}
\centering
\begin{tabular}{|c|c|c|c|}
\hline
%\diagbox{}{}{}
& 
2-point & 
3-point & 
4-point  \\ \hline
all-particle & $\frac{\Lambda^{(2)}_1}{x_L^{1.5}}+\frac{\Lambda^{(2)}_2}{x_L}$ & 
$\frac{\Lambda^{(3)}_1}{x_L^{1.5}}+\frac{\Lambda^{(3)}_2}{x_L}$ &
$\frac{\Lambda^{(4)}_1}{x_L^{1.5}}+\frac{\Lambda^{(4)}_2}{x_L}$      \\ \hline
$\Lambda^{(n)}_1$ & $0.00076\pm 0.00006$ & $0.00044\pm 0.00004$ 
    & $0.000229\pm 0.000013$      \\  
$\Lambda^{(n)}_2$ & $7\times 10^{-9}\pm 0.0005$ & $1.0\times 10^{-9}\pm 0.00028$   
    & $0.000031\pm 0.00011$    \\ \hline
charged-particle & $\frac{\Lambda^{(2)}_{\text{tr},1}}{x_L^{1.5}}+\frac{\Lambda^{(2)}_{\text{tr},2}}{x_L}$ & $\frac{\Lambda^{(3)}_{\text{tr},1}}{x_L^{1.5}}+\frac{\Lambda^{(3)}_{\text{tr},2}}{x_L}$   
    & $\frac{\Lambda^{(4)}_{\text{tr},1}}{x_L^{1.5}}+\frac{\Lambda^{(4)}_{\text{tr},2}}{x_L}$    \\ \hline
$\Lambda^{(n)}_{\text{tr},1}$ & $0.000266\pm 0.000023$ & $0.000106\pm 0.000013$   
    & $0.000044\pm 0.000004$    \\ 
$\Lambda^{(n)}_{\text{tr},2}$ & $0\pm 0.00019$ & $0\pm 0.00011$   
    & $0\pm 0.00004$    \\ 
$\Lambda^{(n)}_{\text{tr},1}/\Lambda^{(n)}_1$ & 0.35 & 0.24 & 0.19
\\ \hline
\end{tabular}
\caption{The results of our fits for the leading non-perturbative power corrections for the projected 2-, 3- and 4-point energy correlators.} \label{tab:fitting_hadcor}
\end{table}
\renewcommand{\arraystretch}{1.0}

The results of our fits for the non-perturbative parameters are shown in table~\ref{tab:fitting_hadcor}. Currently there is no quantitative understanding of the structure of power corrections to the track functions. Therefore, we focus only on understanding the relation between the leading ($1/x_L^{1.5}$) non-perturbative correction for the track-based vs.~all-hadron energy correlator. Intuitively, we expect these to be related to moments of the track function. Indeed, the standard picture of the leading non-perturbative correction is that is comes from the emission of a ``non-perturbative" gluon, on which one of the energy detectors is placed. In the case of the projected energy correlators, only one detector can lie on this low energy gluon, or else it will be further power suppressed. Therefore, one is led to conjecture that
\begin{align}
\Lambda^{(n)}_{\text{tr},1}=T_q(n-1)T_g(1) \Lambda_1^{(n)}.
\label{eq:relate_NP}
\end{align}
This also agrees with what would be obtained by combining the recent renormalon calculation of ref.~\cite{Schindler:2023cww} with track functions. Note that this is the same ratio that governs the rescaling of the perturbative component of the distribution when one changes from all hadrons to tracks, see \Fig{fig:track_evo_compare}. Therefore, if the effect of tracks is to rescale both the perturbative component, and the leading non-perturbative correction by the overall factor, then the dominant difference arises only at the level of the first subleading non-perturbative power correction.

We see that the results of table~\ref{tab:fitting_hadcor} are consistent with the conclusion of \Eq{eq:relate_NP}, providing strong support to this picture. This result is particulary pleasing for the precision study of ratios of the $N$-point/2-point energy correlators for precision measurements of the strong coupling. At the all-hadron level, the dominant non-perturbative power corrections largely cancel in this ratio. Once tracks are incorporated, since the perturbative part of the distribution, and the leading non-perturbative power correction are rescaled by the same factor, this cancellation continues. 

One feature of the non-perturbative power corrections that we do not understand is the results of the fitting for $\Lambda_{2,\text{tr}}^{(n)}$, which were found to be very small (compatible with zero), but with a large uncertainty. It is not clear if this is an artifact of our fitting procedure, namely that it is difficult to disentangle the different power laws, or that they were absorbed elsewhere. Indeed, the values of the moments of the track functions themselves were obtained by fitting with \textsc{Pythia}, and so it is possible that certain non-perturbative parameters were absorbed into that fit.

While this has only been a preliminary exploration, we believe that our results give a general consistent picture of the structure of non-perturbative power corrections for the track-based EEC. In particular, we believe that the leading non-perturbative power corrections with and without track functions are related by moments of the track functions, so that changes in the shape of the ratio are driven by subleading power corrections. It would be interesting to understand these issues in detail.

%%%%%%%%%%%%%%%%%%%%%%
\subsection{Numerical Results with Power Corrections}\label{sec:NP_result}
%%%%%%%%%%%%%%%%%%%%%%  

\begin{figure}
\captionsetup[subfigure]{labelformat=empty}
  \begin{center}
  \subfloat[]{
  \includegraphics[scale=0.34]{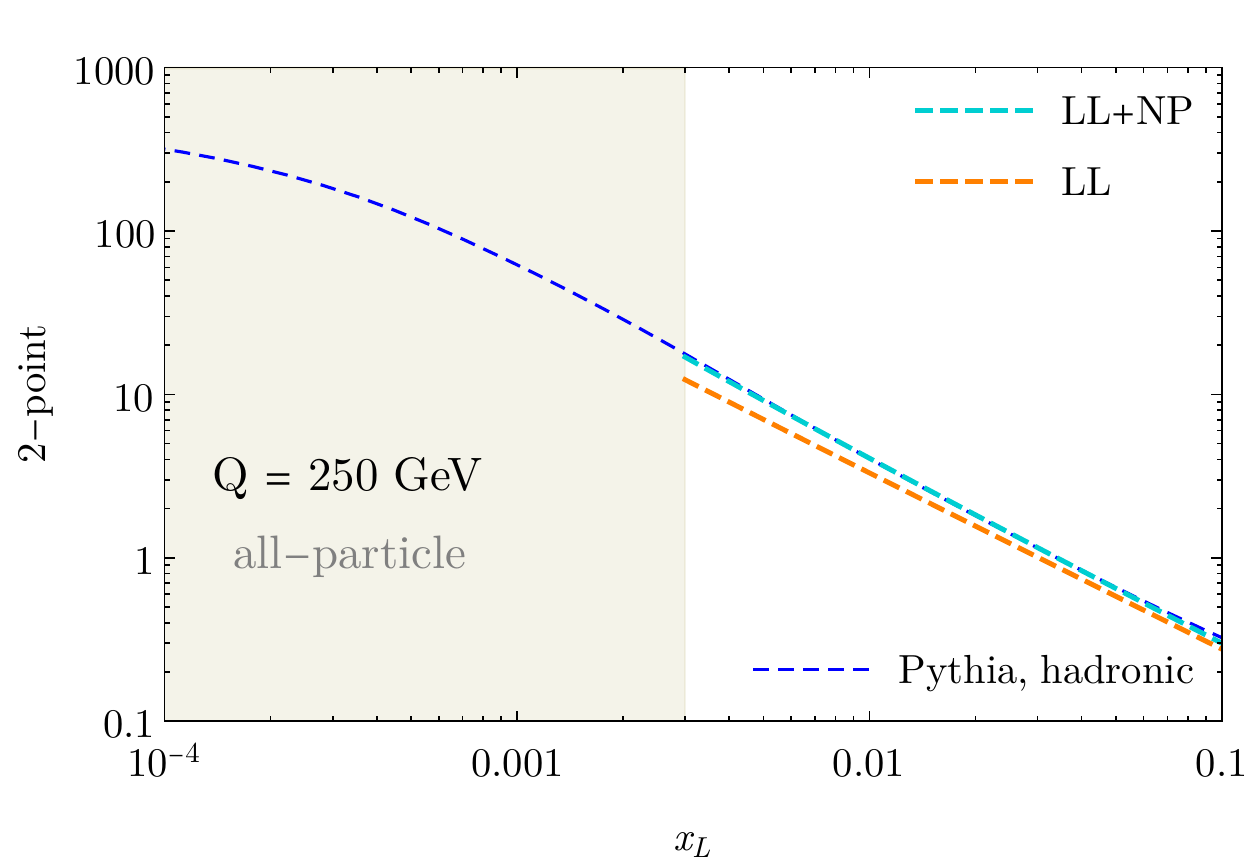}
  \label{fig:NP_2pt_all}
  }
  \quad 
  \subfloat[]{
  \includegraphics[scale=0.34]{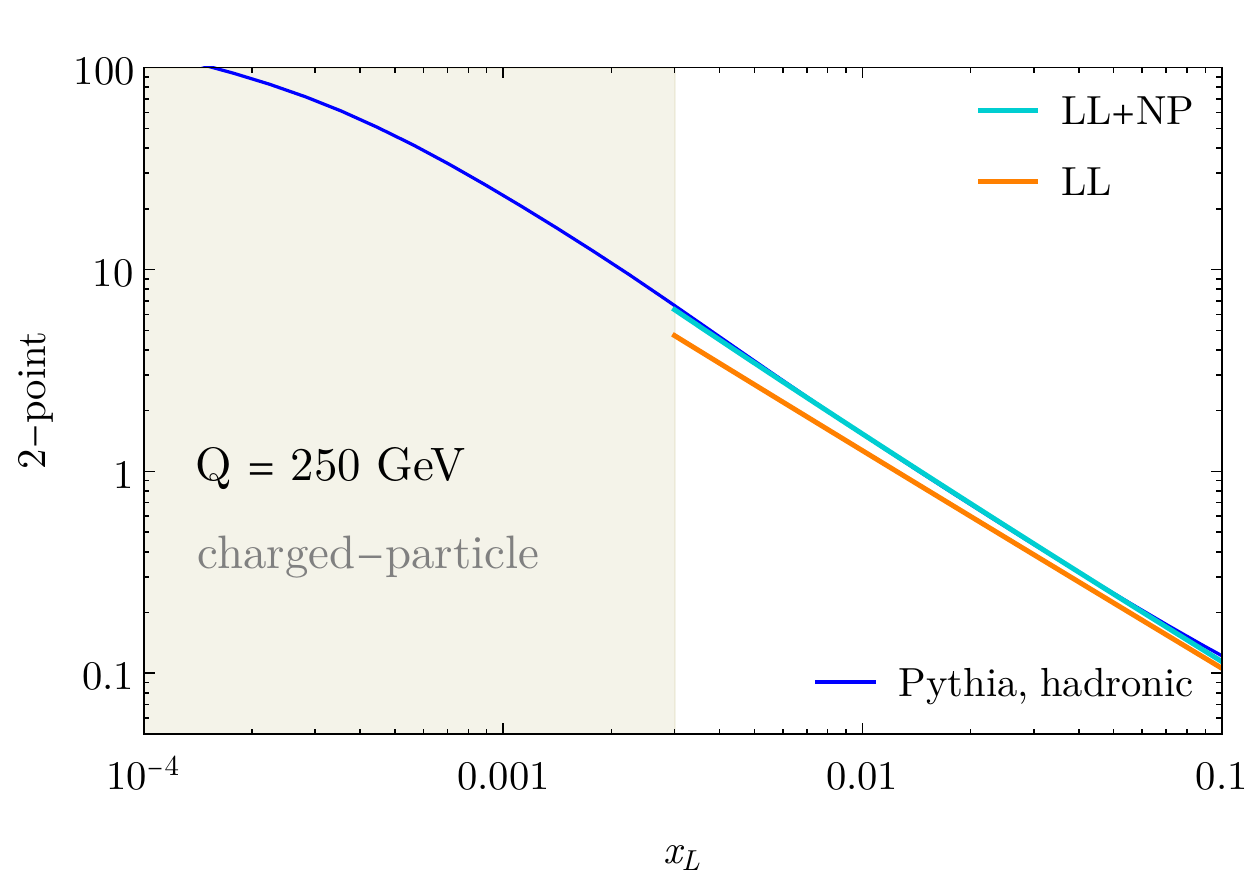}
  \label{fig:NP_2pt_track}
  }
  \\
 \vspace{-0.9cm} 
  \subfloat[]{
  \includegraphics[scale=0.34]{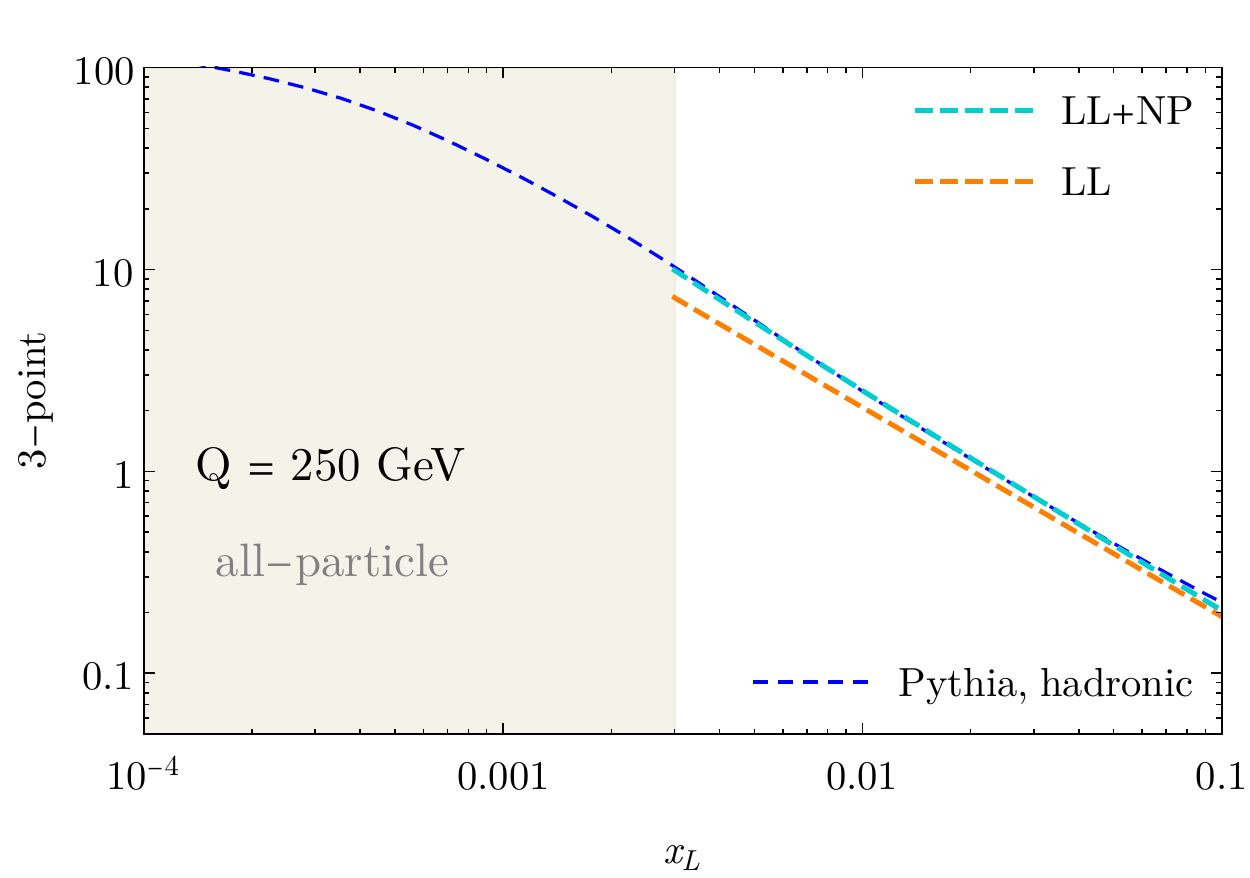}
  \label{fig:NP_3pt_all}
  }
  \quad 
  \subfloat[]{
  \includegraphics[scale=0.34]{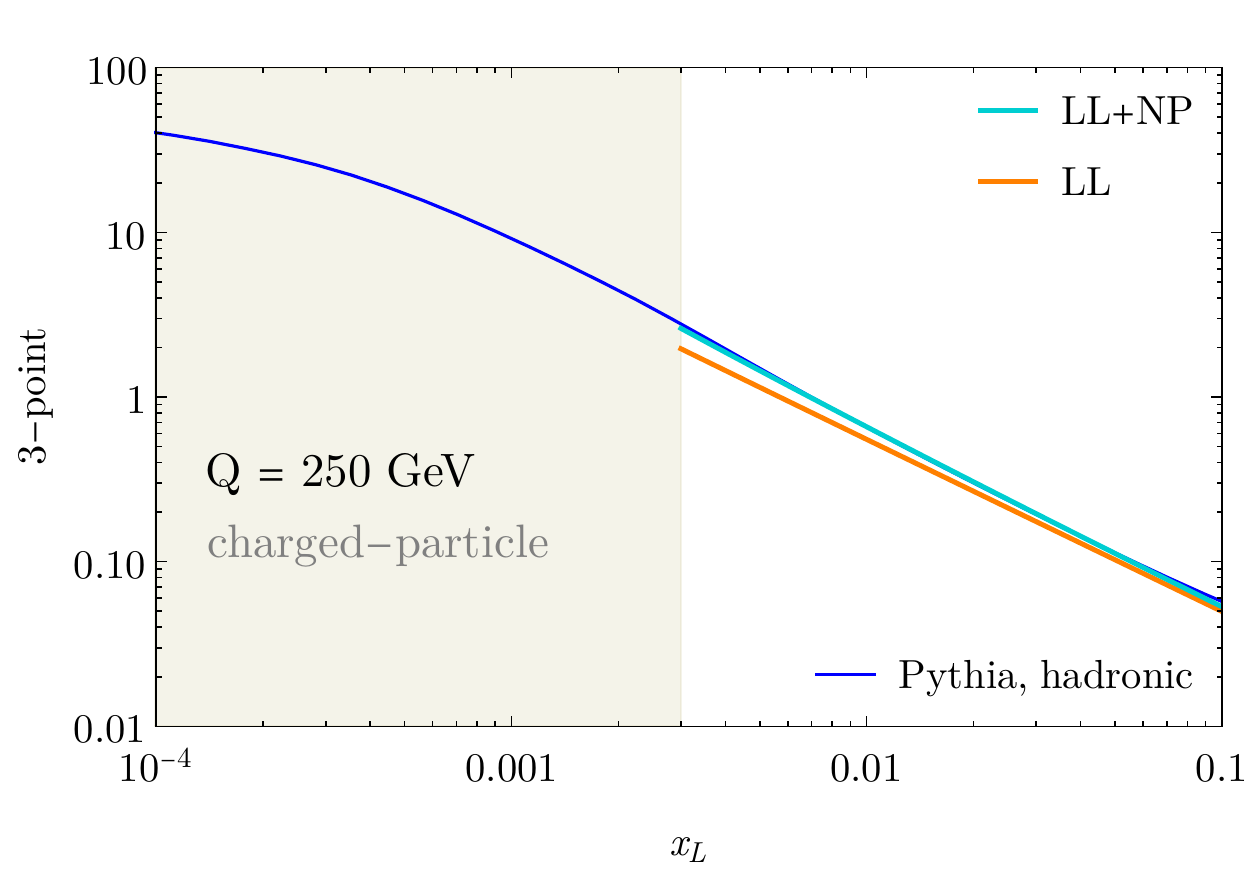}
  \label{fig:NP_3pt_track}
  }
  \\
 \vspace{-0.9cm}   
  \subfloat[]{
  \includegraphics[scale=0.34]{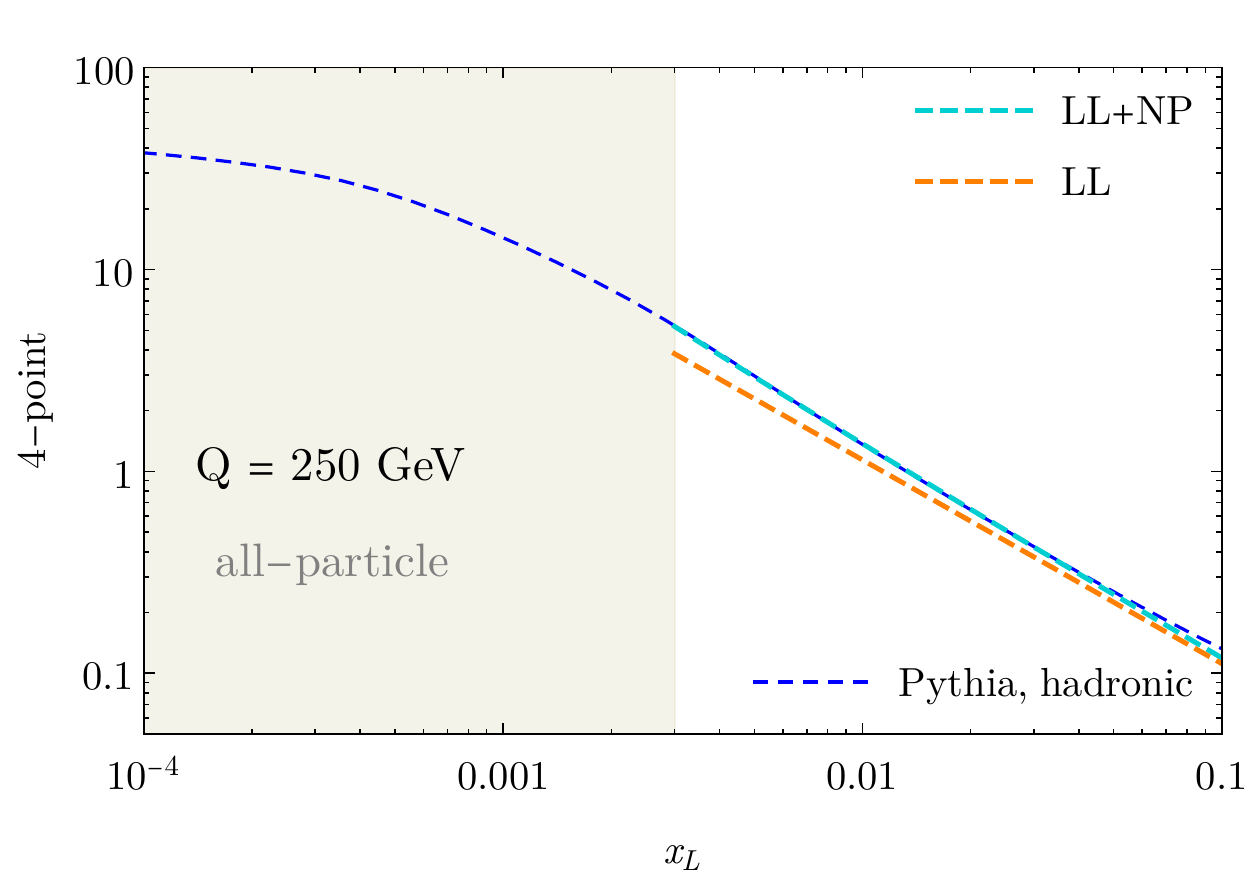}
  \label{fig:NP_4pt_all}
  }
  \quad 
  \subfloat[]{
  \includegraphics[scale=0.34]{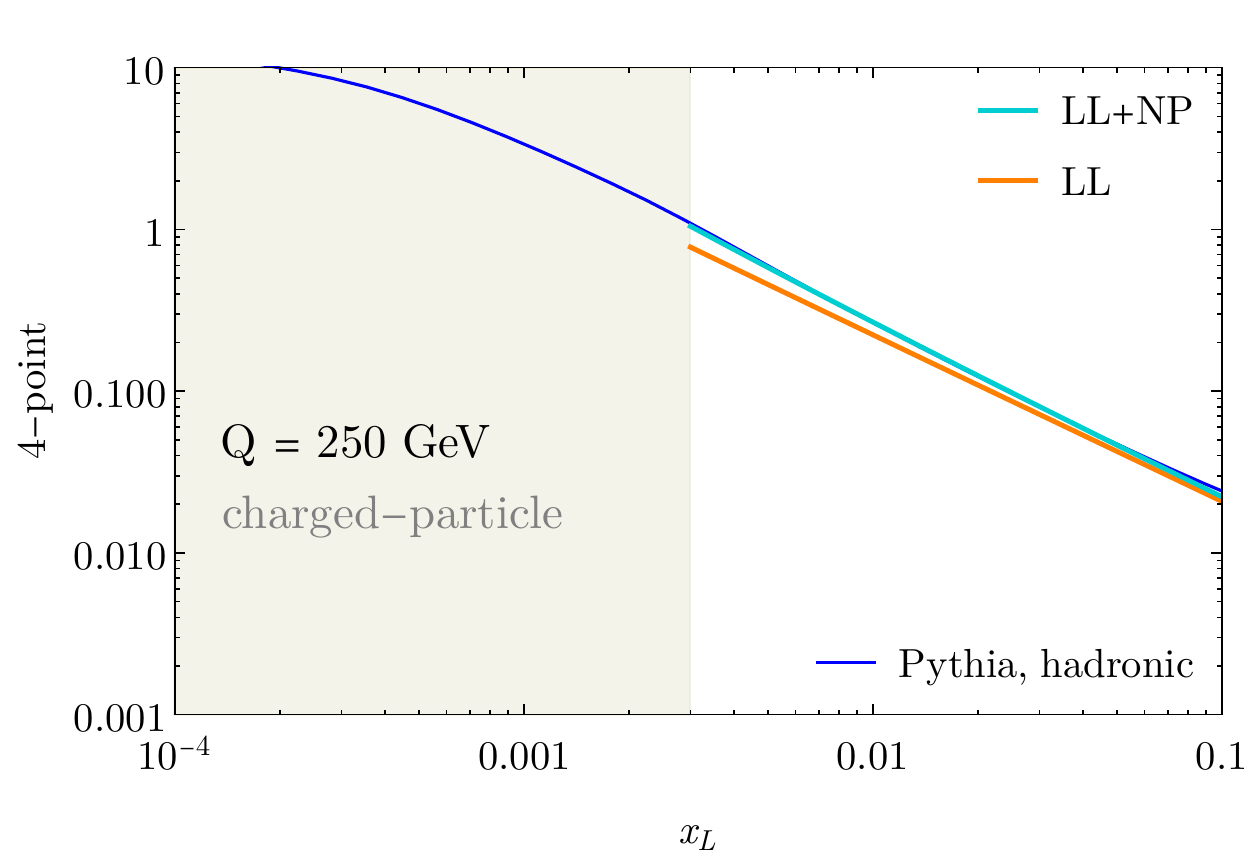}
  \label{fig:NP_4pt_track}
  }
  \end{center}
  \vspace{-1.0cm}
  \caption{The projected energy correlators at LL with the hard scale set to $\mu=Q/10$ as used in our fitting procedure (orange), and incorporating the non-perturbative power correction (green) compared to \textsc{Pythia} for all particles (left) and charged particles (right). In the shaded region the correlators are dominated by non-perturbative effects, and our results should not be trusted. %\WW{Mention $Q/10$? Move ``charged-particle" a bit away from the edge of the plot. Can we use a different color for LL+NP? (we use green for NLL elsewhere)}%\yl{yes, we should mention that here we set the nominal hard scale equals $Q/10$ in order to make the LL results agree with the partonic data in Pythia. For the color, I had the same concern as Wouter's when making the plots, but I'm so into the blue-green-yellow color scheme and thought that green should be regarded as the color for the results more close to the Pythia data. I can try to make new versions since I should move ``charged-particle''. }
  }
  \label{fig:plot_with_NP}
  \end{figure}

Having extracted the leading non-perturbative parameters, we are now able to make resummed predictions for the track-based EEC, including the leading power corrections. Due to the fact that we were only able to do the extraction of the non-perturbative parameters in a self-consistent way at LL accuracy, we will restrict ourselves to LL in this section. It will be interesting to extract the non-perturbative parameter from data in the future when experimental measurements become available.

In \Fig{fig:plot_with_NP} we show results at LL+NP for the two-, three- and four-point projected energy correlators, compared with results from \textsc{Pythia}. By construction the agreement is good, since we have used \textsc{Pythia} for the fit. The predictive power comes from the fact that we have fixed the power law describing the power corrections. Note that we have not included uncertainty bands from scale variations, since we did not incorporate them in our fitting procedure for the NP corrections. We see that the power corrections have a non-negligible effect towards smaller angles, and it will certainly be important to include them when comparing results to experimental measurements.

  \begin{figure}
\captionsetup[subfigure]{labelformat=empty}
  \begin{center}
  \subfloat[]{
  \includegraphics[scale=0.34]{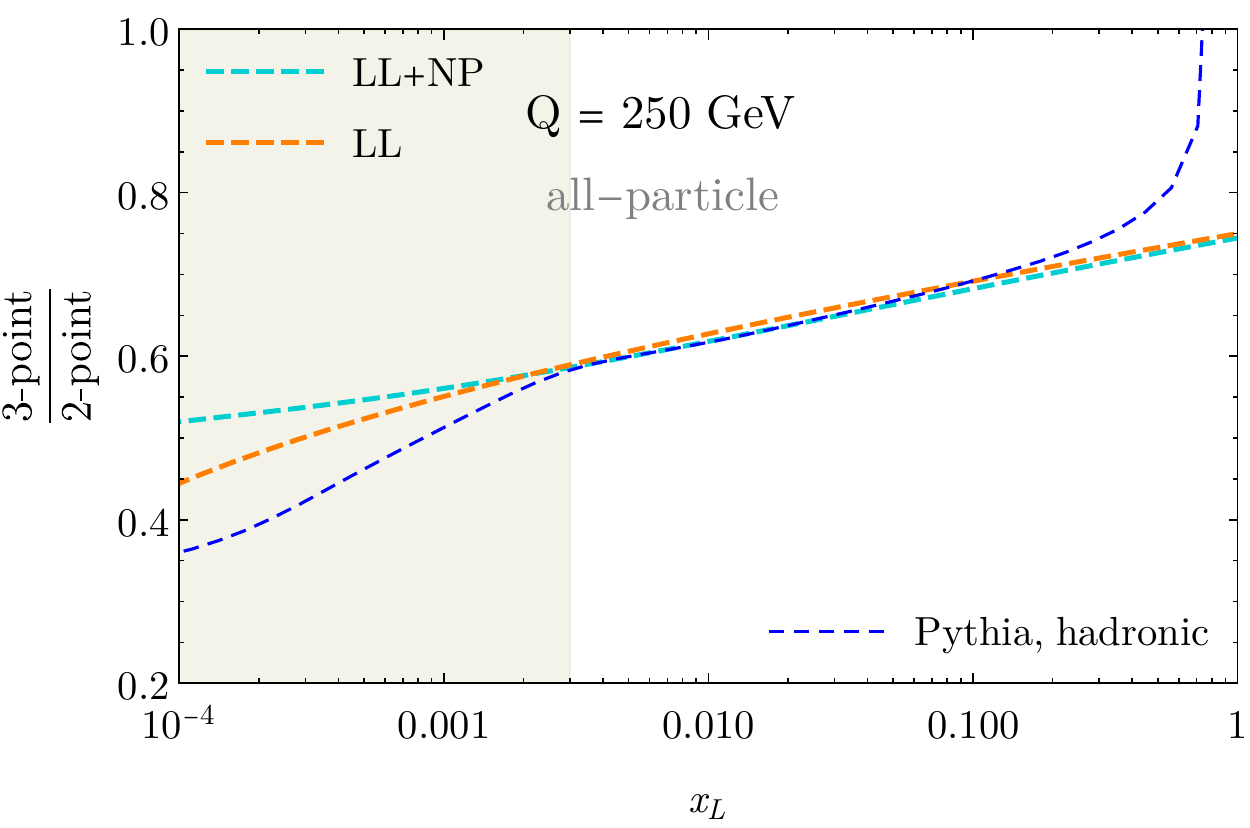}
  \label{fig:NP_2pt_all_ratio}
  }
  \quad 
  \subfloat[]{
  \includegraphics[scale=0.34]{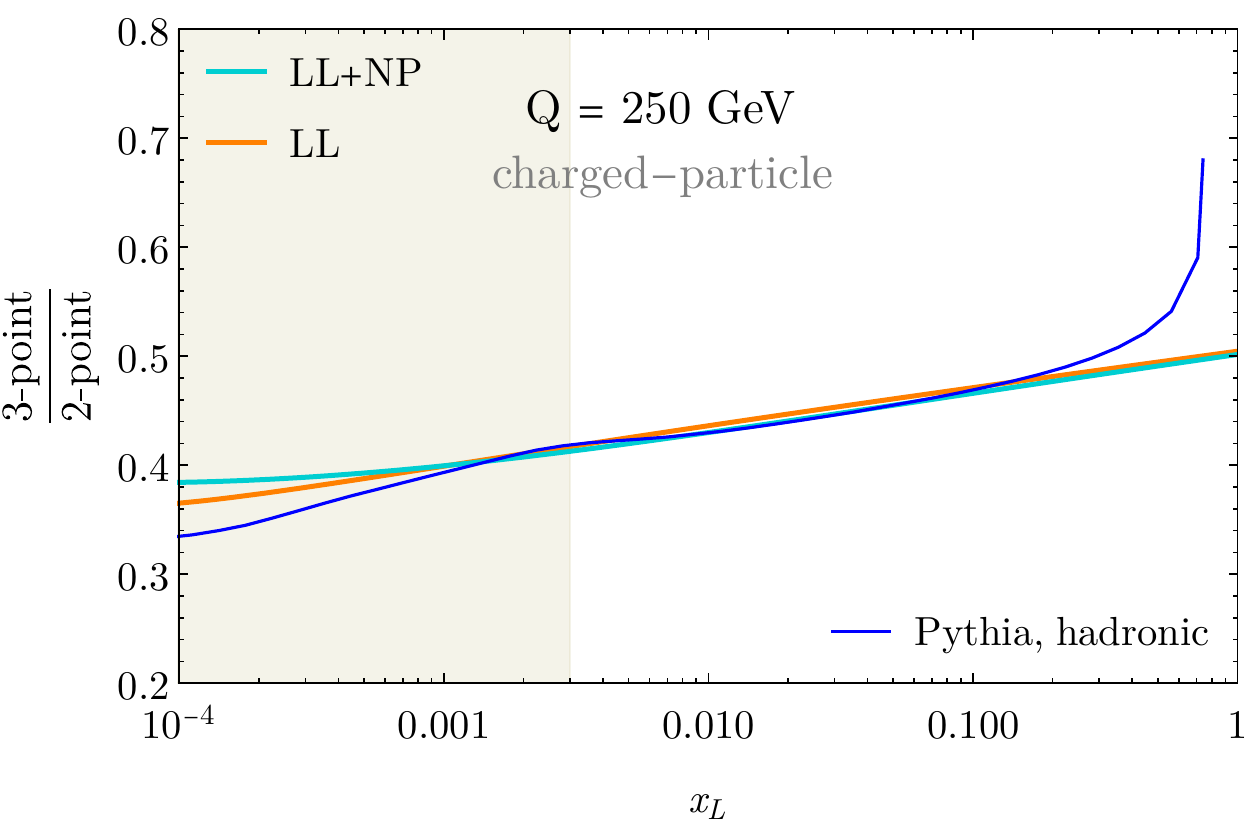}
  \label{fig:NP_2pt_track_ratio}
  }
  \\
 \vspace{-0.9cm} 
  \subfloat[]{
  \includegraphics[scale=0.34]{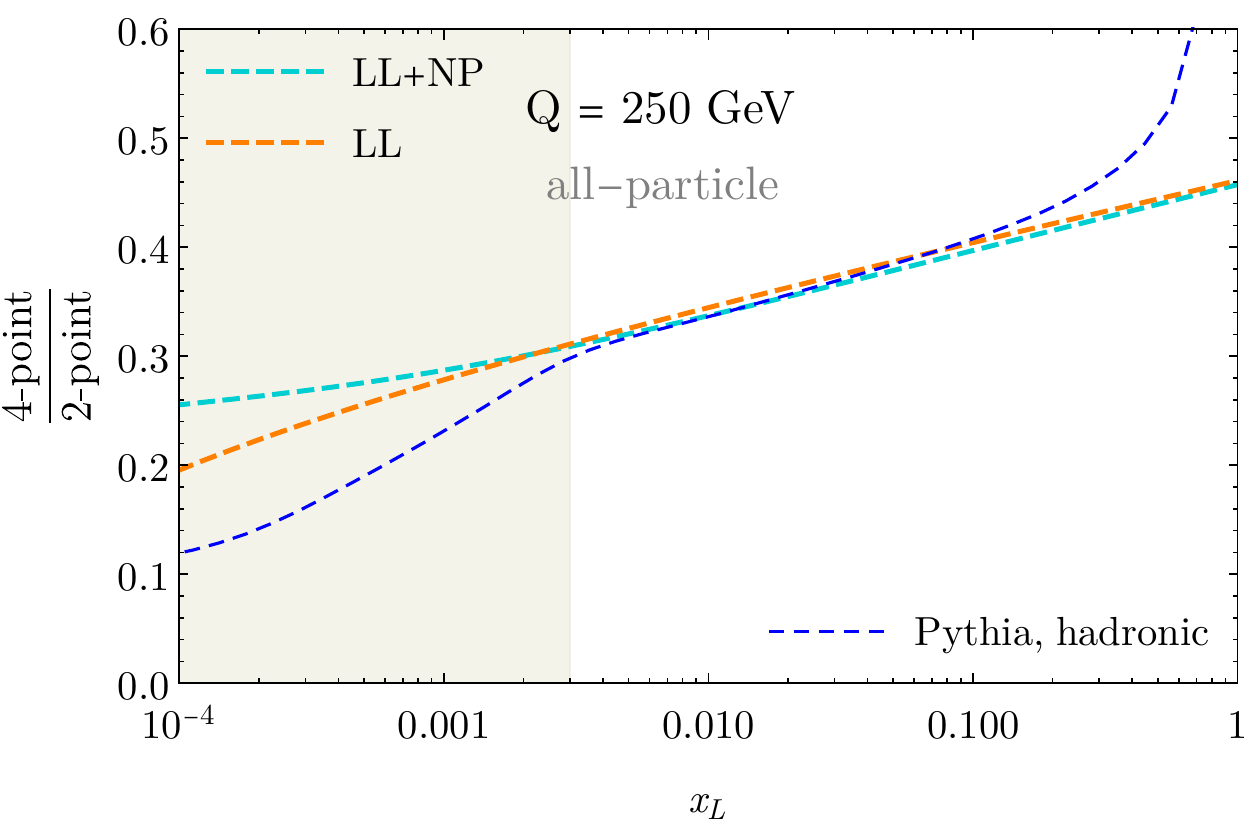}
  \label{fig:NP_3pt_all_ratio}
  }
  \quad 
  \subfloat[]{
  \includegraphics[scale=0.34]{figures/fig8_ratio3to2pt_250GeV_track.pdf}
  \label{fig:NP_3pt_track_ratio}
  }
  \end{center}
  \vspace{-1.0cm}
  \caption{Ratios of the projected energy correlators incorporating the non-perturbative power correction. The leading non-perturbative correction largely cancels in the ratio. In the shaded region the correlators are dominated by non-perturbative effects, and our result should not be trusted. }
  \label{fig:ratio_with_NP}
  \end{figure}

In \Fig{fig:ratio_with_NP} we show results for the ratios of projected energy correlators at  LL,  both with and without the NP power corrections. Here we see a key benefit of the ratio of the energy correlators, namely that the power corrections largely cancel in the ratio, and are hence irrelevant (at least to the perturbative order that we are working). This was highlighted in ref.~\cite{Chen:2020vvp}, where these observables were introduced. Importantly, due to the relation in \Eq{eq:relate_NP}, this continues to hold for the track-based energy correlator (Note that for higher point correlators it also relies on the fact that in QCD the product $T(N-k)T(k)$ is approximately the same for different values of $k$, as discussed previously). We believe that this will be crucial for precision measurements of $\alpha_s$ using jet substructure. A key difficulty in such measurement has been linear power corrections. Here we are able to eliminate these by taking the ratio, and extend the observable to tracks, which should enable extremely precise experimental measurements. This strongly motivates the calculation of these observables at higher perturbative orders. See ref.~\cite{Chen:2023zlx} for a recent extension to NNLL for the all-hadron case for $N=3$.

%%%%%%%%%%%%%%%%%%%%%%
\section{Conclusions}\label{sec:conc}
%%%%%%%%%%%%%%%%%%%%%%  

Energy correlators have provided a variety of new ways to study the dynamics of QCD inside high-energy jets at the LHC.  To maximize their potential will require the use of tracking information to achieve precise angular resolution in experimental measurements. This has motivated significant recent theoretical development of the track function formalism, which enables tracking information to be incorporated into precision perturbative calculations \cite{Chang:2013rca,Chang:2013iba,Li:2021zcf,Jaarsma:2022kdd,Chen:2022pdu,Chen:2022muj}. In this paper, we applied this formalism to understand the interplay of the track function formalism with perturbative resummation and non-perturbative effects in the energy correlator observable. Our results are timely for interpreting recent measurements of the energy correlators at ALICE and STAR \cite{talk1,talk2,talk3}, as well as future measurements. Our results are the first combination of the NLO track functions with resummation in a physical observable, and illustrate that track functions are practical for precision jet substructure calculations.

 A few key takeaways of our study that are important for the interpretation of experimental results are as follows: First, the two-point energy correlator is almost identical whether measured on tracks or all particles throughout the entire distribution. This allows for a clean interpretation of recent measurements of the non-perturbative transition to confinement measured using the two-point energy correlator on tracks \cite{talk1,talk2,talk3}. Second, in the perturbative region, the scaling behavior of energy correlators is slightly modified by the logarithmic running of the track functions, and these modifications can be captured systematically in perturbation theory.  Interestingly, this modification is significantly weaker than naively expected due to coincidences in the non-perturbative values of the track functions.

 We also studied the leading non-perturbative corrections to the track-based energy correlators. We argued that the leading non-perturbative power correction for the projected energy correlators on tracks can be obtained from that on all hadrons by multiplication by a product of track function moments. This relation is important, since it ensures that the leading non-perturbative correction cancels in the ratio of $N$-point/2-point projected energy correlators, even if the measurement is made on tracks. We believe that this will have important applications for measurements of $\alpha_s$ from jet substructure at the LHC.

Our results are a significant step towards precision jet substructure phenomenology at the LHC, exploiting the full angular resolution of modern tracking systems. While here we have focused on the projected correlators, it will be important to extend our results to the full angle-dependent correlators to enable the study of non-gaussianities \cite{Chen:2022swd} on tracks. It would also be interesting to understand in more detail the structure of power corrections to the track function formalism.

The ability to perform precision calculations of observables on tracks significantly expands the scope of the jet substructure program. Recent theory developments strongly motivates the precise measurement of the moments of the track functions, and their associated renormalization group flows, which we hope will be achieved in the near future. 
  
%%%%%%%%%%%%%%%%%%%%%%%%%%%%%%%%%%%%%%%%%%%%
\acknowledgments
%%%%%%%%%%%%%%%%%%%%%%%%%%%%%%%%%%%%%%%%%%%%
We thank Hao Chen for many useful discussions and collaboration on related topics. We thank Jingjing Pan for useful discussions and help with experimental references. We thank Kyle Lee for helpful discussions. Y.L.~and H.X.Z.~are supported by the National Natural Science Foundation of China under contract No.~11975200.
M.J.~is supported by NWO projectruimte 680-91-122. 
I.M.~is supported by start up funds from Yale University.
W.W.~is supported by the D-ITP consortium, a program of NWO that is funded by the Dutch Ministry of Education, Culture and Science (OCW). Y.L. would also like to thank Nikhef and UvA for their hospitality while much of this work was performed.

\bibliographystyle{JHEP}
\bibliography{EEC_ref.bib}

\end{document}